\documentclass[prl,superscriptaddress,twocolumn,longbibliography]{revtex4-2}

\usepackage{dcolumn,array}
\usepackage{bm}
\usepackage{subfigure}
\usepackage{amsmath,amsfonts,amssymb,graphics,graphicx,epsfig,color,times,natbib}
\usepackage{tikz}
\usepackage{pgfplots}
\usepackage{verbatim}
\usepackage{url}
\usepackage[unicode=true,
 bookmarks=true,bookmarksnumbered=false,bookmarksopen=false,
 breaklinks=false,pdfborder={0 0 1},backref=false,colorlinks=true]
 {hyperref}
\hypersetup{
 linkcolor=magenta, urlcolor=blue, citecolor=blue, pdfstartview={FitH}, hyperfootnotes=false, unicode=true}

\date{\today}

\begin{document}
\title{Unsupervised Learning of Non-Hermitian Topological Phases}
\author{Li-Wei Yu}
 \affiliation{Center for Quantum Information, IIIS, Tsinghua University, Beijing 100084, People's Republic of China}
\author{Dong-Ling Deng}
\email{dldeng@tsinghua.edu.cn}
 \affiliation{Center for Quantum Information, IIIS, Tsinghua University, Beijing 100084, People's Republic of China}
\affiliation{Shanghai Qi Zhi Institute, 41th Floor, AI Tower, No. 701 Yunjin Road, Xuhui District, Shanghai 200232, China}

\date{\today}

%%%%%%%%%%%%%%%%%%%%%%%%%%%%%%%%%%%%%%%%%%%%%%%%%%%%%%%%%%%%%%%%%%%

\begin{abstract}
Non-Hermitian topological phases bear a number of exotic properties, such as the non-Hermitian skin effect and the breakdown of conventional bulk-boundary correspondence. In this paper, we introduce an unsupervised machine learning approach to classify non-Hermitian topological phases based on diffusion maps, which are widely used in manifold learning. We find that the non-Hermitian skin effect will pose a notable obstacle, rendering the straightforward extension of unsupervised learning approaches to topological phases for Hermitian systems ineffective in clustering non-Hermitian topological phases.  Through theoretical analysis and numerical simulations of two prototypical models, we show that this difficulty can be circumvented by choosing the ``on-site'' elements of the projective matrix as the input data. Our results provide a valuable guidance for future studies on learning non-Hermitian topological phases in an unsupervised fashion, both in theory and experiment. 

\end{abstract}

\maketitle

\begin{figure}
\centering
\includegraphics[width=\linewidth]{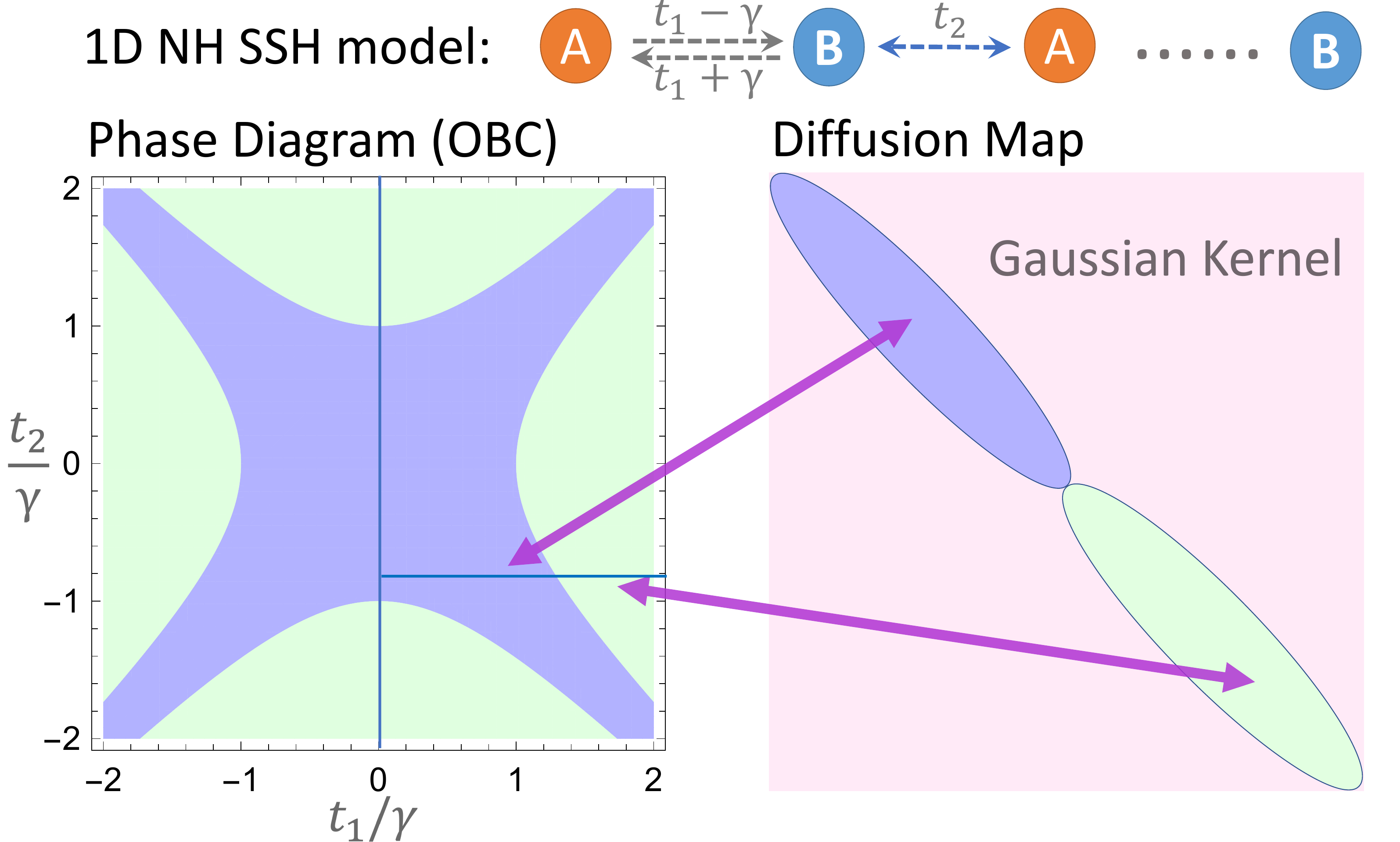}
\caption{A schematic illustration of the 1D non-Hermitian Su-Schrieffer-Heeger (NH SSH)  model, and the unsupervised learning of its harboring topological phases based on the diffusion map. Theoretically, with open boundary condition (OBC) this model entails two distinct phases with phase diagram shown in the lower-left subfigure. Samples from the same phase correspond to the same diagonal blocks of the Gaussian kernel matrix and the boundary between two blocks indicates the topological phase transition point.
}
\label{fig:nonhermitian}
\end{figure}

Non-Hermiticity arises naturally in a wide range of scenarios \cite{moiseyev2011non,RevModPhys.88.035002, ashida2020non},  such as photonic systems with loss and gain \cite{feng2017non, el2018non, miri2019exceptional, ozdemir2019parity,Ozawa2019Topological}, open quantum systems \cite{Rotter2009non-Hermitian,Zhen2015Spawning,Diehl2011Topology,Verstraete2009Quantum}, and quasiparticles with finite lifetimes \cite{kozii2017non,PhysRevB.97.041203,PhysRevLett.121.026403,zhou2018observation,PhysRevB.98.035141}. Recently, the study of non-Hermitian topological phases has attracted tremendous attentions \cite{PhysRevLett.118.045701,PhysRevLett.121.026808,PhysRevB.98.245130,PhysRevLett.123.206404,PhysRevLett.116.133903, PhysRevB.99.081103, PhysRevB.98.035141,PhysRevB.97.115453,PhysRevB.99.201103,PhysRevLett.118.040401,PhysRevA.97.052115,kawabata2019topological,PhysRevX.8.031079,PhysRevX.9.041015,PhysRevLett.120.146402,PhysRevLett.123.066404,PhysRevB.100.054105,PhysRevLett.120.146601,PhysRevLett.124.236403,PhysRevB.99.041406,yoshida2019symmetry,PhysRevB.99.081102,Okuma2020Topological,li2020critical,PhysRevLett.121.086803,PhysRevLett.121.136802,PhysRevLett.123.246801,PhysRevLett.124.186402,bessho2020topological,zhou2020non,PhysRevResearch.2.023235,PhysRevB.101.205417,li2019homotopical,Liu2019Second,Deng2019Non-Bloch, Xiao2020non,PhysRevLett.115.040402,poli2015selective, weimann2017topologically,chen2017exceptional,Zhou1009,PhysRevLett.124.046401,cerjan2019experimental,Bandreseaar4005,PhysRevLett.124.250402}.  Exciting progresses have been made in both theory \cite{PhysRevLett.118.045701,PhysRevLett.121.026808,PhysRevB.98.245130,PhysRevLett.123.206404,PhysRevLett.116.133903, PhysRevB.99.081103, PhysRevB.98.035141,PhysRevB.97.115453,PhysRevB.99.201103,PhysRevLett.118.040401,PhysRevA.97.052115,kawabata2019topological,PhysRevX.8.031079,PhysRevX.9.041015,PhysRevLett.120.146402,PhysRevLett.123.066404,PhysRevB.100.054105,PhysRevLett.120.146601,PhysRevLett.124.236403,PhysRevB.99.041406,yoshida2019symmetry,PhysRevB.99.081102,Okuma2020Topological,li2020critical,PhysRevLett.121.086803,PhysRevLett.121.136802,PhysRevLett.123.246801,PhysRevLett.124.186402,bessho2020topological,zhou2020non,PhysRevResearch.2.023235,PhysRevB.101.205417,li2019homotopical,Liu2019Second,Deng2019Non-Bloch} and experiment \cite{Xiao2020non,PhysRevLett.115.040402,poli2015selective, weimann2017topologically,chen2017exceptional,Zhou1009,PhysRevLett.124.046401,cerjan2019experimental,Bandreseaar4005,PhysRevLett.124.250402}. One of the prominent phenomena of non-Hermitian systems is the so-called non-Hermitian skin effect (NHSE) \cite{PhysRevLett.121.086803,PhysRevLett.121.136802,Okuma2020Topological,li2020critical}, where the majority of the eigenstates of a non-Hermitian operator are exponentially localized at boundaries.   This leads to the breakdown of the conventional bulk-boundary correspondence (a guiding principle  for topological phases of Hermitian systems) and calls for the non-Bloch band theory based on the generalized Brillouin zone \cite{PhysRevLett.121.086803,PhysRevLett.121.136802,PhysRevLett.123.066404,Liu2019Second,Deng2019Non-Bloch}. The NHSE has been observed in recent experiments \cite{Helbig2020Generalized,Xiao2020non,Ghatak2019Observation}, and its physical implications and consequences are still under active studies at the current stage \cite{PhysRevLett.121.136802,Liu2019Second,PhysRevB.99.081103,zirnstein2019bulk,Wang2019Non-Hermitian,Jiang2019Interplay,PhysRevB.99.201103,Lee2019Hybrid,Edvardsson2019Non-Hermitian,Borgnia2020Non-Hermitian,Ezawa2019Non-Hermitian,Yang2019Non-Hermitian}. Here, we introduce an unsupervised machine learning  approach based on diffusion maps to clustering non-Hermitian topological phases, with a focus on these exhibiting NHSE  that are drastically distinct from their Hermitian counterparts (see Fig.~\ref{fig:nonhermitian} for a pictorial illustration). 

Machine learning techniques \cite{goodfellow2016deep,Jordan2015Machine,Lecun2015Deep} are exquisitely tailored to identify hidden patterns in complex data and their applications to physics have recently been invoked in various contexts \cite{Dunjko2018,Sarma2019Machine,RevModPhys.91.045002}, ranging from black hole detection \cite{Pasquato2016Detecting},  gravitational lenses \cite{Hezaveh2017Fast} and wave analysis \cite{Rahul2013Application,Abbott2016Observation}, and quantum nonlocality detection \cite{Deng2017MachineBN}, to  glassy dynamics \cite{Schoenholz2016Structural} and material design \cite{Kalinin2015Big}, etc. Within the vein of learning different phases of matter and phase transitions, a number of different approaches have been proposed, with some been demonstrated in recent experiments. In particular, for learning topological phases both supervised \cite{PhysRevLett.118.216401,PhysRevB.96.245119,PhysRevB.97.205110,PhysRevLett.120.066401,PhysRevB.102.054107,narayan2020machine,zhang2020machine,PhysRevLett.122.210503}  and unsupervised \cite{rodriguez2019identifying,PhysRevLett.124.226401,che2020topological,PhysRevLett.124.185501,lidiak2020unsupervised,fukushima2019featuring,PhysRevE.99.062107,PhysRevResearch.2.013354,alexandrou2020critical,greplova2020unsupervised,arnold2020interpretable,Kottmann2020Unsupervised} methods have been introduced, despite the fact that topological phases are typically more difficult to learn than conventional symmetry-breaking ones due to their lack of local order parameters \cite{Beach2018Machine}. Supervised methods require prior labeling of the data samples, whereas  unsupervised learning can detect and classify topological phases from unlabeled raw data,  without  {\it a priori} knowledge of the underlying topological mechanism. Therefore, to some extent unsupervised approaches are more powerful and practical in detecting and identifying new topological phases. An intriguing unsupervised approach  is based on diffusion maps \cite{Coifman7426,Coifman7432,coifman2006diffusion}, which naturally implements the notion of continuous deformation (homotopy) and thus is particularly suitable for classifying topological objects. Along this line, notable works have demonstrated that  diffusion maps are strikingly effective in clustering topological orders in the Ising gauge theory \cite{rodriguez2019identifying}, symmetry protected topological phases \cite{PhysRevLett.124.226401,che2020topological},  valence-bond solid \cite{lidiak2020unsupervised}, and topological phononics \cite{PhysRevLett.124.185501}. Nevertheless, most of these existing works focus on Hermitian systems and learning of non-Hermitian topological phases, especially for these with NHSE, remains largely unexplored.

In this paper, we use diffusion maps to cluster non-Hermitian topological phases in an unsupervised fashion. Through theoretical analysis, we find that the NHSE gives rise to unwanted singularities that result in vanishing diffusion probabilities even for data samples from the same topological phase, thus making the direct application of the diffusion map method  impotent for classifying non-Hermitian topological phases with skin effect. To overcome this obstacle, we propose to use only the ``on-site'' elements $\{P_{iA,iB}| i\in[1,N]\}$ of the projective matrices $P$, which are typically used for defining topological indices for non-Hermitian systems, as input data and show, through concrete examples involving one-dimensional (1D) non-Hermitian Su-Schrieffer-Heeger (SSH) and 2D Qi-Wu-Zhang (QWZ) models, that the diffusion map method is indeed  capable of classifying non-Hermitian topological phases without supervision, even for these with the NHSE if we use the adjusted input data.

\textit{Theoretical analysis.}---Suppose we are given a set of input data $\mathbf{D}=\{\mathbf{x}^{(1)},\mathbf{x}^{(2)},\cdots, \mathbf{x}^{(L)} \}$ coming from different non-Hermitian topological phases. Our goal is to classify these samples topologically, namely, determine the number of different topological phases and for each sample identify which category it belongs to.  To measure the local similarity between samples $\mathbf{x}^{(l)}$ and $\mathbf{x}^{(l')}$, we use the Gaussian kernel function with variance controlled by the parameter  $\epsilon$ ($0<\epsilon\ll 1$):  
%\begin{equation}\label{GK_L1}
%\mathcal{K}_{l,l'} = \exp\left(-\frac{\|{\bf x}^{(l)}-{\bf x}^{(l')}\|_{\mathbb{L}_1}^2}{2\epsilon N^2}\right),
%\end{equation}
\begin{equation}\label{GK_L1}
\mathcal{K}_{l,l'} = \exp\left(-\|{\bf x}^{(l)}-{\bf x}^{(l')}\|_{\mathbb{L}_1}^2/(2\epsilon N^2)\right),
\end{equation}
where $\|{\bf x}^{(l)}-{\bf x}^{(l')}\|_{\mathbb{L}_1}$ denotes the Taxicab $\mathbb{L}_1$-norm distance, i.e. $\|\vec{A}\|_{\mathbb{L}_1}=\sum_{i}|A_{i}|$. $N$ denotes the number of unit cells of the Hamiltonian. The one-step diffusion probability $\mathcal{P}_{l,l'}$ from sample $\mathbf{x}^{(l)}$ to $\mathbf{x}^{(l')}$ is defined as: $\mathcal{P}_{l,l'}=\frac{\mathcal{K}_{l,l'}}{\sum_{l'}\mathcal{K}_{l,l'}}$. After $2t$ steps, the diffusion distance between $\mathbf{x}^{(j)}$ and $\mathbf{x}^{(j')}$ is $D_t(j,j')=\sum_k \frac{(\mathcal{P}^t_{j,k}-\mathcal{P}^t_{j',k})^2}{\sum_l \mathcal{K}_{k,l}}=\sum_k \lambda_k^{2t}[(\psi_k)_j-(\psi_k)_{j'}]^2$, where $\{\psi_k\}$ are the right eigenvectors of $\mathcal{P}$ and $\{\lambda_k\}$ denotes their corresponding eigenvalues. From $D_t$, it is clear that in the long-time  limit $t\rightarrow \infty$, only the few components with largest $|\lambda|\approx 1$ will dominate, and these few components can be used for dimensional reduction and clustering non-Hermintian topological phases. 

The success of the diffusion map method relies crucially on the input data samples. For learning topological phases in an unsupervised way, two types of data samples have been considered in previous works \cite{rodriguez2019identifying,PhysRevLett.124.226401,che2020topological,PhysRevLett.124.185501,lidiak2020unsupervised}: bulk Hamiltonian vectors (or equivalently bulk states) in the momentum space and full projective matrices in real space. For Hermitian systems, these two types of data should lead to the same classification due to the bulk-boundary correspondence.  However, for non-Hermitian systems the phase transition points might be boundary sensitive and the conventional bulk-boundary correspondence may not hold. As a result,  the choice of input data becomes subtle, especially for these topological phases with NHSE.  In particular, in the following discussion we  will show that the data set of full projective matrices cannot be employed for clustering non-Hermitian topological phases with NHSE via the diffusion map method.

For simplicity and concreteness, we consider spinless non-Hermitian topological models with both periodic and open boundary conditions (abbreviated as PBC and OBC, respectively).  To apply the first-order perturbation theory, here we only focus on the lattice models with discrete eigenvalues. 

We start with a general two-band model with PBC in the momentum space: 
%\begin{equation}
$\hat{H} = \vec{\mathbf{d}}\cdot\vec{\bf \sigma} = d_x\sigma_x+d_y\sigma_y + d_z\sigma_z$,
%\end{equation}
where $\sigma_{x,y,z}$ denotes the usual Pauli matrices and $d_{x,y,z}$ is complex for non-Hermitian systems. As discussed in previous papers \cite{rodriguez2019identifying,PhysRevLett.124.226401,che2020topological}, we may choose the input data sample to be  ${\bf x}^{(l)} = \{\hat{\bf d}^{(l)}(\vec{k})| \vec{k}\in \text{BZ}\}$ with $\hat{\bf d} = \frac{ \vec{\mathbf{d}}}{\sqrt{d_x^2+d_y^2+d_z^2}}$ and BZ denoting the first Brillouin zone. By varying the model parameters $\vec{t} = (t_1,t_2,...)$ contained implicitly in $\hat{\bf d}$, one obtains the input data set $\{{\bf x}^{(l)}\}$. 
We now analyze how the diffusion map method can classify these samples into different categories. From the definition of the one-step diffusion probability, it is clear that the dominant terms are these corresponding to the nearest samples labeled by $(l,\,l+\delta l)$ \footnote{Note that for the  diffusion map approach, two samples far from each other may have considerable diffusion probability with the assistance of  symmetry, as discussed in Ref.\cite{PhysRevLett.124.226401}.  However,  in this work we do not assume that the model Hamiltonian has certain symmetry and thus the effect of symmetries will not be discussed for simplicity.}. 
%The  $\mathbb{L}_1$-norm distance  between $\mathbf{x}^{(l)}$ and $\mathbf{x}^{(l+\delta l)} $ can be approximately recast to $\|{\bf x}^{(l)}-{\bf x}^{(l+\delta l)}\|_{\mathbb{L}_1} \approx \sum_i\sum_{\alpha} \|\nabla_{\vec{t}}\,(\hat{d}_\alpha^{(l)}(k_i))\|_{\mathbb{L}_1}\cdot \delta{\vec{t}}$,
% and consequently the Gaussian kernel is reduced to $\mathcal{K}_{l,l+\delta l}\approx \exp\left(-\left(\sum_{i,\alpha} \|\nabla_{\vec{t}}\,(\hat{d}_\alpha^{(l)}(k_i))\|_{\mathbb{L}_1}\cdot \delta{\vec{t}}\,\right)^2 /2\epsilon N^2\right)$. Here, 
%where the summation of $i$ and $\alpha$  runs over all discretized momentum points and $\alpha=x,y,z$ respectively,  and $\|\nabla_{\vec{t}} \,d\|_{\mathbb{L}_1} = \left(\|\partial_{t_1} d\|_{\mathbb{L}_1}, \|\partial_{t_2} d\|_{\mathbb{L}_1},\dots \|\partial_{t_n} d\|_{\mathbb{L}_1}\right)$. 
By adjusting the hyper parameter $\epsilon$, one can show that
%As long as the term $\frac{1}{N}\sum_{i,\alpha} \|\nabla_{\vec{t}}\,(\hat{d}_\alpha^{(l)}(k_i))\|_{\mathbb{L}_1}$ is finite, the constant $\delta {\vec{t}\,}^{2}/\epsilon$ can always be adjusted so as to keep $\mathcal{K}_{l,l+\delta l}\approx 1$.  Hence
 the connectivity between $\mathbf{x}^{(l)}$ and $\mathbf{x}^{(l+\delta l)}$ depends on the derivability of the unit vector $\hat{\bf d}^{(l)} = { \vec{\mathbf{d}}^{(l)}}/{E_+^{(l)}}$ on $\vec{t}$ for all $k_i \in [-\pi,\pi]$, where $E_+^{(l)}=\sqrt{d_x^2+d_y^2+d_z^2}$. The gap closure points $E_{\pm}^{(l)}=0$ typically break the derivability. Hence, two samples $(l,l+\delta l)$ separated by the gap closure point should result in  $\mathcal{K}_{l,l+\delta l}\approx 0$, i.e. vanishing one-step diffusion probability between them. Combined with the approximation that only the nearest samples dominantly contribute to the diffusion, the  gap closure points divide the  kernel matrix into blocks, and samples corresponding the same block are connected via diffusion, hence belonging to the same topological phase. Thus for those models with phase transition occurring at the gap closure points,  the data samples in the same phase should be clustered into the same category via the diffusion map.

With OBC, things become tricky due to the possible existence of the NHSE. In this case, a straightforward choice for the input data would be the projective matrices defined as  $P = \sum_{\text{Re}[E_m]<0} |m_R\rangle\langle m_L|$ \cite{PhysRevLett.123.246801}, 
%\begin{equation}
%P = \sum_{\text{Re}[E_m]<0} |m_R\rangle\langle m_L|,
%\end{equation} 
where $ |m_R\rangle$ and $\langle m_L|$ are the right and left eigenstates (with the corresponding eigenenergy $E_m$ and $E_m^*$ respectively) of the non-Hermitian Hamiltonian in real space. We mention that $P$ can be used to define the topological invariants for non-Hermitian systems in real space \cite{PhysRevLett.123.246801}. The $\mathbb{L}_1$-norm of a matrix is defined by $\|P\|_{\mathbb{L}_1}=\sum_{i,j}|P_{ij}|$. %Different choices of orthonormal bases lead to different $\mathbb{L}_1$-norm values, but have no influence on the singularity we concern.  For consistency, here we choose the free fermionic bases $\{C_{1,A},C_{1,B},C_{2,A},..., C_{N,B}\}$. 
As was mentioned above,  the nearest samples dominantly contribute to the diffusion probability with proper hyper parameter $\epsilon$.
%In the first-order perturbation theory,  $\delta P = P ^ {(l+\delta l)} -P^{(l)} \approx 
%\sum_{\text{Re}[E_{m}]<0, n\neq m}\frac{\langle n_L|\delta \hat{H}|m_R\rangle |n_R\rangle\langle m_L| +(n\leftrightarrow m) }{E_m-E_n} $. 
Then the Gaussian kernel can be expressed as %$\mathcal{K}_{l,l+\delta l}= e^{-\tfrac{\|\delta P\|^2_{\mathbb{L}_1}}{2\epsilon N^2}}= e^{ -\tfrac{(\|\nabla_{\vec{t}} P\|_{\mathbb{L}_1}\cdot \delta \vec{t}\,)^2}{2\epsilon N^2}}$.
$\mathcal{K}_{l,l+\delta l}= \exp\left( -\tfrac{\|\delta P\|^2_{\mathbb{L}_1}}{2\epsilon N^2}\right)= \exp\left( -\frac{(\|\nabla_{\vec{t}} P\|_{\mathbb{L}_1}\cdot \delta \vec{t}\,)^2}{2\epsilon N^2}\right)$. 
The singularity of $\|\nabla_{\vec{t}} P\|_{\mathbb{L}_1}$ is crucial to the kernel values. For convenient illustration, let us take the  one-dimensional (1D) non-Hermitian SSH model with OBC (in Fig. ~\ref{fig:nonhermitian}) as an example,
%: $\hat{H}_o^{1D}=\sum_i(t_1+\gamma)C^\dag_{i,A}C_{i,B}+(t_1-\gamma)C^\dag_{i,B}C_{i,A} +t_2C^\dag_{i,B}C_{i+1,A}+t_2C^\dag_{i+1,A}C_{i,B}$,
%\begin{equation*}
%\begin{aligned}
%\hat{H}_o^{1D}&=\sum_i(t_1+\gamma)C^\dag_{i,A}C_{i,B}+(t_1-\gamma)C^\dag_{i,B}C_{i,A} \\
%&+t_2C^\dag_{i,B}C_{i+1,A}+t_2C^\dag_{i+1,A}C_{i,B},
%\end{aligned}
%\end{equation*}
where the fermion annihilation (creation) operators on the A and B sublattices are denoted by $C_A$ ($C_A^\dagger$) and $C_B$ ($C_B^\dagger$), respectively; $t_1$, $t_2$, and $\gamma$ are model parameters characterizing the hopping strength. A sketch of the phase diagram is shown in Fig.~\ref{GK_L1}. We consider the parameter region $t_1>|\gamma|$, where the non-Hermitian SSH Hamiltonian  $\hat{H}_o^{1D}$   in orthonormal bases can be transformed into a Hermitian matrix $\bar{H}_o^{1D}$ in non-orthonormal bases \cite{PhysRevLett.121.086803}: $\bar{H}_o^{1D} = \Gamma^{-1}\hat{H}_o^{1D}\Gamma $ with $\Gamma = \text{diag} (1, r, r, r^2,r^2,\cdots, r^{N-1}, r^N)$, and $r=\sqrt{|(t_1-\gamma)/(t_1+\gamma)|}$.
Then $\partial_{t_1} P$ reduces to \cite{USMLNonHTopSupp}: 
\begin{equation*}%\label{nabla_p_nh}
\partial_{t_1} P=  \sum_{\substack{{\rm Re}[E_{m}]<0 \\ {n\neq m}}}\frac{\langle n|\Gamma\partial_{t_1} \hat{H}\Gamma^{-1}|m\rangle \Gamma^{-1}|n\rangle\langle m|\Gamma+ \, (n\leftrightarrow m) }{E_m-E_n}.
\end{equation*}

Direct calculations show that all  terms ${\langle n|\Gamma\partial_{t_1} \hat{H}_o^{1D}\Gamma^{-1}|m\rangle}/{N}$ are finite \cite{USMLNonHTopSupp}.  Suppose the system is initially in the topologically trivial phase, then when the system approaches the phase transition point $|t_1|=\sqrt{t_2^2+\gamma^2}$, the two levels $E_{\pm1}$ approach zero and become the zero modes, which will be eliminated in the $P$-matrix after passing through the transition point.  The only singularity ($|t_1|\neq|\gamma|$) of $1/(E_m-E_n)$ occurs  at the gap closing point, where the two nearest levels $|E_{-1}-E_{1}|\rightarrow0$. So far, everything looks similar to the Hermitian SSH model. However, the term $\Gamma^{-1}|n\rangle\langle m|\Gamma$, which basically gives rise to the NHSE, involves matrix elements with values $r^{\pm(i-j+1)}$ that are singular when $|i-j|\rightarrow \infty$. This singularity is boundary condition sensitive and unique to non-Hermitian systems. It persists for a wide range of parameter space, independent of the phase transitions. As a consequence, it will render the diffusion map method invalid and should be removed from the input data. To address this problem, we propose to  use the ``on-site'' part of the $P$-matrix elements  as the raw input data: $\{P_{iA,iB}|i\in[1,N]\}$, where $r^{\pm(i-j+1)}$ factors cancel out  and no singularity shows up. Such ``on-site'' extraction should be suitable to those non-Hermitian models where the NHSE exhibits an exponential function of lattice site. With this adjusted $P$-matrix as the input data, the kernel values $\mathcal{K}_{l,l+\delta l}\approx 1$ for all the data samples from the same phase, but two samples $\{l,\,l+\delta l\}$ crossing the phase transition point have no connectivity  $\mathcal{K}_{l,l+\delta l}\approx 0$.  This restores  the capability of the diffusion map method in classifying non-Hermitian topological phases with NHSE in the unsupervised fashion. 

The above discussion explains in theory why and when the diffusion map method can be applied to classify non-Hermitian topological phases, and how to overcome the obstacles due to the NHSE. To illustrate how this method works in practice, in the following we apply it to a couple of concrete examples, including the cases with and without the NHSE.

\begin{figure}[tp!]
\centering
\includegraphics[width=\linewidth]{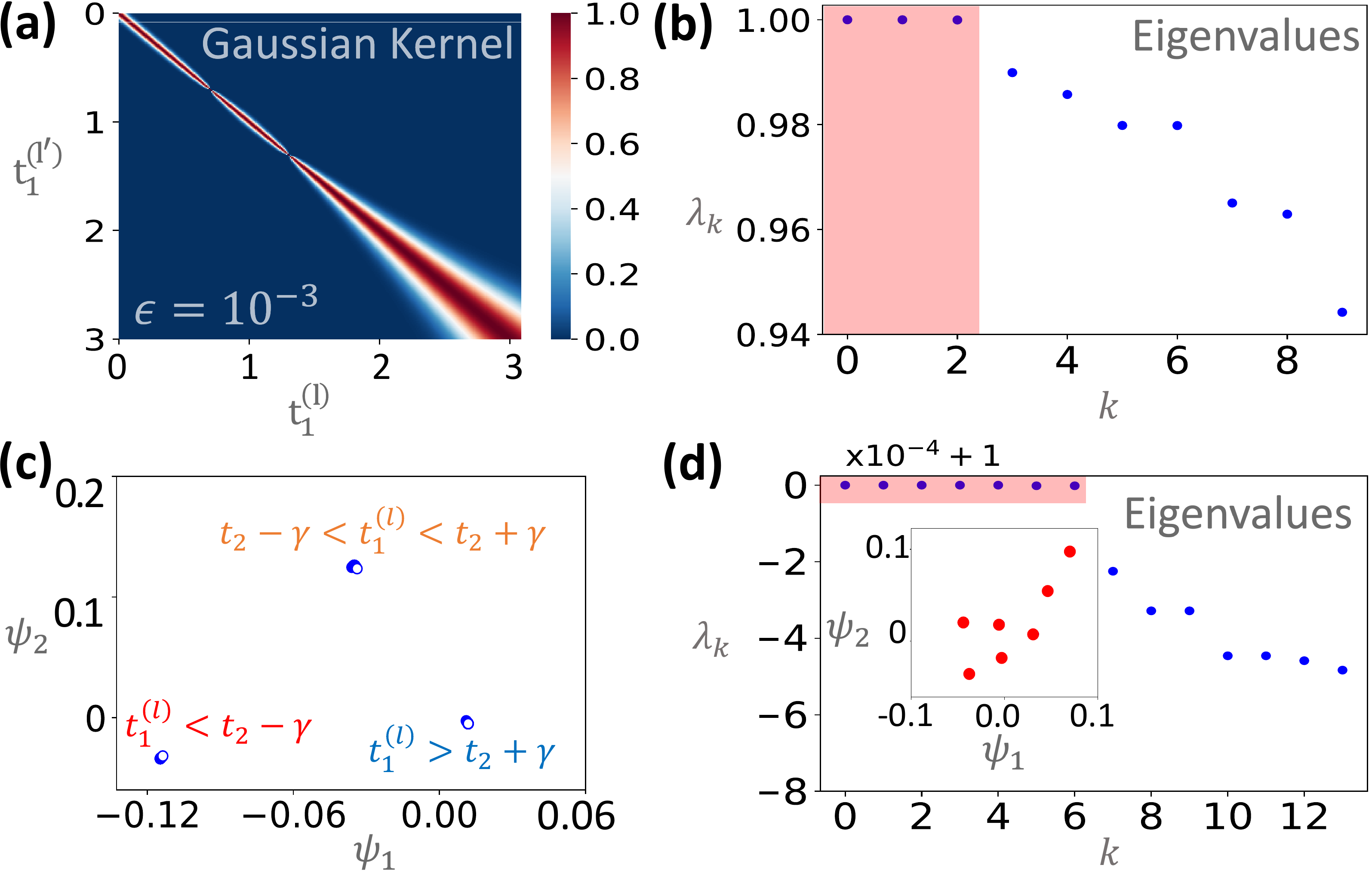}
\caption{Numerical results of unsupervised learning without NHSE for the 1D non-Hermitian SSH [subfigures (a-c)] and 2D QWZ model [subfigure (d)]. (a) Heatmap for Gaussian kernel value distribution between samples with varying $t_1$. (b) Eigenvalues of the one-step diffusion matrix $\mathcal{P}$. (c) Scatter diagram of eigenvectors $\{\psi_1,\, \psi_2\}$ with the corresponding eigenvalues $\lambda_{1,2}\approx1$, where the samples are clustered into three topological phases.  (d) For the  QWZ model, the input samples are classified into seven categories \cite{USMLNonHTopSupp}.}
\label{NHSSH: PBC}
\end{figure}

\textit{Unsupervised learning without NHSE.}---The first example we consider is the non-Hermitian SSH model with PBC. In the momentum space, this model reads $H_p(k)=\vec{\bf d}\cdot \vec{\bf \sigma} = (t_1+t_2\cos k)\sigma_x + (t_2\sin k+ i\gamma)\sigma_y$. The input data set can be chosen as $\{{\bf x}^{(l)}| {\bf x}^{(l)} = \{\hat{\bf d}(k_i),| k_i =\frac{2i-N-2}{N}\pi, \, i\in[1,N] \}\}$ with varying $t_1$, while fixing $t_2$ and $\gamma$.  Our numerical results are shown in Fig.~\ref{NHSSH: PBC} (a-c). From Fig.~\ref{NHSSH: PBC} (a), the kernel matrix $\mathcal{K}$ is separated into three blocks, which correspond to the three largest eigenvalues $\lambda_{0,1,2}\approx1$ of the one-step diffusion  matrix $\mathcal{P}$, as shown in Fig.~\ref{NHSSH: PBC} (b). As a result, the input samples are classified into three different topological phases. This is also clearly indicated in Fig.~\ref{NHSSH: PBC} (c), where we show the scatter diagram of eigenvectors $\{\psi_1,\, \psi_2\}$ corresponding to $\lambda_{1,2}$. In addition, the phase boundaries can also be obtained from Fig.~\ref{NHSSH: PBC} (a), which match exactly with the theoretical one  \cite{USMLNonHTopSupp}. 
 %In addition, the phase boundaries obtained from Fig.~\ref{NHSSH: PBC} (a) match exactly with the theoretical prediction  \cite{USMLNonHTopSupp}. 

Another example we consider is the 2D non-Hermitian QWZ model with PBC:  $\hat{H}_p^{2D}({\bf k})= (v_x\sin k_x +i\gamma_x)\sigma_x+(v_y\sin k_y +i\gamma_y)\sigma_y+(M-0.5\cos k_x-0.5\cos k_y)\sigma_z$. The phase boundaries occur at $M^{(\nu)}_\pm =(2-\nu)\pm \sqrt{\gamma_x^2+\gamma_y^2}$, $\nu=1,2,3$.
The Hamiltonian in the region $M\in (-\infty, M^{(3)}_-) \cup   (M^{(3)}_+,M^{(2)}_-)\cup  (M^{(2)}_+,M^{(1)}_-)\cup  (M^{(1)}_+,+\infty)$ is gapped and the topological Chern number is well-defined. However, in  regions $M\in (M_{-}^{(\nu)},\,M_{+}^{(\nu)})$ the Hamiltonian is gapless and the topological indices are not well-defined. These gapless regions lead to  singularities and hence no diffusion probability among samples from such regions. 
 Nevertheless, one can still utilize the diffusion map method to locate the phase boundaries $M^{(1,2,3)}_{\pm}$ based on the ``effective'' kernel matrix, which is the average of a set of kernel matrices with different lattice sizes (different discrete $\vec{k}$ configurations) as the input data. By such a construction, the diffusion map method still works even with  input samples in gapless regions. We choose the bulk Hamiltonian as the input samples by varying the parameter $M$ and our numerical results are shown in Fig.~\ref{NHSSH: PBC} (d). From this figure, the samples are classified into seven categories. In addition, careful examinations of the heatmap for the  kernel matrix yield that the phase boundaries identified by the diffusion map method coincide with the theoretical ones \cite{USMLNonHTopSupp}. 
%For details, see Supplementary Materials (SM), Sec. VI.} 

%Nevertheless, one can still utilize the diffusion map method to locate the phase boundaries $M^{(1,2,3)}_{\pm}$, as long as all of the discrete $(k_x,k_y)$ points for the raw data evade  the gapless momentum for  $M\in (M_{-}^{(1,2,3)},\,M_{+}^{(1,2,3)})$ by choosing judiciously proper parameter values so as to avoid the singularities.   We choose the bulk Hamiltonian as the raw data samples $\{{\bf x}^{(l)}\}$ by varying the parameter $M$ and our numerical results are shown in Fig. ~\ref{NHSSH: PBC} (d). From this figure, the samples are classified into seven different groups. In addition, careful examinations of the heatmap for the Gaussian kernel yield that the phase boundaries identified by the diffusion map method coincide with the theoretical ones \cite{USMLNonHTopSupp}. 

\begin{figure}[tp!]
\centering
\includegraphics[width=\linewidth]{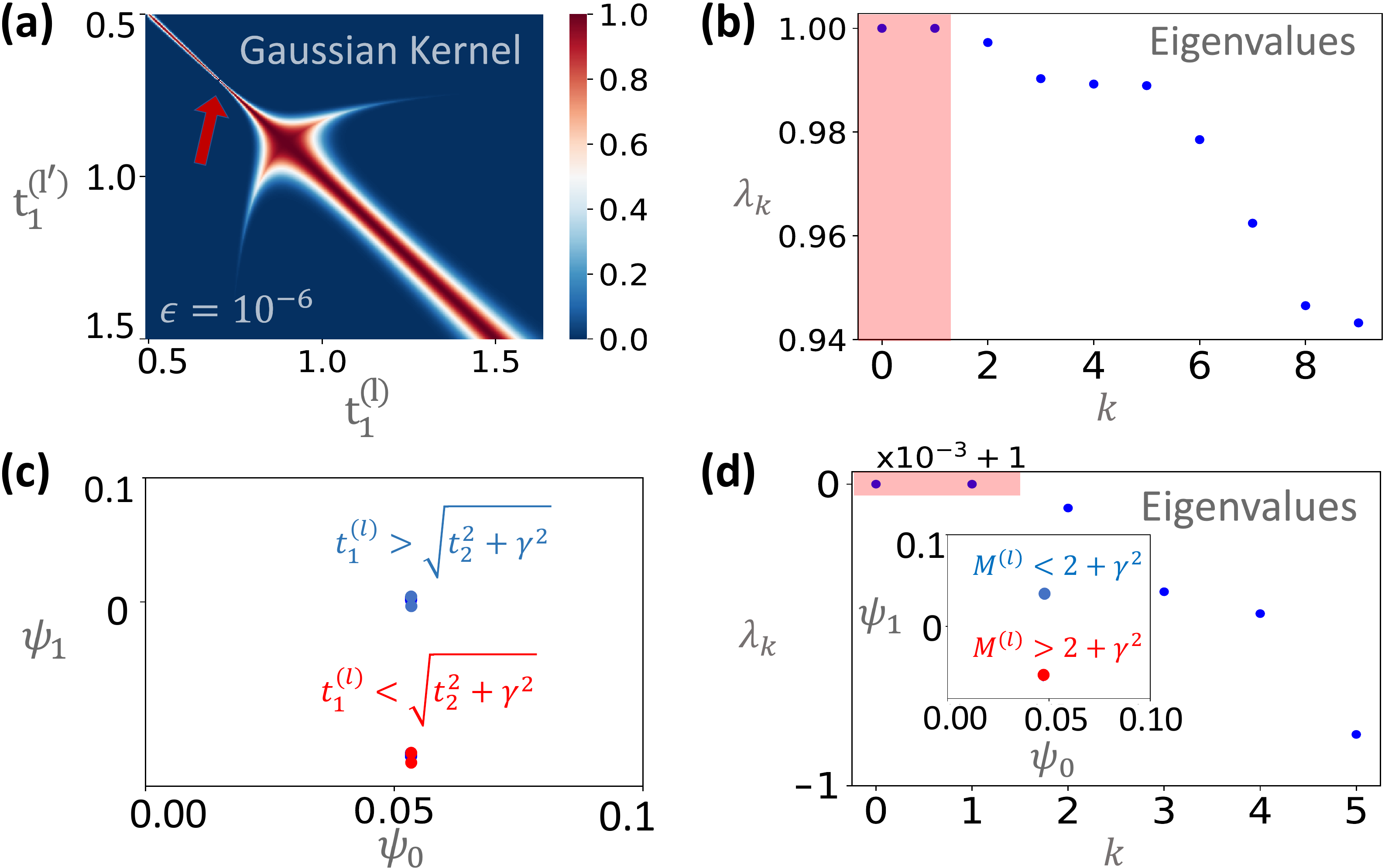}
%\vspace{-0.8cm}
\caption{Numerical results of unsupervised learning with NHSE. (a-c) show respectively the heatmap of the Gaussian kernel, the eigenvalues of the one-step diffusion matrix, and the scatter diagram of eigenvectors for the non-Hermitian SSH model. The red arrow in (a) indicates the phase transition point $t_1\approx 0.6993$. The input data samples are clustered into two (rather than three) topological categories, in sharp contrast to the case of learning without NHSE shown in Fig. \ref{NHSSH: PBC}.  (d) The eigenvalues of the one-step diffusion matrix for the non-Hermitian QWZ model, with the inset showing the scatter diagram of eigenvectors corresponding to the largest two eigenvalues.   The input samples are classified into two (rather than seven) categories \cite{USMLNonHTopSupp}. 
 }
\label{NHSSH: OBC}
\end{figure}

\textit{Unsupervised learning with NHSE.}---The above numerical results show clearly that the diffusion map method is indeed capable of classifying non-Hermitian topological phases  without NHSE. Yet, as discussed in the beginning the presence of the NHSE may handicap the performance of this approach.  A possible way out of this is to choose different input data. Now, we turn to this case and examine the applicability of the diffusion map method for clustering non-Hermitian topological phases with NHSE. We still focus on the non-Hermitian SSH and QWZ models, but with OBCs this time.  

We start with the 1D non-Hermitian SSH model \cite{PhysRevLett.121.086803}. With OBC, the phase boundaries are theoretically predicted to be  $t_1 = \pm \sqrt{t_2^2+\gamma^2}$. Topological non-trivial phase with ground state degeneracy occurs in the interval $|t_1|< \sqrt{t_2^2+\gamma^2}$.  We numerically diagonalize the real space Hamiltonian $\hat{H}^{1D}_{o}$ for given system sizes and parameters, and choose partial elements of the projective matrices $\{P^{(l)}_{iA,iB}|i\in[1,N]\}$ as the raw input data $\{{\bf x}^{(l)}\}$.
Our numerical results are shown in Fig.~\ref{NHSSH: OBC} (a-c). From this figure, the input samples are classified into two categories and the learnt phase transition point occurs at  $t_1\approx 0.6993$, which is consistent with theoretically predicted values of $t_1=\sqrt{t_2^2+\gamma^2}\approx 0.7211$ \cite{USMLNonHTopSupp}.   The small discrepancy is mainly due to the fact that only partial information of the projective matrices are used and the finite size effect in our numerical simulations (see \cite{USMLNonHTopSupp} for details). 
%We have made a more detailed discussions about such discrepancy in supplementary materials. }

For the 2D non-Hermitian QWZ model with OBC,  it has been shown from  both analytical and numerical aspects that one topological phase boundary occurs approximately at
$ M=M_0 = t_x + t_y + \frac{t_x\gamma_x^2}{2v_x^2} + \frac{t_y\gamma_y^2}{2v_y^2}$
for small $\frac{\gamma_{x(y)}}{v_{x(y)}}$ \cite{PhysRevLett.121.136802}.  When $M < M_0$, the corresponding Chern number in generalized Brillouin zone for the valence band ($\text{Re}(E)<0$) is $C=1$. Whereas for $M > M_0$, the Chern number is $C=0$. With partial elements of the projective matrices as input data, the diffusion map method works as well and
our numerical results are shown in Fig.~\ref{NHSSH: OBC} (d). It is clear from this figure that the input samples are classified into two distinct categories and the learnt phase boundary occurs at $M\approx 2.1731$, which matches the theoretical value $M=2.2$ with a desirable accuracy \cite{USMLNonHTopSupp}. 

%{\color{red}Usually the NHSE can be regarded as multiplying an exponential  factor on each site, and extracting the ``on-site'' part of  $P$-matrix as the input data should be an efficient approach for unsupervised learning most of the unknown systems with NHSE.}
%{\color{red}Based on the above two successful examples with NHSE, we believe that the principle of extracting feature should be suitable for those models with NHSE in diffusion map. }%Because the NHSE can be regarded as multiplying an exponential factor on each site, the ``on-site'' extraction of right and left eigenvectors in projective matrix just removes the singularity caused by NHSE.}

 We mention that non-Hermitian systems are extremely sensitive to boundary conditions.   A change of boundary conditions could alter drastically both the eigenspectra and eigenstates \cite{Bergholtz2019Exceptional}. This is in sharp contrast to the case of Hermitian systems. From the unsupervised learning results shown above, it is also clear that different boundary conditions lead to completely different clustering results of the input samples. For instance, with PBC the data samples of the QWZ model are clustered into seven groups via the diffusion map method, whereas with OBC one obtains only two categories.  Owing to the NHSE, the relevant features of input data could be dramatically suppressed and consequently are harder to extract. Here we note that the obstacle induced by the NHSE may also exist in other machine learning approaches, e.g., the CNN-based ones \cite{zhang2020machine}, and the solution we provide here should carry over to these algorithms as well \cite{USMLNonHTopSupp}. 
We also remark that finding the appropriate topological invariants based on the projective matrix is a highly non-trivial task. In fact, a number of works have been reported recently in the literature to deal with this problem \cite{PhysRevLett.123.246801,PhysRevX.9.041015, Zhang2020Correspondence,Okuma2020Topological}. Although the diffusion map approach still requires full diagonalization of Hamiltonians in real space, it does not rely on any {\it a priori} information about the underlying topological invariants. As a result, this approach is also applicable to non-Hermitian systems whose characteristic topological invariants have not yet been discovered. In addition, the requirement of the costly diagonalization might be circumvented by replacing the projective matrix with observables (e.g., correlation functions)  that can be measured in experiment.  We leave this interesting and important problem for future studies.

\textit{Discussion and conclusion.}---Symmetries play a crucial role in the study of topological phases and, analogous to the Hermitian case \cite{Kitaev2009Periodic}, a periodic table for non-Hermitian Hamiltonians has also been established from the $K$ theory perspective recently \cite{PhysRevX.8.031079}. Yet, incorporating symmetry constraints into unsupervised learning approaches to topological phases is highly nontrivial \cite{PhysRevLett.124.226401}. In the future, it would be interesting and desirable to extend our results to symmetry protected or enriched non-Hermitian topological phases, especially those predicted in the periodic table.  In addition, non-Hermitian topological phases for interacting systems remain elusive and we expect that unsupervised learning  will provide valuable wisdom in studying such phases as well.

In summary, we have introduced an unsupervised machine learning approach to classify non-Hermitian topological phases based on diffusion maps. We show that the NHSE can result in a  critical handicap for the straightforward extension of the unsupervised method of learning Hermitian topological phases to the non-Hermitian case. 
% {\color{red}
%Besides, a recent paper \cite{zhang2020machine} points out that the convolutional neural network (CNN) has not yet been capable of learning models with NHSE,  which indicates that the obstacle of NHSE may also appear in other unsupervised (at least CNN-based) machine learning methods \cite{fukushima2019featuring,PhysRevE.99.062107,PhysRevResearch.2.013354,alexandrou2020critical,greplova2020unsupervised,arnold2020interpretable}.} 
%{\color{red}Besides, a recent paper \cite{zhang2020machine} points out that the convolutional neural network (CNN) has not yet been capable of learning models with NHSE,  which indicates that the obstacle of NHSE may also appear in other unsupervised (at least CNN-based) machine learning methods \cite{fukushima2019featuring,PhysRevE.99.062107,PhysRevResearch.2.013354,alexandrou2020critical,greplova2020unsupervised,arnold2020interpretable} besides the diffusion map.}
Through theoretical analysis and numerical simulations, we have demonstrated that this obstacle can be avoided by appropriately choosing the input data, such as the ``on-site'' elements of the projective matrices.  Our results reveal a new consequence of the NHSE and would benefit future studies across non-Hermitian topological phases and machine learning.

We acknowledge Mathias S. Scheurer for sharing his previous programming code  on diffusion map with us. This work was supported by the start-up fund from Tsinghua University (Grant No. 53330300320) and the National Natural Science Foundation of China (Grant No. 11905108).  DLD also would like to acknowledge additional support from the Shanghai Qi Zhi Institute.

\bibliography{DMNH_Man_R2}

%apsrev4-2.bst 2019-01-14 (MD) hand-edited version of apsrev4-1.bst
%Control: key (0)
%Control: author (8) initials jnrlst
%Control: editor formatted (1) identically to author
%Control: production of article title (0) allowed
%Control: page (0) single
%Control: year (1) truncated
%Control: production of eprint (0) enabled
\begin{thebibliography}{115}%
\makeatletter
\providecommand \@ifxundefined [1]{%
 \@ifx{#1\undefined}
}%
\providecommand \@ifnum [1]{%
 \ifnum #1\expandafter \@firstoftwo
 \else \expandafter \@secondoftwo
 \fi
}%
\providecommand \@ifx [1]{%
 \ifx #1\expandafter \@firstoftwo
 \else \expandafter \@secondoftwo
 \fi
}%
\providecommand \natexlab [1]{#1}%
\providecommand \enquote  [1]{``#1''}%
\providecommand \bibnamefont  [1]{#1}%
\providecommand \bibfnamefont [1]{#1}%
\providecommand \citenamefont [1]{#1}%
\providecommand \href@noop [0]{\@secondoftwo}%
\providecommand \href [0]{\begingroup \@sanitize@url \@href}%
\providecommand \@href[1]{\@@startlink{#1}\@@href}%
\providecommand \@@href[1]{\endgroup#1\@@endlink}%
\providecommand \@sanitize@url [0]{\catcode `\\12\catcode `\$12\catcode
  `\&12\catcode `\#12\catcode `\^12\catcode `\_12\catcode `\%12\relax}%
\providecommand \@@startlink[1]{}%
\providecommand \@@endlink[0]{}%
\providecommand \url  [0]{\begingroup\@sanitize@url \@url }%
\providecommand \@url [1]{\endgroup\@href {#1}{\urlprefix }}%
\providecommand \urlprefix  [0]{URL }%
\providecommand \Eprint [0]{\href }%
\providecommand \doibase [0]{https://doi.org/}%
\providecommand \selectlanguage [0]{\@gobble}%
\providecommand \bibinfo  [0]{\@secondoftwo}%
\providecommand \bibfield  [0]{\@secondoftwo}%
\providecommand \translation [1]{[#1]}%
\providecommand \BibitemOpen [0]{}%
\providecommand \bibitemStop [0]{}%
\providecommand \bibitemNoStop [0]{.\EOS\space}%
\providecommand \EOS [0]{\spacefactor3000\relax}%
\providecommand \BibitemShut  [1]{\csname bibitem#1\endcsname}%
\let\auto@bib@innerbib\@empty
%</preamble>
\bibitem [{\citenamefont {Moiseyev}(2011)}]{moiseyev2011non}%
  \BibitemOpen
  \bibfield  {author} {\bibinfo {author} {\bibfnamefont {N.}~\bibnamefont
  {Moiseyev}},\ }\href@noop {} {\emph {\bibinfo {title} {Non-Hermitian quantum
  mechanics}}}\ (\bibinfo  {publisher} {Cambridge University Press},\ \bibinfo
  {year} {2011})\BibitemShut {NoStop}%
\bibitem [{\citenamefont {Konotop}\ \emph {et~al.}(2016)\citenamefont
  {Konotop}, \citenamefont {Yang},\ and\ \citenamefont
  {Zezyulin}}]{RevModPhys.88.035002}%
  \BibitemOpen
  \bibfield  {author} {\bibinfo {author} {\bibfnamefont {V.~V.}\ \bibnamefont
  {Konotop}}, \bibinfo {author} {\bibfnamefont {J.}~\bibnamefont {Yang}},\ and\
  \bibinfo {author} {\bibfnamefont {D.~A.}\ \bibnamefont {Zezyulin}},\
  }\bibfield  {title} {\bibinfo {title} {{Nonlinear waves in
  $\mathcal{PT}$-symmetric systems}},\ }\href
  {https://doi.org/10.1103/RevModPhys.88.035002} {\bibfield  {journal}
  {\bibinfo  {journal} {Rev. Mod. Phys.}\ }\textbf {\bibinfo {volume} {88}},\
  \bibinfo {pages} {035002} (\bibinfo {year} {2016})}\BibitemShut {NoStop}%
\bibitem [{\citenamefont {Ashida}\ \emph {et~al.}(2020)\citenamefont {Ashida},
  \citenamefont {Gong},\ and\ \citenamefont {Ueda}}]{ashida2020non}%
  \BibitemOpen
  \bibfield  {author} {\bibinfo {author} {\bibfnamefont {Y.}~\bibnamefont
  {Ashida}}, \bibinfo {author} {\bibfnamefont {Z.}~\bibnamefont {Gong}},\ and\
  \bibinfo {author} {\bibfnamefont {M.}~\bibnamefont {Ueda}},\ }\bibfield
  {title} {\bibinfo {title} {{Non-Hermitian Physics}},\ }\href
  {https://arxiv.org/pdf/2006.01837.pdf} {\bibfield  {journal} {\bibinfo
  {journal} {arXiv:2006.01837}\ } (\bibinfo {year} {2020})}\BibitemShut
  {NoStop}%
\bibitem [{\citenamefont {Feng}\ \emph {et~al.}(2017)\citenamefont {Feng},
  \citenamefont {El-Ganainy},\ and\ \citenamefont {Ge}}]{feng2017non}%
  \BibitemOpen
  \bibfield  {author} {\bibinfo {author} {\bibfnamefont {L.}~\bibnamefont
  {Feng}}, \bibinfo {author} {\bibfnamefont {R.}~\bibnamefont {El-Ganainy}},\
  and\ \bibinfo {author} {\bibfnamefont {L.}~\bibnamefont {Ge}},\ }\bibfield
  {title} {\bibinfo {title} {{Non-Hermitian photonics based on parity--time
  symmetry}},\ }\href
  {https://www.nature.com/articles/s41566-017-0031-1/briefing/signup/}
  {\bibfield  {journal} {\bibinfo  {journal} {Nat. Photon.}\ }\textbf {\bibinfo
  {volume} {11}},\ \bibinfo {pages} {752} (\bibinfo {year} {2017})}\BibitemShut
  {NoStop}%
\bibitem [{\citenamefont {El-Ganainy}\ \emph {et~al.}(2018)\citenamefont
  {El-Ganainy}, \citenamefont {Makris}, \citenamefont {Khajavikhan},
  \citenamefont {Musslimani}, \citenamefont {Rotter},\ and\ \citenamefont
  {Christodoulides}}]{el2018non}%
  \BibitemOpen
  \bibfield  {author} {\bibinfo {author} {\bibfnamefont {R.}~\bibnamefont
  {El-Ganainy}}, \bibinfo {author} {\bibfnamefont {K.~G.}\ \bibnamefont
  {Makris}}, \bibinfo {author} {\bibfnamefont {M.}~\bibnamefont {Khajavikhan}},
  \bibinfo {author} {\bibfnamefont {Z.~H.}\ \bibnamefont {Musslimani}},
  \bibinfo {author} {\bibfnamefont {S.}~\bibnamefont {Rotter}},\ and\ \bibinfo
  {author} {\bibfnamefont {D.~N.}\ \bibnamefont {Christodoulides}},\ }\bibfield
   {title} {\bibinfo {title} {{Non-Hermitian physics and PT symmetry}},\ }\href
  {https://www.nature.com/articles/nphys4323} {\bibfield  {journal} {\bibinfo
  {journal} {Nat. Phys.}\ }\textbf {\bibinfo {volume} {14}},\ \bibinfo {pages}
  {11} (\bibinfo {year} {2018})}\BibitemShut {NoStop}%
\bibitem [{\citenamefont {Miri}\ and\ \citenamefont
  {Alu}(2019)}]{miri2019exceptional}%
  \BibitemOpen
  \bibfield  {author} {\bibinfo {author} {\bibfnamefont {M.-A.}\ \bibnamefont
  {Miri}}\ and\ \bibinfo {author} {\bibfnamefont {A.}~\bibnamefont {Alu}},\
  }\bibfield  {title} {\bibinfo {title} {{Exceptional points in optics and
  photonics}},\ }\href
  {https://science.sciencemag.org/content/363/6422/eaar7709.abstract}
  {\bibfield  {journal} {\bibinfo  {journal} {Science}\ }\textbf {\bibinfo
  {volume} {363}},\ \bibinfo {pages} {eaar7709} (\bibinfo {year}
  {2019})}\BibitemShut {NoStop}%
\bibitem [{\citenamefont {{\"O}zdemir}\ \emph {et~al.}(2019)\citenamefont
  {{\"O}zdemir}, \citenamefont {Rotter}, \citenamefont {Nori},\ and\
  \citenamefont {Yang}}]{ozdemir2019parity}%
  \BibitemOpen
  \bibfield  {author} {\bibinfo {author} {\bibfnamefont {{\c{S}}.}~\bibnamefont
  {{\"O}zdemir}}, \bibinfo {author} {\bibfnamefont {S.}~\bibnamefont {Rotter}},
  \bibinfo {author} {\bibfnamefont {F.}~\bibnamefont {Nori}},\ and\ \bibinfo
  {author} {\bibfnamefont {L.}~\bibnamefont {Yang}},\ }\bibfield  {title}
  {\bibinfo {title} {{Parity--time symmetry and exceptional points in
  photonics}},\ }\href {https://www.nature.com/articles/s41563-019-0304-9}
  {\bibfield  {journal} {\bibinfo  {journal} {Nat. Mater.}\ }\textbf {\bibinfo
  {volume} {18}},\ \bibinfo {pages} {783} (\bibinfo {year} {2019})}\BibitemShut
  {NoStop}%
\bibitem [{\citenamefont {Ozawa}\ \emph {et~al.}(2019)\citenamefont {Ozawa},
  \citenamefont {Price}, \citenamefont {Amo}, \citenamefont {Goldman},
  \citenamefont {Hafezi}, \citenamefont {Lu}, \citenamefont {Rechtsman},
  \citenamefont {Schuster}, \citenamefont {Simon}, \citenamefont {Zilberberg},\
  and\ \citenamefont {Carusotto}}]{Ozawa2019Topological}%
  \BibitemOpen
  \bibfield  {author} {\bibinfo {author} {\bibfnamefont {T.}~\bibnamefont
  {Ozawa}}, \bibinfo {author} {\bibfnamefont {H.~M.}\ \bibnamefont {Price}},
  \bibinfo {author} {\bibfnamefont {A.}~\bibnamefont {Amo}}, \bibinfo {author}
  {\bibfnamefont {N.}~\bibnamefont {Goldman}}, \bibinfo {author} {\bibfnamefont
  {M.}~\bibnamefont {Hafezi}}, \bibinfo {author} {\bibfnamefont
  {L.}~\bibnamefont {Lu}}, \bibinfo {author} {\bibfnamefont {M.~C.}\
  \bibnamefont {Rechtsman}}, \bibinfo {author} {\bibfnamefont {D.}~\bibnamefont
  {Schuster}}, \bibinfo {author} {\bibfnamefont {J.}~\bibnamefont {Simon}},
  \bibinfo {author} {\bibfnamefont {O.}~\bibnamefont {Zilberberg}},\ and\
  \bibinfo {author} {\bibfnamefont {I.}~\bibnamefont {Carusotto}},\ }\bibfield
  {title} {\bibinfo {title} {Topological photonics},\ }\href
  {https://doi.org/10.1103/RevModPhys.91.015006} {\bibfield  {journal}
  {\bibinfo  {journal} {Rev. Mod. Phys.}\ }\textbf {\bibinfo {volume} {91}},\
  \bibinfo {pages} {015006} (\bibinfo {year} {2019})}\BibitemShut {NoStop}%
\bibitem [{\citenamefont {Rotter}(2009)}]{Rotter2009non-Hermitian}%
  \BibitemOpen
  \bibfield  {author} {\bibinfo {author} {\bibfnamefont {I.}~\bibnamefont
  {Rotter}},\ }\bibfield  {title} {\bibinfo {title} {{A non-Hermitian Hamilton
  operator and the physics of open quantum systems}},\ }\href
  {https://iopscience.iop.org/article/10.1088/1751-8113/42/15/153001/pdf}
  {\bibfield  {journal} {\bibinfo  {journal} {J. Phys. A: Math. Theor.}\
  }\textbf {\bibinfo {volume} {42}},\ \bibinfo {pages} {153001} (\bibinfo
  {year} {2009})}\BibitemShut {NoStop}%
\bibitem [{\citenamefont {Zhen}\ \emph {et~al.}(2015)\citenamefont {Zhen},
  \citenamefont {Hsu}, \citenamefont {Igarashi}, \citenamefont {Lu},
  \citenamefont {Kaminer}, \citenamefont {Pick}, \citenamefont {Chua},
  \citenamefont {Joannopoulos},\ and\ \citenamefont
  {Solja{\v{c}}i{\'c}}}]{Zhen2015Spawning}%
  \BibitemOpen
  \bibfield  {author} {\bibinfo {author} {\bibfnamefont {B.}~\bibnamefont
  {Zhen}}, \bibinfo {author} {\bibfnamefont {C.~W.}\ \bibnamefont {Hsu}},
  \bibinfo {author} {\bibfnamefont {Y.}~\bibnamefont {Igarashi}}, \bibinfo
  {author} {\bibfnamefont {L.}~\bibnamefont {Lu}}, \bibinfo {author}
  {\bibfnamefont {I.}~\bibnamefont {Kaminer}}, \bibinfo {author} {\bibfnamefont
  {A.}~\bibnamefont {Pick}}, \bibinfo {author} {\bibfnamefont {S.-L.}\
  \bibnamefont {Chua}}, \bibinfo {author} {\bibfnamefont {J.~D.}\ \bibnamefont
  {Joannopoulos}},\ and\ \bibinfo {author} {\bibfnamefont {M.}~\bibnamefont
  {Solja{\v{c}}i{\'c}}},\ }\bibfield  {title} {\bibinfo {title} {{Spawning
  rings of exceptional points out of Dirac cones}},\ }\href
  {https://www.nature.com/articles/nature14889} {\bibfield  {journal} {\bibinfo
   {journal} {Nature}\ }\textbf {\bibinfo {volume} {525}},\ \bibinfo {pages}
  {354} (\bibinfo {year} {2015})}\BibitemShut {NoStop}%
\bibitem [{\citenamefont {Diehl}\ \emph {et~al.}(2011)\citenamefont {Diehl},
  \citenamefont {Rico}, \citenamefont {Baranov},\ and\ \citenamefont
  {Zoller}}]{Diehl2011Topology}%
  \BibitemOpen
  \bibfield  {author} {\bibinfo {author} {\bibfnamefont {S.}~\bibnamefont
  {Diehl}}, \bibinfo {author} {\bibfnamefont {E.}~\bibnamefont {Rico}},
  \bibinfo {author} {\bibfnamefont {M.~A.}\ \bibnamefont {Baranov}},\ and\
  \bibinfo {author} {\bibfnamefont {P.}~\bibnamefont {Zoller}},\ }\bibfield
  {title} {\bibinfo {title} {Topology by dissipation in atomic quantum wires},\
  }\href {https://www.nature.com/articles/nphys2106?page=5} {\bibfield
  {journal} {\bibinfo  {journal} {Nat. Phys.}\ }\textbf {\bibinfo {volume}
  {7}},\ \bibinfo {pages} {971} (\bibinfo {year} {2011})}\BibitemShut {NoStop}%
\bibitem [{\citenamefont {Verstraete}\ \emph {et~al.}(2009)\citenamefont
  {Verstraete}, \citenamefont {Wolf},\ and\ \citenamefont
  {Cirac}}]{Verstraete2009Quantum}%
  \BibitemOpen
  \bibfield  {author} {\bibinfo {author} {\bibfnamefont {F.}~\bibnamefont
  {Verstraete}}, \bibinfo {author} {\bibfnamefont {M.~M.}\ \bibnamefont
  {Wolf}},\ and\ \bibinfo {author} {\bibfnamefont {J.~I.}\ \bibnamefont
  {Cirac}},\ }\bibfield  {title} {\bibinfo {title} {Quantum computation and
  quantum-state engineering driven by dissipation},\ }\href
  {https://www.nature.com/articles/nphys1342/} {\bibfield  {journal} {\bibinfo
  {journal} {Nat. Phys.}\ }\textbf {\bibinfo {volume} {5}},\ \bibinfo {pages}
  {633} (\bibinfo {year} {2009})}\BibitemShut {NoStop}%
\bibitem [{\citenamefont {Kozii}\ and\ \citenamefont
  {Fu}(2017)}]{kozii2017non}%
  \BibitemOpen
  \bibfield  {author} {\bibinfo {author} {\bibfnamefont {V.}~\bibnamefont
  {Kozii}}\ and\ \bibinfo {author} {\bibfnamefont {L.}~\bibnamefont {Fu}},\
  }\bibfield  {title} {\bibinfo {title} {{Non-Hermitian topological theory of
  finite-lifetime quasiparticles: prediction of bulk Fermi arc due to
  exceptional point}},\ }\href {https://arxiv.org/pdf/1708.05841.pdf}
  {\bibfield  {journal} {\bibinfo  {journal} {arXiv:1708.05841}\ } (\bibinfo
  {year} {2017})}\BibitemShut {NoStop}%
\bibitem [{\citenamefont {Zyuzin}\ and\ \citenamefont
  {Zyuzin}(2018)}]{PhysRevB.97.041203}%
  \BibitemOpen
  \bibfield  {author} {\bibinfo {author} {\bibfnamefont {A.~A.}\ \bibnamefont
  {Zyuzin}}\ and\ \bibinfo {author} {\bibfnamefont {A.~Y.}\ \bibnamefont
  {Zyuzin}},\ }\bibfield  {title} {\bibinfo {title} {{Flat band in
  disorder-driven non-Hermitian Weyl semimetals}},\ }\href
  {https://doi.org/10.1103/PhysRevB.97.041203} {\bibfield  {journal} {\bibinfo
  {journal} {Phys. Rev. B}\ }\textbf {\bibinfo {volume} {97}},\ \bibinfo
  {pages} {041203(R)} (\bibinfo {year} {2018})}\BibitemShut {NoStop}%
\bibitem [{\citenamefont {Shen}\ and\ \citenamefont
  {Fu}(2018)}]{PhysRevLett.121.026403}%
  \BibitemOpen
  \bibfield  {author} {\bibinfo {author} {\bibfnamefont {H.}~\bibnamefont
  {Shen}}\ and\ \bibinfo {author} {\bibfnamefont {L.}~\bibnamefont {Fu}},\
  }\bibfield  {title} {\bibinfo {title} {{Quantum Oscillation from In-Gap
  States and a Non-Hermitian Landau Level Problem}},\ }\href
  {https://doi.org/10.1103/PhysRevLett.121.026403} {\bibfield  {journal}
  {\bibinfo  {journal} {Phys. Rev. Lett.}\ }\textbf {\bibinfo {volume} {121}},\
  \bibinfo {pages} {026403} (\bibinfo {year} {2018})}\BibitemShut {NoStop}%
\bibitem [{\citenamefont {Zhou}\ \emph
  {et~al.}(2018{\natexlab{a}})\citenamefont {Zhou}, \citenamefont {Peng},
  \citenamefont {Yoon}, \citenamefont {Hsu}, \citenamefont {Nelson},
  \citenamefont {Fu}, \citenamefont {Joannopoulos}, \citenamefont
  {Solja{\v{c}}i{\'c}},\ and\ \citenamefont {Zhen}}]{zhou2018observation}%
  \BibitemOpen
  \bibfield  {author} {\bibinfo {author} {\bibfnamefont {H.}~\bibnamefont
  {Zhou}}, \bibinfo {author} {\bibfnamefont {C.}~\bibnamefont {Peng}}, \bibinfo
  {author} {\bibfnamefont {Y.}~\bibnamefont {Yoon}}, \bibinfo {author}
  {\bibfnamefont {C.~W.}\ \bibnamefont {Hsu}}, \bibinfo {author} {\bibfnamefont
  {K.~A.}\ \bibnamefont {Nelson}}, \bibinfo {author} {\bibfnamefont
  {L.}~\bibnamefont {Fu}}, \bibinfo {author} {\bibfnamefont {J.~D.}\
  \bibnamefont {Joannopoulos}}, \bibinfo {author} {\bibfnamefont
  {M.}~\bibnamefont {Solja{\v{c}}i{\'c}}},\ and\ \bibinfo {author}
  {\bibfnamefont {B.}~\bibnamefont {Zhen}},\ }\bibfield  {title} {\bibinfo
  {title} {{Observation of bulk Fermi arc and polarization half charge from
  paired exceptional points}},\ }\href
  {https://science.sciencemag.org/content/359/6379/1009.abstract} {\bibfield
  {journal} {\bibinfo  {journal} {Science}\ }\textbf {\bibinfo {volume}
  {359}},\ \bibinfo {pages} {1009} (\bibinfo {year}
  {2018}{\natexlab{a}})}\BibitemShut {NoStop}%
\bibitem [{\citenamefont {Yoshida}\ \emph {et~al.}(2018)\citenamefont
  {Yoshida}, \citenamefont {Peters},\ and\ \citenamefont
  {Kawakami}}]{PhysRevB.98.035141}%
  \BibitemOpen
  \bibfield  {author} {\bibinfo {author} {\bibfnamefont {T.}~\bibnamefont
  {Yoshida}}, \bibinfo {author} {\bibfnamefont {R.}~\bibnamefont {Peters}},\
  and\ \bibinfo {author} {\bibfnamefont {N.}~\bibnamefont {Kawakami}},\
  }\bibfield  {title} {\bibinfo {title} {{Non-Hermitian perspective of the band
  structure in heavy-fermion systems}},\ }\href
  {https://doi.org/10.1103/PhysRevB.98.035141} {\bibfield  {journal} {\bibinfo
  {journal} {Phys. Rev. B}\ }\textbf {\bibinfo {volume} {98}},\ \bibinfo
  {pages} {035141} (\bibinfo {year} {2018})}\BibitemShut {NoStop}%
\bibitem [{\citenamefont {Xu}\ \emph {et~al.}(2017)\citenamefont {Xu},
  \citenamefont {Wang},\ and\ \citenamefont {Duan}}]{PhysRevLett.118.045701}%
  \BibitemOpen
  \bibfield  {author} {\bibinfo {author} {\bibfnamefont {Y.}~\bibnamefont
  {Xu}}, \bibinfo {author} {\bibfnamefont {S.-T.}\ \bibnamefont {Wang}},\ and\
  \bibinfo {author} {\bibfnamefont {L.-M.}\ \bibnamefont {Duan}},\ }\bibfield
  {title} {\bibinfo {title} {{Weyl Exceptional Rings in a Three-Dimensional
  Dissipative Cold Atomic Gas}},\ }\href
  {https://doi.org/10.1103/PhysRevLett.118.045701} {\bibfield  {journal}
  {\bibinfo  {journal} {Phys. Rev. Lett.}\ }\textbf {\bibinfo {volume} {118}},\
  \bibinfo {pages} {045701} (\bibinfo {year} {2017})}\BibitemShut {NoStop}%
\bibitem [{\citenamefont {Kunst}\ \emph {et~al.}(2018)\citenamefont {Kunst},
  \citenamefont {Edvardsson}, \citenamefont {Budich},\ and\ \citenamefont
  {Bergholtz}}]{PhysRevLett.121.026808}%
  \BibitemOpen
  \bibfield  {author} {\bibinfo {author} {\bibfnamefont {F.~K.}\ \bibnamefont
  {Kunst}}, \bibinfo {author} {\bibfnamefont {E.}~\bibnamefont {Edvardsson}},
  \bibinfo {author} {\bibfnamefont {J.~C.}\ \bibnamefont {Budich}},\ and\
  \bibinfo {author} {\bibfnamefont {E.~J.}\ \bibnamefont {Bergholtz}},\
  }\bibfield  {title} {\bibinfo {title} {{Biorthogonal Bulk-Boundary
  Correspondence in Non-Hermitian Systems}},\ }\href
  {https://doi.org/10.1103/PhysRevLett.121.026808} {\bibfield  {journal}
  {\bibinfo  {journal} {Phys. Rev. Lett.}\ }\textbf {\bibinfo {volume} {121}},\
  \bibinfo {pages} {026808} (\bibinfo {year} {2018})}\BibitemShut {NoStop}%
\bibitem [{\citenamefont {Chen}\ and\ \citenamefont
  {Zhai}(2018)}]{PhysRevB.98.245130}%
  \BibitemOpen
  \bibfield  {author} {\bibinfo {author} {\bibfnamefont {Y.}~\bibnamefont
  {Chen}}\ and\ \bibinfo {author} {\bibfnamefont {H.}~\bibnamefont {Zhai}},\
  }\bibfield  {title} {\bibinfo {title} {{Hall conductance of a non-Hermitian
  Chern insulator}},\ }\href {https://doi.org/10.1103/PhysRevB.98.245130}
  {\bibfield  {journal} {\bibinfo  {journal} {Phys. Rev. B}\ }\textbf {\bibinfo
  {volume} {98}},\ \bibinfo {pages} {245130} (\bibinfo {year}
  {2018})}\BibitemShut {NoStop}%
\bibitem [{\citenamefont {Lee}\ \emph {et~al.}(2019{\natexlab{a}})\citenamefont
  {Lee}, \citenamefont {Ahn}, \citenamefont {Zhou},\ and\ \citenamefont
  {Vishwanath}}]{PhysRevLett.123.206404}%
  \BibitemOpen
  \bibfield  {author} {\bibinfo {author} {\bibfnamefont {J.~Y.}\ \bibnamefont
  {Lee}}, \bibinfo {author} {\bibfnamefont {J.}~\bibnamefont {Ahn}}, \bibinfo
  {author} {\bibfnamefont {H.}~\bibnamefont {Zhou}},\ and\ \bibinfo {author}
  {\bibfnamefont {A.}~\bibnamefont {Vishwanath}},\ }\bibfield  {title}
  {\bibinfo {title} {{Topological Correspondence between Hermitian and
  Non-Hermitian Systems: Anomalous Dynamics}},\ }\href
  {https://doi.org/10.1103/PhysRevLett.123.206404} {\bibfield  {journal}
  {\bibinfo  {journal} {Phys. Rev. Lett.}\ }\textbf {\bibinfo {volume} {123}},\
  \bibinfo {pages} {206404} (\bibinfo {year} {2019}{\natexlab{a}})}\BibitemShut
  {NoStop}%
\bibitem [{\citenamefont {Lee}(2016)}]{PhysRevLett.116.133903}%
  \BibitemOpen
  \bibfield  {author} {\bibinfo {author} {\bibfnamefont {T.~E.}\ \bibnamefont
  {Lee}},\ }\bibfield  {title} {\bibinfo {title} {{Anomalous Edge State in a
  Non-Hermitian Lattice}},\ }\href
  {https://doi.org/10.1103/PhysRevLett.116.133903} {\bibfield  {journal}
  {\bibinfo  {journal} {Phys. Rev. Lett.}\ }\textbf {\bibinfo {volume} {116}},\
  \bibinfo {pages} {133903} (\bibinfo {year} {2016})}\BibitemShut {NoStop}%
\bibitem [{\citenamefont {Jin}\ and\ \citenamefont
  {Song}(2019)}]{PhysRevB.99.081103}%
  \BibitemOpen
  \bibfield  {author} {\bibinfo {author} {\bibfnamefont {L.}~\bibnamefont
  {Jin}}\ and\ \bibinfo {author} {\bibfnamefont {Z.}~\bibnamefont {Song}},\
  }\bibfield  {title} {\bibinfo {title} {{Bulk-boundary correspondence in a
  non-Hermitian system in one dimension with chiral inversion symmetry}},\
  }\href {https://doi.org/10.1103/PhysRevB.99.081103} {\bibfield  {journal}
  {\bibinfo  {journal} {Phys. Rev. B}\ }\textbf {\bibinfo {volume} {99}},\
  \bibinfo {pages} {081103(R)} (\bibinfo {year} {2019})}\BibitemShut {NoStop}%
\bibitem [{\citenamefont {Carvalho}\ \emph {et~al.}(2018)\citenamefont
  {Carvalho}, \citenamefont {Garc\'{\i}a-Mart\'{\i}nez}, \citenamefont {Lado},\
  and\ \citenamefont {Fern\'andez-Rossier}}]{PhysRevB.97.115453}%
  \BibitemOpen
  \bibfield  {author} {\bibinfo {author} {\bibfnamefont {D.}~\bibnamefont
  {Carvalho}}, \bibinfo {author} {\bibfnamefont {N.~A.}\ \bibnamefont
  {Garc\'{\i}a-Mart\'{\i}nez}}, \bibinfo {author} {\bibfnamefont {J.~L.}\
  \bibnamefont {Lado}},\ and\ \bibinfo {author} {\bibfnamefont
  {J.}~\bibnamefont {Fern\'andez-Rossier}},\ }\bibfield  {title} {\bibinfo
  {title} {{Real-space mapping of topological invariants using artificial
  neural networks}},\ }\href {https://doi.org/10.1103/PhysRevB.97.115453}
  {\bibfield  {journal} {\bibinfo  {journal} {Phys. Rev. B}\ }\textbf {\bibinfo
  {volume} {97}},\ \bibinfo {pages} {115453} (\bibinfo {year}
  {2018})}\BibitemShut {NoStop}%
\bibitem [{\citenamefont {Lee}\ and\ \citenamefont
  {Thomale}(2019)}]{PhysRevB.99.201103}%
  \BibitemOpen
  \bibfield  {author} {\bibinfo {author} {\bibfnamefont {C.~H.}\ \bibnamefont
  {Lee}}\ and\ \bibinfo {author} {\bibfnamefont {R.}~\bibnamefont {Thomale}},\
  }\bibfield  {title} {\bibinfo {title} {{Anatomy of skin modes and topology in
  non-Hermitian systems}},\ }\href {https://doi.org/10.1103/PhysRevB.99.201103}
  {\bibfield  {journal} {\bibinfo  {journal} {Phys. Rev. B}\ }\textbf {\bibinfo
  {volume} {99}},\ \bibinfo {pages} {201103(R)} (\bibinfo {year}
  {2019})}\BibitemShut {NoStop}%
\bibitem [{\citenamefont {Leykam}\ \emph {et~al.}(2017)\citenamefont {Leykam},
  \citenamefont {Bliokh}, \citenamefont {Huang}, \citenamefont {Chong},\ and\
  \citenamefont {Nori}}]{PhysRevLett.118.040401}%
  \BibitemOpen
  \bibfield  {author} {\bibinfo {author} {\bibfnamefont {D.}~\bibnamefont
  {Leykam}}, \bibinfo {author} {\bibfnamefont {K.~Y.}\ \bibnamefont {Bliokh}},
  \bibinfo {author} {\bibfnamefont {C.}~\bibnamefont {Huang}}, \bibinfo
  {author} {\bibfnamefont {Y.~D.}\ \bibnamefont {Chong}},\ and\ \bibinfo
  {author} {\bibfnamefont {F.}~\bibnamefont {Nori}},\ }\bibfield  {title}
  {\bibinfo {title} {{Edge Modes, Degeneracies, and Topological Numbers in
  Non-Hermitian Systems}},\ }\href
  {https://doi.org/10.1103/PhysRevLett.118.040401} {\bibfield  {journal}
  {\bibinfo  {journal} {Phys. Rev. Lett.}\ }\textbf {\bibinfo {volume} {118}},\
  \bibinfo {pages} {040401} (\bibinfo {year} {2017})}\BibitemShut {NoStop}%
\bibitem [{\citenamefont {Yin}\ \emph {et~al.}(2018)\citenamefont {Yin},
  \citenamefont {Jiang}, \citenamefont {Li}, \citenamefont {L\"u},\ and\
  \citenamefont {Chen}}]{PhysRevA.97.052115}%
  \BibitemOpen
  \bibfield  {author} {\bibinfo {author} {\bibfnamefont {C.}~\bibnamefont
  {Yin}}, \bibinfo {author} {\bibfnamefont {H.}~\bibnamefont {Jiang}}, \bibinfo
  {author} {\bibfnamefont {L.}~\bibnamefont {Li}}, \bibinfo {author}
  {\bibfnamefont {R.}~\bibnamefont {L\"u}},\ and\ \bibinfo {author}
  {\bibfnamefont {S.}~\bibnamefont {Chen}},\ }\bibfield  {title} {\bibinfo
  {title} {{Geometrical meaning of winding number and its characterization of
  topological phases in one-dimensional chiral non-Hermitian systems}},\ }\href
  {https://doi.org/10.1103/PhysRevA.97.052115} {\bibfield  {journal} {\bibinfo
  {journal} {Phys. Rev. A}\ }\textbf {\bibinfo {volume} {97}},\ \bibinfo
  {pages} {052115} (\bibinfo {year} {2018})}\BibitemShut {NoStop}%
\bibitem [{\citenamefont {Kawabata}\ \emph
  {et~al.}(2019{\natexlab{a}})\citenamefont {Kawabata}, \citenamefont
  {Higashikawa}, \citenamefont {Gong}, \citenamefont {Ashida},\ and\
  \citenamefont {Ueda}}]{kawabata2019topological}%
  \BibitemOpen
  \bibfield  {author} {\bibinfo {author} {\bibfnamefont {K.}~\bibnamefont
  {Kawabata}}, \bibinfo {author} {\bibfnamefont {S.}~\bibnamefont
  {Higashikawa}}, \bibinfo {author} {\bibfnamefont {Z.}~\bibnamefont {Gong}},
  \bibinfo {author} {\bibfnamefont {Y.}~\bibnamefont {Ashida}},\ and\ \bibinfo
  {author} {\bibfnamefont {M.}~\bibnamefont {Ueda}},\ }\bibfield  {title}
  {\bibinfo {title} {{Topological unification of time-reversal and
  particle-hole symmetries in non-Hermitian physics}},\ }\href
  {https://www.nature.com/articles/s41467-018-08254-y#citeas} {\bibfield
  {journal} {\bibinfo  {journal} {Nat. Commun.}\ }\textbf {\bibinfo {volume}
  {10}},\ \bibinfo {pages} {1} (\bibinfo {year}
  {2019}{\natexlab{a}})}\BibitemShut {NoStop}%
\bibitem [{\citenamefont {Gong}\ \emph {et~al.}(2018)\citenamefont {Gong},
  \citenamefont {Ashida}, \citenamefont {Kawabata}, \citenamefont {Takasan},
  \citenamefont {Higashikawa},\ and\ \citenamefont {Ueda}}]{PhysRevX.8.031079}%
  \BibitemOpen
  \bibfield  {author} {\bibinfo {author} {\bibfnamefont {Z.}~\bibnamefont
  {Gong}}, \bibinfo {author} {\bibfnamefont {Y.}~\bibnamefont {Ashida}},
  \bibinfo {author} {\bibfnamefont {K.}~\bibnamefont {Kawabata}}, \bibinfo
  {author} {\bibfnamefont {K.}~\bibnamefont {Takasan}}, \bibinfo {author}
  {\bibfnamefont {S.}~\bibnamefont {Higashikawa}},\ and\ \bibinfo {author}
  {\bibfnamefont {M.}~\bibnamefont {Ueda}},\ }\bibfield  {title} {\bibinfo
  {title} {{Topological Phases of Non-Hermitian Systems}},\ }\href
  {https://doi.org/10.1103/PhysRevX.8.031079} {\bibfield  {journal} {\bibinfo
  {journal} {Phys. Rev. X}\ }\textbf {\bibinfo {volume} {8}},\ \bibinfo {pages}
  {031079} (\bibinfo {year} {2018})}\BibitemShut {NoStop}%
\bibitem [{\citenamefont {Kawabata}\ \emph
  {et~al.}(2019{\natexlab{b}})\citenamefont {Kawabata}, \citenamefont
  {Shiozaki}, \citenamefont {Ueda},\ and\ \citenamefont
  {Sato}}]{PhysRevX.9.041015}%
  \BibitemOpen
  \bibfield  {author} {\bibinfo {author} {\bibfnamefont {K.}~\bibnamefont
  {Kawabata}}, \bibinfo {author} {\bibfnamefont {K.}~\bibnamefont {Shiozaki}},
  \bibinfo {author} {\bibfnamefont {M.}~\bibnamefont {Ueda}},\ and\ \bibinfo
  {author} {\bibfnamefont {M.}~\bibnamefont {Sato}},\ }\bibfield  {title}
  {\bibinfo {title} {{Symmetry and Topology in Non-Hermitian Physics}},\ }\href
  {https://doi.org/10.1103/PhysRevX.9.041015} {\bibfield  {journal} {\bibinfo
  {journal} {Phys. Rev. X}\ }\textbf {\bibinfo {volume} {9}},\ \bibinfo {pages}
  {041015} (\bibinfo {year} {2019}{\natexlab{b}})}\BibitemShut {NoStop}%
\bibitem [{\citenamefont {Shen}\ \emph {et~al.}(2018)\citenamefont {Shen},
  \citenamefont {Zhen},\ and\ \citenamefont {Fu}}]{PhysRevLett.120.146402}%
  \BibitemOpen
  \bibfield  {author} {\bibinfo {author} {\bibfnamefont {H.}~\bibnamefont
  {Shen}}, \bibinfo {author} {\bibfnamefont {B.}~\bibnamefont {Zhen}},\ and\
  \bibinfo {author} {\bibfnamefont {L.}~\bibnamefont {Fu}},\ }\bibfield
  {title} {\bibinfo {title} {{Topological Band Theory for Non-Hermitian
  Hamiltonians}},\ }\href {https://doi.org/10.1103/PhysRevLett.120.146402}
  {\bibfield  {journal} {\bibinfo  {journal} {Phys. Rev. Lett.}\ }\textbf
  {\bibinfo {volume} {120}},\ \bibinfo {pages} {146402} (\bibinfo {year}
  {2018})}\BibitemShut {NoStop}%
\bibitem [{\citenamefont {Yokomizo}\ and\ \citenamefont
  {Murakami}(2019)}]{PhysRevLett.123.066404}%
  \BibitemOpen
  \bibfield  {author} {\bibinfo {author} {\bibfnamefont {K.}~\bibnamefont
  {Yokomizo}}\ and\ \bibinfo {author} {\bibfnamefont {S.}~\bibnamefont
  {Murakami}},\ }\bibfield  {title} {\bibinfo {title} {{Non-Bloch Band Theory
  of Non-Hermitian Systems}},\ }\href
  {https://doi.org/10.1103/PhysRevLett.123.066404} {\bibfield  {journal}
  {\bibinfo  {journal} {Phys. Rev. Lett.}\ }\textbf {\bibinfo {volume} {123}},\
  \bibinfo {pages} {066404} (\bibinfo {year} {2019})}\BibitemShut {NoStop}%
\bibitem [{\citenamefont {Ge}\ \emph {et~al.}(2019)\citenamefont {Ge},
  \citenamefont {Zhang}, \citenamefont {Liu}, \citenamefont {Li}, \citenamefont
  {Fan},\ and\ \citenamefont {Nori}}]{PhysRevB.100.054105}%
  \BibitemOpen
  \bibfield  {author} {\bibinfo {author} {\bibfnamefont {Z.-Y.}\ \bibnamefont
  {Ge}}, \bibinfo {author} {\bibfnamefont {Y.-R.}\ \bibnamefont {Zhang}},
  \bibinfo {author} {\bibfnamefont {T.}~\bibnamefont {Liu}}, \bibinfo {author}
  {\bibfnamefont {S.-W.}\ \bibnamefont {Li}}, \bibinfo {author} {\bibfnamefont
  {H.}~\bibnamefont {Fan}},\ and\ \bibinfo {author} {\bibfnamefont
  {F.}~\bibnamefont {Nori}},\ }\bibfield  {title} {\bibinfo {title}
  {{Topological band theory for non-Hermitian systems from the Dirac
  equation}},\ }\href {https://doi.org/10.1103/PhysRevB.100.054105} {\bibfield
  {journal} {\bibinfo  {journal} {Phys. Rev. B}\ }\textbf {\bibinfo {volume}
  {100}},\ \bibinfo {pages} {054105} (\bibinfo {year} {2019})}\BibitemShut
  {NoStop}%
\bibitem [{\citenamefont {Molina}\ and\ \citenamefont
  {Gonz\'alez}(2018)}]{PhysRevLett.120.146601}%
  \BibitemOpen
  \bibfield  {author} {\bibinfo {author} {\bibfnamefont {R.~A.}\ \bibnamefont
  {Molina}}\ and\ \bibinfo {author} {\bibfnamefont {J.}~\bibnamefont
  {Gonz\'alez}},\ }\bibfield  {title} {\bibinfo {title} {{Surface and 3D
  Quantum Hall Effects from Engineering of Exceptional Points in Nodal-Line
  Semimetals}},\ }\href {https://doi.org/10.1103/PhysRevLett.120.146601}
  {\bibfield  {journal} {\bibinfo  {journal} {Phys. Rev. Lett.}\ }\textbf
  {\bibinfo {volume} {120}},\ \bibinfo {pages} {146601} (\bibinfo {year}
  {2018})}\BibitemShut {NoStop}%
\bibitem [{\citenamefont {Xue}\ \emph {et~al.}(2020)\citenamefont {Xue},
  \citenamefont {Wang}, \citenamefont {Zhang},\ and\ \citenamefont
  {Chong}}]{PhysRevLett.124.236403}%
  \BibitemOpen
  \bibfield  {author} {\bibinfo {author} {\bibfnamefont {H.}~\bibnamefont
  {Xue}}, \bibinfo {author} {\bibfnamefont {Q.}~\bibnamefont {Wang}}, \bibinfo
  {author} {\bibfnamefont {B.}~\bibnamefont {Zhang}},\ and\ \bibinfo {author}
  {\bibfnamefont {Y.~D.}\ \bibnamefont {Chong}},\ }\bibfield  {title} {\bibinfo
  {title} {{Non-Hermitian Dirac Cones}},\ }\href
  {https://doi.org/10.1103/PhysRevLett.124.236403} {\bibfield  {journal}
  {\bibinfo  {journal} {Phys. Rev. Lett.}\ }\textbf {\bibinfo {volume} {124}},\
  \bibinfo {pages} {236403} (\bibinfo {year} {2020})}\BibitemShut {NoStop}%
\bibitem [{\citenamefont {Budich}\ \emph {et~al.}(2019)\citenamefont {Budich},
  \citenamefont {Carlstr\"om}, \citenamefont {Kunst},\ and\ \citenamefont
  {Bergholtz}}]{PhysRevB.99.041406}%
  \BibitemOpen
  \bibfield  {author} {\bibinfo {author} {\bibfnamefont {J.~C.}\ \bibnamefont
  {Budich}}, \bibinfo {author} {\bibfnamefont {J.}~\bibnamefont {Carlstr\"om}},
  \bibinfo {author} {\bibfnamefont {F.~K.}\ \bibnamefont {Kunst}},\ and\
  \bibinfo {author} {\bibfnamefont {E.~J.}\ \bibnamefont {Bergholtz}},\
  }\bibfield  {title} {\bibinfo {title} {{Symmetry-protected nodal phases in
  non-Hermitian systems}},\ }\href {https://doi.org/10.1103/PhysRevB.99.041406}
  {\bibfield  {journal} {\bibinfo  {journal} {Phys. Rev. B}\ }\textbf {\bibinfo
  {volume} {99}},\ \bibinfo {pages} {041406(R)} (\bibinfo {year}
  {2019})}\BibitemShut {NoStop}%
\bibitem [{\citenamefont {Yoshida}\ \emph {et~al.}(2019)\citenamefont
  {Yoshida}, \citenamefont {Peters}, \citenamefont {Kawakami},\ and\
  \citenamefont {Hatsugai}}]{yoshida2019symmetry}%
  \BibitemOpen
  \bibfield  {author} {\bibinfo {author} {\bibfnamefont {T.}~\bibnamefont
  {Yoshida}}, \bibinfo {author} {\bibfnamefont {R.}~\bibnamefont {Peters}},
  \bibinfo {author} {\bibfnamefont {N.}~\bibnamefont {Kawakami}},\ and\
  \bibinfo {author} {\bibfnamefont {Y.}~\bibnamefont {Hatsugai}},\ }\bibfield
  {title} {\bibinfo {title} {Symmetry-protected exceptional rings in
  two-dimensional correlated systems with chiral symmetry},\ }\href
  {https://doi.org/10.1103/PhysRevB.99.121101} {\bibfield  {journal} {\bibinfo
  {journal} {Phys. Rev. B}\ }\textbf {\bibinfo {volume} {99}},\ \bibinfo
  {pages} {121101(R)} (\bibinfo {year} {2019})}\BibitemShut {NoStop}%
\bibitem [{\citenamefont {Yang}\ and\ \citenamefont
  {Hu}(2019)}]{PhysRevB.99.081102}%
  \BibitemOpen
  \bibfield  {author} {\bibinfo {author} {\bibfnamefont {Z.}~\bibnamefont
  {Yang}}\ and\ \bibinfo {author} {\bibfnamefont {J.}~\bibnamefont {Hu}},\
  }\bibfield  {title} {\bibinfo {title} {{Non-Hermitian Hopf-link exceptional
  line semimetals}},\ }\href {https://doi.org/10.1103/PhysRevB.99.081102}
  {\bibfield  {journal} {\bibinfo  {journal} {Phys. Rev. B}\ }\textbf {\bibinfo
  {volume} {99}},\ \bibinfo {pages} {081102(R)} (\bibinfo {year}
  {2019})}\BibitemShut {NoStop}%
\bibitem [{\citenamefont {Okuma}\ \emph {et~al.}(2020)\citenamefont {Okuma},
  \citenamefont {Kawabata}, \citenamefont {Shiozaki},\ and\ \citenamefont
  {Sato}}]{Okuma2020Topological}%
  \BibitemOpen
  \bibfield  {author} {\bibinfo {author} {\bibfnamefont {N.}~\bibnamefont
  {Okuma}}, \bibinfo {author} {\bibfnamefont {K.}~\bibnamefont {Kawabata}},
  \bibinfo {author} {\bibfnamefont {K.}~\bibnamefont {Shiozaki}},\ and\
  \bibinfo {author} {\bibfnamefont {M.}~\bibnamefont {Sato}},\ }\bibfield
  {title} {\bibinfo {title} {{Topological Origin of Non-Hermitian Skin
  Effects}},\ }\href {https://doi.org/10.1103/PhysRevLett.124.086801}
  {\bibfield  {journal} {\bibinfo  {journal} {Phys. Rev. Lett.}\ }\textbf
  {\bibinfo {volume} {124}},\ \bibinfo {pages} {086801} (\bibinfo {year}
  {2020})}\BibitemShut {NoStop}%
\bibitem [{\citenamefont {Li}\ \emph {et~al.}(2020{\natexlab{a}})\citenamefont
  {Li}, \citenamefont {Lee}, \citenamefont {Mu},\ and\ \citenamefont
  {Gong}}]{li2020critical}%
  \BibitemOpen
  \bibfield  {author} {\bibinfo {author} {\bibfnamefont {L.}~\bibnamefont
  {Li}}, \bibinfo {author} {\bibfnamefont {C.~H.}\ \bibnamefont {Lee}},
  \bibinfo {author} {\bibfnamefont {S.}~\bibnamefont {Mu}},\ and\ \bibinfo
  {author} {\bibfnamefont {J.}~\bibnamefont {Gong}},\ }\bibfield  {title}
  {\bibinfo {title} {{Critical non-Hermitian Skin Effect}},\ }\href
  {https://arxiv.org/abs/2003.03039} {\bibfield  {journal} {\bibinfo  {journal}
  {arXiv:2003.03039}\ } (\bibinfo {year} {2020}{\natexlab{a}})}\BibitemShut
  {NoStop}%
\bibitem [{\citenamefont {Yao}\ and\ \citenamefont
  {Wang}(2018)}]{PhysRevLett.121.086803}%
  \BibitemOpen
  \bibfield  {author} {\bibinfo {author} {\bibfnamefont {S.}~\bibnamefont
  {Yao}}\ and\ \bibinfo {author} {\bibfnamefont {Z.}~\bibnamefont {Wang}},\
  }\bibfield  {title} {\bibinfo {title} {{Edge States and Topological
  Invariants of Non-Hermitian Systems}},\ }\href
  {https://doi.org/10.1103/PhysRevLett.121.086803} {\bibfield  {journal}
  {\bibinfo  {journal} {Phys. Rev. Lett.}\ }\textbf {\bibinfo {volume} {121}},\
  \bibinfo {pages} {086803} (\bibinfo {year} {2018})}\BibitemShut {NoStop}%
\bibitem [{\citenamefont {Yao}\ \emph {et~al.}(2018)\citenamefont {Yao},
  \citenamefont {Song},\ and\ \citenamefont {Wang}}]{PhysRevLett.121.136802}%
  \BibitemOpen
  \bibfield  {author} {\bibinfo {author} {\bibfnamefont {S.}~\bibnamefont
  {Yao}}, \bibinfo {author} {\bibfnamefont {F.}~\bibnamefont {Song}},\ and\
  \bibinfo {author} {\bibfnamefont {Z.}~\bibnamefont {Wang}},\ }\bibfield
  {title} {\bibinfo {title} {{Non-Hermitian Chern Bands}},\ }\href
  {https://doi.org/10.1103/PhysRevLett.121.136802} {\bibfield  {journal}
  {\bibinfo  {journal} {Phys. Rev. Lett.}\ }\textbf {\bibinfo {volume} {121}},\
  \bibinfo {pages} {136802} (\bibinfo {year} {2018})}\BibitemShut {NoStop}%
\bibitem [{\citenamefont {Song}\ \emph {et~al.}(2019)\citenamefont {Song},
  \citenamefont {Yao},\ and\ \citenamefont {Wang}}]{PhysRevLett.123.246801}%
  \BibitemOpen
  \bibfield  {author} {\bibinfo {author} {\bibfnamefont {F.}~\bibnamefont
  {Song}}, \bibinfo {author} {\bibfnamefont {S.}~\bibnamefont {Yao}},\ and\
  \bibinfo {author} {\bibfnamefont {Z.}~\bibnamefont {Wang}},\ }\bibfield
  {title} {\bibinfo {title} {{Non-Hermitian Topological Invariants in Real
  Space}},\ }\href {https://doi.org/10.1103/PhysRevLett.123.246801} {\bibfield
  {journal} {\bibinfo  {journal} {Phys. Rev. Lett.}\ }\textbf {\bibinfo
  {volume} {123}},\ \bibinfo {pages} {246801} (\bibinfo {year}
  {2019})}\BibitemShut {NoStop}%
\bibitem [{\citenamefont {Yang}\ \emph {et~al.}(2020)\citenamefont {Yang},
  \citenamefont {Chiu}, \citenamefont {Fang},\ and\ \citenamefont
  {Hu}}]{PhysRevLett.124.186402}%
  \BibitemOpen
  \bibfield  {author} {\bibinfo {author} {\bibfnamefont {Z.}~\bibnamefont
  {Yang}}, \bibinfo {author} {\bibfnamefont {C.-K.}\ \bibnamefont {Chiu}},
  \bibinfo {author} {\bibfnamefont {C.}~\bibnamefont {Fang}},\ and\ \bibinfo
  {author} {\bibfnamefont {J.}~\bibnamefont {Hu}},\ }\bibfield  {title}
  {\bibinfo {title} {{Jones Polynomial and Knot Transitions in Hermitian and
  non-Hermitian Topological Semimetals}},\ }\href
  {https://doi.org/10.1103/PhysRevLett.124.186402} {\bibfield  {journal}
  {\bibinfo  {journal} {Phys. Rev. Lett.}\ }\textbf {\bibinfo {volume} {124}},\
  \bibinfo {pages} {186402} (\bibinfo {year} {2020})}\BibitemShut {NoStop}%
\bibitem [{\citenamefont {Bessho}\ and\ \citenamefont
  {Sato}(2020)}]{bessho2020topological}%
  \BibitemOpen
  \bibfield  {author} {\bibinfo {author} {\bibfnamefont {T.}~\bibnamefont
  {Bessho}}\ and\ \bibinfo {author} {\bibfnamefont {M.}~\bibnamefont {Sato}},\
  }\bibfield  {title} {\bibinfo {title} {{Topological Duality in Floquet and
  Non-Hermitian Dynamical Anomalies: Extended Nielsen-Ninomiya Theorem and
  Chiral Magnetic Effect}},\ }\href {https://arxiv.org/abs/2006.04204}
  {\bibfield  {journal} {\bibinfo  {journal} {arXiv:2006.04204}\ } (\bibinfo
  {year} {2020})}\BibitemShut {NoStop}%
\bibitem [{\citenamefont {Zhou}(2020)}]{zhou2020non}%
  \BibitemOpen
  \bibfield  {author} {\bibinfo {author} {\bibfnamefont {L.}~\bibnamefont
  {Zhou}},\ }\bibfield  {title} {\bibinfo {title} {{Non-Hermitian Floquet
  phases with even-integer topological invariants in a periodically quenched
  two-leg ladder}},\ }\href {https://arxiv.org/abs/2006.08897} {\bibfield
  {journal} {\bibinfo  {journal} {arXiv:2006.08897}\ } (\bibinfo {year}
  {2020})}\BibitemShut {NoStop}%
\bibitem [{\citenamefont {H\"ockendorf}\ \emph {et~al.}(2020)\citenamefont
  {H\"ockendorf}, \citenamefont {Alvermann},\ and\ \citenamefont
  {Fehske}}]{PhysRevResearch.2.023235}%
  \BibitemOpen
  \bibfield  {author} {\bibinfo {author} {\bibfnamefont {B.}~\bibnamefont
  {H\"ockendorf}}, \bibinfo {author} {\bibfnamefont {A.}~\bibnamefont
  {Alvermann}},\ and\ \bibinfo {author} {\bibfnamefont {H.}~\bibnamefont
  {Fehske}},\ }\bibfield  {title} {\bibinfo {title} {{Topological origin of
  quantized transport in non-Hermitian Floquet chains}},\ }\href
  {https://doi.org/10.1103/PhysRevResearch.2.023235} {\bibfield  {journal}
  {\bibinfo  {journal} {Phys. Rev. Research}\ }\textbf {\bibinfo {volume}
  {2}},\ \bibinfo {pages} {023235} (\bibinfo {year} {2020})}\BibitemShut
  {NoStop}%
\bibitem [{\citenamefont {Wojcik}\ \emph {et~al.}(2020)\citenamefont {Wojcik},
  \citenamefont {Sun}, \citenamefont {Bzdu\ifmmode~\check{s}\else
  \v{s}\fi{}ek},\ and\ \citenamefont {Fan}}]{PhysRevB.101.205417}%
  \BibitemOpen
  \bibfield  {author} {\bibinfo {author} {\bibfnamefont {C.~C.}\ \bibnamefont
  {Wojcik}}, \bibinfo {author} {\bibfnamefont {X.-Q.}\ \bibnamefont {Sun}},
  \bibinfo {author} {\bibfnamefont {T.~c.~v.}\ \bibnamefont
  {Bzdu\ifmmode~\check{s}\else \v{s}\fi{}ek}},\ and\ \bibinfo {author}
  {\bibfnamefont {S.}~\bibnamefont {Fan}},\ }\bibfield  {title} {\bibinfo
  {title} {{Homotopy characterization of non-Hermitian Hamiltonians}},\ }\href
  {https://doi.org/10.1103/PhysRevB.101.205417} {\bibfield  {journal} {\bibinfo
   {journal} {Phys. Rev. B}\ }\textbf {\bibinfo {volume} {101}},\ \bibinfo
  {pages} {205417} (\bibinfo {year} {2020})}\BibitemShut {NoStop}%
\bibitem [{\citenamefont {Li}\ and\ \citenamefont
  {Mong}(2019)}]{li2019homotopical}%
  \BibitemOpen
  \bibfield  {author} {\bibinfo {author} {\bibfnamefont {Z.}~\bibnamefont
  {Li}}\ and\ \bibinfo {author} {\bibfnamefont {R.~S.}\ \bibnamefont {Mong}},\
  }\bibfield  {title} {\bibinfo {title} {{Homotopical classification of
  non-Hermitian band structures}},\ }\href {https://arxiv.org/abs/1911.02697}
  {\bibfield  {journal} {\bibinfo  {journal} {arXiv:1911.02697}\ } (\bibinfo
  {year} {2019})}\BibitemShut {NoStop}%
\bibitem [{\citenamefont {Liu}\ \emph {et~al.}(2019)\citenamefont {Liu},
  \citenamefont {Zhang}, \citenamefont {Ai}, \citenamefont {Gong},
  \citenamefont {Kawabata}, \citenamefont {Ueda},\ and\ \citenamefont
  {Nori}}]{Liu2019Second}%
  \BibitemOpen
  \bibfield  {author} {\bibinfo {author} {\bibfnamefont {T.}~\bibnamefont
  {Liu}}, \bibinfo {author} {\bibfnamefont {Y.-R.}\ \bibnamefont {Zhang}},
  \bibinfo {author} {\bibfnamefont {Q.}~\bibnamefont {Ai}}, \bibinfo {author}
  {\bibfnamefont {Z.}~\bibnamefont {Gong}}, \bibinfo {author} {\bibfnamefont
  {K.}~\bibnamefont {Kawabata}}, \bibinfo {author} {\bibfnamefont
  {M.}~\bibnamefont {Ueda}},\ and\ \bibinfo {author} {\bibfnamefont
  {F.}~\bibnamefont {Nori}},\ }\bibfield  {title} {\bibinfo {title}
  {{Second-Order Topological Phases in Non-Hermitian Systems}},\ }\href
  {https://doi.org/10.1103/PhysRevLett.122.076801} {\bibfield  {journal}
  {\bibinfo  {journal} {Phys. Rev. Lett.}\ }\textbf {\bibinfo {volume} {122}},\
  \bibinfo {pages} {076801} (\bibinfo {year} {2019})}\BibitemShut {NoStop}%
\bibitem [{\citenamefont {Deng}\ and\ \citenamefont
  {Yi}(2019)}]{Deng2019Non-Bloch}%
  \BibitemOpen
  \bibfield  {author} {\bibinfo {author} {\bibfnamefont {T.-S.}\ \bibnamefont
  {Deng}}\ and\ \bibinfo {author} {\bibfnamefont {W.}~\bibnamefont {Yi}},\
  }\bibfield  {title} {\bibinfo {title} {{Non-Bloch topological invariants in a
  non-Hermitian domain wall system}},\ }\href
  {https://doi.org/10.1103/PhysRevB.100.035102} {\bibfield  {journal} {\bibinfo
   {journal} {Phys. Rev. B}\ }\textbf {\bibinfo {volume} {100}},\ \bibinfo
  {pages} {035102} (\bibinfo {year} {2019})}\BibitemShut {NoStop}%
\bibitem [{\citenamefont {Xiao}\ \emph {et~al.}(2020)\citenamefont {Xiao},
  \citenamefont {Deng}, \citenamefont {Wang}, \citenamefont {Zhu},
  \citenamefont {Wang}, \citenamefont {Yi},\ and\ \citenamefont
  {Xue}}]{Xiao2020non}%
  \BibitemOpen
  \bibfield  {author} {\bibinfo {author} {\bibfnamefont {L.}~\bibnamefont
  {Xiao}}, \bibinfo {author} {\bibfnamefont {T.}~\bibnamefont {Deng}}, \bibinfo
  {author} {\bibfnamefont {K.}~\bibnamefont {Wang}}, \bibinfo {author}
  {\bibfnamefont {G.}~\bibnamefont {Zhu}}, \bibinfo {author} {\bibfnamefont
  {Z.}~\bibnamefont {Wang}}, \bibinfo {author} {\bibfnamefont {W.}~\bibnamefont
  {Yi}},\ and\ \bibinfo {author} {\bibfnamefont {P.}~\bibnamefont {Xue}},\
  }\bibfield  {title} {\bibinfo {title} {{Non-Hermitian bulk--boundary
  correspondence in quantum dynamics}},\ }\href
  {https://www.nature.com/articles/s41567-020-0836-6} {\bibfield  {journal}
  {\bibinfo  {journal} {Nat. Phys.}\ }\textbf {\bibinfo {volume} {16}},\
  \bibinfo {pages} {761} (\bibinfo {year} {2020})}\BibitemShut {NoStop}%
\bibitem [{\citenamefont {Zeuner}\ \emph {et~al.}(2015)\citenamefont {Zeuner},
  \citenamefont {Rechtsman}, \citenamefont {Plotnik}, \citenamefont {Lumer},
  \citenamefont {Nolte}, \citenamefont {Rudner}, \citenamefont {Segev},\ and\
  \citenamefont {Szameit}}]{PhysRevLett.115.040402}%
  \BibitemOpen
  \bibfield  {author} {\bibinfo {author} {\bibfnamefont {J.~M.}\ \bibnamefont
  {Zeuner}}, \bibinfo {author} {\bibfnamefont {M.~C.}\ \bibnamefont
  {Rechtsman}}, \bibinfo {author} {\bibfnamefont {Y.}~\bibnamefont {Plotnik}},
  \bibinfo {author} {\bibfnamefont {Y.}~\bibnamefont {Lumer}}, \bibinfo
  {author} {\bibfnamefont {S.}~\bibnamefont {Nolte}}, \bibinfo {author}
  {\bibfnamefont {M.~S.}\ \bibnamefont {Rudner}}, \bibinfo {author}
  {\bibfnamefont {M.}~\bibnamefont {Segev}},\ and\ \bibinfo {author}
  {\bibfnamefont {A.}~\bibnamefont {Szameit}},\ }\bibfield  {title} {\bibinfo
  {title} {{Observation of a Topological Transition in the Bulk of a
  Non-Hermitian System}},\ }\href
  {https://doi.org/10.1103/PhysRevLett.115.040402} {\bibfield  {journal}
  {\bibinfo  {journal} {Phys. Rev. Lett.}\ }\textbf {\bibinfo {volume} {115}},\
  \bibinfo {pages} {040402} (\bibinfo {year} {2015})}\BibitemShut {NoStop}%
\bibitem [{\citenamefont {Poli}\ \emph {et~al.}(2015)\citenamefont {Poli},
  \citenamefont {Bellec}, \citenamefont {Kuhl}, \citenamefont {Mortessagne},\
  and\ \citenamefont {Schomerus}}]{poli2015selective}%
  \BibitemOpen
  \bibfield  {author} {\bibinfo {author} {\bibfnamefont {C.}~\bibnamefont
  {Poli}}, \bibinfo {author} {\bibfnamefont {M.}~\bibnamefont {Bellec}},
  \bibinfo {author} {\bibfnamefont {U.}~\bibnamefont {Kuhl}}, \bibinfo {author}
  {\bibfnamefont {F.}~\bibnamefont {Mortessagne}},\ and\ \bibinfo {author}
  {\bibfnamefont {H.}~\bibnamefont {Schomerus}},\ }\bibfield  {title} {\bibinfo
  {title} {{Selective enhancement of topologically induced interface states in
  a dielectric resonator chain}},\ }\href
  {https://www.nature.com/articles/ncomms7710/} {\bibfield  {journal} {\bibinfo
   {journal} {Nat. Commun.}\ }\textbf {\bibinfo {volume} {6}},\ \bibinfo
  {pages} {6710} (\bibinfo {year} {2015})}\BibitemShut {NoStop}%
\bibitem [{\citenamefont {Weimann}\ \emph {et~al.}(2017)\citenamefont
  {Weimann}, \citenamefont {Kremer}, \citenamefont {Plotnik}, \citenamefont
  {Lumer}, \citenamefont {Nolte}, \citenamefont {Makris}, \citenamefont
  {Segev}, \citenamefont {Rechtsman},\ and\ \citenamefont
  {Szameit}}]{weimann2017topologically}%
  \BibitemOpen
  \bibfield  {author} {\bibinfo {author} {\bibfnamefont {S.}~\bibnamefont
  {Weimann}}, \bibinfo {author} {\bibfnamefont {M.}~\bibnamefont {Kremer}},
  \bibinfo {author} {\bibfnamefont {Y.}~\bibnamefont {Plotnik}}, \bibinfo
  {author} {\bibfnamefont {Y.}~\bibnamefont {Lumer}}, \bibinfo {author}
  {\bibfnamefont {S.}~\bibnamefont {Nolte}}, \bibinfo {author} {\bibfnamefont
  {K.~G.}\ \bibnamefont {Makris}}, \bibinfo {author} {\bibfnamefont
  {M.}~\bibnamefont {Segev}}, \bibinfo {author} {\bibfnamefont {M.~C.}\
  \bibnamefont {Rechtsman}},\ and\ \bibinfo {author} {\bibfnamefont
  {A.}~\bibnamefont {Szameit}},\ }\bibfield  {title} {\bibinfo {title}
  {{Topologically protected bound states in photonic parity--time-symmetric
  crystals}},\ }\href {https://www.nature.com/articles/s41566-019-0453-z}
  {\bibfield  {journal} {\bibinfo  {journal} {Nat. Mater.}\ }\textbf {\bibinfo
  {volume} {16}},\ \bibinfo {pages} {433} (\bibinfo {year} {2017})}\BibitemShut
  {NoStop}%
\bibitem [{\citenamefont {Chen}\ \emph {et~al.}(2017)\citenamefont {Chen},
  \citenamefont {{\"O}zdemir}, \citenamefont {Zhao}, \citenamefont {Wiersig},\
  and\ \citenamefont {Yang}}]{chen2017exceptional}%
  \BibitemOpen
  \bibfield  {author} {\bibinfo {author} {\bibfnamefont {W.}~\bibnamefont
  {Chen}}, \bibinfo {author} {\bibfnamefont {{\c{S}}.~K.}\ \bibnamefont
  {{\"O}zdemir}}, \bibinfo {author} {\bibfnamefont {G.}~\bibnamefont {Zhao}},
  \bibinfo {author} {\bibfnamefont {J.}~\bibnamefont {Wiersig}},\ and\ \bibinfo
  {author} {\bibfnamefont {L.}~\bibnamefont {Yang}},\ }\bibfield  {title}
  {\bibinfo {title} {{Exceptional points enhance sensing in an optical
  microcavity}},\ }\href {https://www.nature.com/articles/nature23281}
  {\bibfield  {journal} {\bibinfo  {journal} {Nature}\ }\textbf {\bibinfo
  {volume} {548}},\ \bibinfo {pages} {192} (\bibinfo {year}
  {2017})}\BibitemShut {NoStop}%
\bibitem [{\citenamefont {Zhou}\ \emph
  {et~al.}(2018{\natexlab{b}})\citenamefont {Zhou}, \citenamefont {Peng},
  \citenamefont {Yoon}, \citenamefont {Hsu}, \citenamefont {Nelson},
  \citenamefont {Fu}, \citenamefont {Joannopoulos}, \citenamefont {Solja{\v
  c}i{\'c}},\ and\ \citenamefont {Zhen}}]{Zhou1009}%
  \BibitemOpen
  \bibfield  {author} {\bibinfo {author} {\bibfnamefont {H.}~\bibnamefont
  {Zhou}}, \bibinfo {author} {\bibfnamefont {C.}~\bibnamefont {Peng}}, \bibinfo
  {author} {\bibfnamefont {Y.}~\bibnamefont {Yoon}}, \bibinfo {author}
  {\bibfnamefont {C.~W.}\ \bibnamefont {Hsu}}, \bibinfo {author} {\bibfnamefont
  {K.~A.}\ \bibnamefont {Nelson}}, \bibinfo {author} {\bibfnamefont
  {L.}~\bibnamefont {Fu}}, \bibinfo {author} {\bibfnamefont {J.~D.}\
  \bibnamefont {Joannopoulos}}, \bibinfo {author} {\bibfnamefont
  {M.}~\bibnamefont {Solja{\v c}i{\'c}}},\ and\ \bibinfo {author}
  {\bibfnamefont {B.}~\bibnamefont {Zhen}},\ }\bibfield  {title} {\bibinfo
  {title} {{Observation of bulk Fermi arc and polarization half charge from
  paired exceptional points}},\ }\href
  {https://doi.org/10.1126/science.aap9859} {\bibfield  {journal} {\bibinfo
  {journal} {Science}\ }\textbf {\bibinfo {volume} {359}},\ \bibinfo {pages}
  {1009} (\bibinfo {year} {2018}{\natexlab{b}})}\BibitemShut {NoStop}%
\bibitem [{\citenamefont {Zhang}\ and\ \citenamefont
  {Franz}(2020)}]{PhysRevLett.124.046401}%
  \BibitemOpen
  \bibfield  {author} {\bibinfo {author} {\bibfnamefont {X.-X.}\ \bibnamefont
  {Zhang}}\ and\ \bibinfo {author} {\bibfnamefont {M.}~\bibnamefont {Franz}},\
  }\bibfield  {title} {\bibinfo {title} {{Non-Hermitian Exceptional Landau
  Quantization in Electric Circuits}},\ }\href
  {https://doi.org/10.1103/PhysRevLett.124.046401} {\bibfield  {journal}
  {\bibinfo  {journal} {Phys. Rev. Lett.}\ }\textbf {\bibinfo {volume} {124}},\
  \bibinfo {pages} {046401} (\bibinfo {year} {2020})}\BibitemShut {NoStop}%
\bibitem [{\citenamefont {Cerjan}\ \emph {et~al.}(2019)\citenamefont {Cerjan},
  \citenamefont {Huang}, \citenamefont {Wang}, \citenamefont {Chen},
  \citenamefont {Chong},\ and\ \citenamefont
  {Rechtsman}}]{cerjan2019experimental}%
  \BibitemOpen
  \bibfield  {author} {\bibinfo {author} {\bibfnamefont {A.}~\bibnamefont
  {Cerjan}}, \bibinfo {author} {\bibfnamefont {S.}~\bibnamefont {Huang}},
  \bibinfo {author} {\bibfnamefont {M.}~\bibnamefont {Wang}}, \bibinfo {author}
  {\bibfnamefont {K.~P.}\ \bibnamefont {Chen}}, \bibinfo {author}
  {\bibfnamefont {Y.}~\bibnamefont {Chong}},\ and\ \bibinfo {author}
  {\bibfnamefont {M.~C.}\ \bibnamefont {Rechtsman}},\ }\bibfield  {title}
  {\bibinfo {title} {{Experimental realization of a Weyl exceptional ring}},\
  }\href {https://www.nature.com/articles/s41566-019-0453-z} {\bibfield
  {journal} {\bibinfo  {journal} {Nat. Photon.}\ }\textbf {\bibinfo {volume}
  {13}},\ \bibinfo {pages} {623} (\bibinfo {year} {2019})}\BibitemShut
  {NoStop}%
\bibitem [{\citenamefont {Bandres}\ \emph {et~al.}(2018)\citenamefont
  {Bandres}, \citenamefont {Wittek}, \citenamefont {Harari}, \citenamefont
  {Parto}, \citenamefont {Ren}, \citenamefont {Segev}, \citenamefont
  {Christodoulides},\ and\ \citenamefont {Khajavikhan}}]{Bandreseaar4005}%
  \BibitemOpen
  \bibfield  {author} {\bibinfo {author} {\bibfnamefont {M.~A.}\ \bibnamefont
  {Bandres}}, \bibinfo {author} {\bibfnamefont {S.}~\bibnamefont {Wittek}},
  \bibinfo {author} {\bibfnamefont {G.}~\bibnamefont {Harari}}, \bibinfo
  {author} {\bibfnamefont {M.}~\bibnamefont {Parto}}, \bibinfo {author}
  {\bibfnamefont {J.}~\bibnamefont {Ren}}, \bibinfo {author} {\bibfnamefont
  {M.}~\bibnamefont {Segev}}, \bibinfo {author} {\bibfnamefont {D.~N.}\
  \bibnamefont {Christodoulides}},\ and\ \bibinfo {author} {\bibfnamefont
  {M.}~\bibnamefont {Khajavikhan}},\ }\bibfield  {title} {\bibinfo {title}
  {{Topological insulator laser: Experiments}},\ }\href
  {https://doi.org/10.1126/science.aar4005} {\bibfield  {journal} {\bibinfo
  {journal} {Science}\ }\textbf {\bibinfo {volume} {359}},\ \bibinfo {pages}
  {eaar4005} (\bibinfo {year} {2018})}\BibitemShut {NoStop}%
\bibitem [{\citenamefont {Li}\ \emph {et~al.}(2020{\natexlab{b}})\citenamefont
  {Li}, \citenamefont {Lee},\ and\ \citenamefont
  {Gong}}]{PhysRevLett.124.250402}%
  \BibitemOpen
  \bibfield  {author} {\bibinfo {author} {\bibfnamefont {L.}~\bibnamefont
  {Li}}, \bibinfo {author} {\bibfnamefont {C.~H.}\ \bibnamefont {Lee}},\ and\
  \bibinfo {author} {\bibfnamefont {J.}~\bibnamefont {Gong}},\ }\bibfield
  {title} {\bibinfo {title} {{Topological Switch for Non-Hermitian Skin Effect
  in Cold-Atom Systems with Loss}},\ }\href
  {https://doi.org/10.1103/PhysRevLett.124.250402} {\bibfield  {journal}
  {\bibinfo  {journal} {Phys. Rev. Lett.}\ }\textbf {\bibinfo {volume} {124}},\
  \bibinfo {pages} {250402} (\bibinfo {year} {2020}{\natexlab{b}})}\BibitemShut
  {NoStop}%
\bibitem [{\citenamefont {Helbig}\ \emph {et~al.}(2020)\citenamefont {Helbig},
  \citenamefont {Hofmann}, \citenamefont {Imhof}, \citenamefont {Abdelghany},
  \citenamefont {Kiessling}, \citenamefont {Molenkamp}, \citenamefont {Lee},
  \citenamefont {Szameit}, \citenamefont {Greiter},\ and\ \citenamefont
  {Thomale}}]{Helbig2020Generalized}%
  \BibitemOpen
  \bibfield  {author} {\bibinfo {author} {\bibfnamefont {T.}~\bibnamefont
  {Helbig}}, \bibinfo {author} {\bibfnamefont {T.}~\bibnamefont {Hofmann}},
  \bibinfo {author} {\bibfnamefont {S.}~\bibnamefont {Imhof}}, \bibinfo
  {author} {\bibfnamefont {M.}~\bibnamefont {Abdelghany}}, \bibinfo {author}
  {\bibfnamefont {T.}~\bibnamefont {Kiessling}}, \bibinfo {author}
  {\bibfnamefont {L.}~\bibnamefont {Molenkamp}}, \bibinfo {author}
  {\bibfnamefont {C.}~\bibnamefont {Lee}}, \bibinfo {author} {\bibfnamefont
  {A.}~\bibnamefont {Szameit}}, \bibinfo {author} {\bibfnamefont
  {M.}~\bibnamefont {Greiter}},\ and\ \bibinfo {author} {\bibfnamefont
  {R.}~\bibnamefont {Thomale}},\ }\bibfield  {title} {\bibinfo {title}
  {{Generalized bulk--boundary correspondence in non-Hermitian topolectrical
  circuits}},\ }\href {https://www.nature.com/articles/s41567-020-0922-9}
  {\bibfield  {journal} {\bibinfo  {journal} {Nat. Phys.}\ }\textbf {\bibinfo
  {volume} {16}},\ \bibinfo {pages} {747} (\bibinfo {year} {2020})}\BibitemShut
  {NoStop}%
\bibitem [{\citenamefont {Ghatak}\ \emph {et~al.}(2019)\citenamefont {Ghatak},
  \citenamefont {Brandenbourger}, \citenamefont {van Wezel},\ and\
  \citenamefont {Coulais}}]{Ghatak2019Observation}%
  \BibitemOpen
  \bibfield  {author} {\bibinfo {author} {\bibfnamefont {A.}~\bibnamefont
  {Ghatak}}, \bibinfo {author} {\bibfnamefont {M.}~\bibnamefont
  {Brandenbourger}}, \bibinfo {author} {\bibfnamefont {J.}~\bibnamefont {van
  Wezel}},\ and\ \bibinfo {author} {\bibfnamefont {C.}~\bibnamefont
  {Coulais}},\ }\bibfield  {title} {\bibinfo {title} {{Observation of
  non-Hermitian topology and its bulk-edge correspondence}},\ }\href
  {https://arxiv.org/abs/1907.11619} {\bibfield  {journal} {\bibinfo  {journal}
  {arXiv:1907.11619}\ } (\bibinfo {year} {2019})}\BibitemShut {NoStop}%
\bibitem [{\citenamefont {Zirnstein}\ \emph {et~al.}(2019)\citenamefont
  {Zirnstein}, \citenamefont {Refael},\ and\ \citenamefont
  {Rosenow}}]{zirnstein2019bulk}%
  \BibitemOpen
  \bibfield  {author} {\bibinfo {author} {\bibfnamefont {H.-G.}\ \bibnamefont
  {Zirnstein}}, \bibinfo {author} {\bibfnamefont {G.}~\bibnamefont {Refael}},\
  and\ \bibinfo {author} {\bibfnamefont {B.}~\bibnamefont {Rosenow}},\
  }\bibfield  {title} {\bibinfo {title} {{Bulk-boundary correspondence for
  non-Hermitian Hamiltonians via Green functions}},\ }\href
  {https://arxiv.org/abs/1901.11241} {\bibfield  {journal} {\bibinfo  {journal}
  {arXiv:1901.11241}\ } (\bibinfo {year} {2019})}\BibitemShut {NoStop}%
\bibitem [{\citenamefont {Wang}\ \emph {et~al.}(2019)\citenamefont {Wang},
  \citenamefont {Ruan},\ and\ \citenamefont {Zhang}}]{Wang2019Non-Hermitian}%
  \BibitemOpen
  \bibfield  {author} {\bibinfo {author} {\bibfnamefont {H.}~\bibnamefont
  {Wang}}, \bibinfo {author} {\bibfnamefont {J.}~\bibnamefont {Ruan}},\ and\
  \bibinfo {author} {\bibfnamefont {H.}~\bibnamefont {Zhang}},\ }\bibfield
  {title} {\bibinfo {title} {{Non-Hermitian nodal-line semimetals with an
  anomalous bulk-boundary correspondence}},\ }\href
  {https://doi.org/10.1103/PhysRevB.99.075130} {\bibfield  {journal} {\bibinfo
  {journal} {Phys. Rev. B}\ }\textbf {\bibinfo {volume} {99}},\ \bibinfo
  {pages} {075130} (\bibinfo {year} {2019})}\BibitemShut {NoStop}%
\bibitem [{\citenamefont {Jiang}\ \emph {et~al.}(2019)\citenamefont {Jiang},
  \citenamefont {Lang}, \citenamefont {Yang}, \citenamefont {Zhu},\ and\
  \citenamefont {Chen}}]{Jiang2019Interplay}%
  \BibitemOpen
  \bibfield  {author} {\bibinfo {author} {\bibfnamefont {H.}~\bibnamefont
  {Jiang}}, \bibinfo {author} {\bibfnamefont {L.-J.}\ \bibnamefont {Lang}},
  \bibinfo {author} {\bibfnamefont {C.}~\bibnamefont {Yang}}, \bibinfo {author}
  {\bibfnamefont {S.-L.}\ \bibnamefont {Zhu}},\ and\ \bibinfo {author}
  {\bibfnamefont {S.}~\bibnamefont {Chen}},\ }\bibfield  {title} {\bibinfo
  {title} {{Interplay of non-Hermitian skin effects and Anderson localization
  in nonreciprocal quasiperiodic lattices}},\ }\href
  {https://doi.org/10.1103/PhysRevB.100.054301} {\bibfield  {journal} {\bibinfo
   {journal} {Phys. Rev. B}\ }\textbf {\bibinfo {volume} {100}},\ \bibinfo
  {pages} {054301} (\bibinfo {year} {2019})}\BibitemShut {NoStop}%
\bibitem [{\citenamefont {Lee}\ \emph {et~al.}(2019{\natexlab{b}})\citenamefont
  {Lee}, \citenamefont {Li},\ and\ \citenamefont {Gong}}]{Lee2019Hybrid}%
  \BibitemOpen
  \bibfield  {author} {\bibinfo {author} {\bibfnamefont {C.~H.}\ \bibnamefont
  {Lee}}, \bibinfo {author} {\bibfnamefont {L.}~\bibnamefont {Li}},\ and\
  \bibinfo {author} {\bibfnamefont {J.}~\bibnamefont {Gong}},\ }\bibfield
  {title} {\bibinfo {title} {Hybrid higher-order skin-topological modes in
  nonreciprocal systems},\ }\href
  {https://doi.org/10.1103/PhysRevLett.123.016805} {\bibfield  {journal}
  {\bibinfo  {journal} {Phys. Rev. Lett.}\ }\textbf {\bibinfo {volume} {123}},\
  \bibinfo {pages} {016805} (\bibinfo {year} {2019}{\natexlab{b}})}\BibitemShut
  {NoStop}%
\bibitem [{\citenamefont {Edvardsson}\ \emph {et~al.}(2019)\citenamefont
  {Edvardsson}, \citenamefont {Kunst},\ and\ \citenamefont
  {Bergholtz}}]{Edvardsson2019Non-Hermitian}%
  \BibitemOpen
  \bibfield  {author} {\bibinfo {author} {\bibfnamefont {E.}~\bibnamefont
  {Edvardsson}}, \bibinfo {author} {\bibfnamefont {F.~K.}\ \bibnamefont
  {Kunst}},\ and\ \bibinfo {author} {\bibfnamefont {E.~J.}\ \bibnamefont
  {Bergholtz}},\ }\bibfield  {title} {\bibinfo {title} {{Non-Hermitian
  extensions of higher-order topological phases and their biorthogonal
  bulk-boundary correspondence}},\ }\href
  {https://doi.org/10.1103/PhysRevB.99.081302} {\bibfield  {journal} {\bibinfo
  {journal} {Phys. Rev. B}\ }\textbf {\bibinfo {volume} {99}},\ \bibinfo
  {pages} {081302(R)} (\bibinfo {year} {2019})}\BibitemShut {NoStop}%
\bibitem [{\citenamefont {Borgnia}\ \emph {et~al.}(2020)\citenamefont
  {Borgnia}, \citenamefont {Kruchkov},\ and\ \citenamefont
  {Slager}}]{Borgnia2020Non-Hermitian}%
  \BibitemOpen
  \bibfield  {author} {\bibinfo {author} {\bibfnamefont {D.~S.}\ \bibnamefont
  {Borgnia}}, \bibinfo {author} {\bibfnamefont {A.~J.}\ \bibnamefont
  {Kruchkov}},\ and\ \bibinfo {author} {\bibfnamefont {R.-J.}\ \bibnamefont
  {Slager}},\ }\bibfield  {title} {\bibinfo {title} {{Non-Hermitian Boundary
  Modes and Topology}},\ }\href
  {https://doi.org/10.1103/PhysRevLett.124.056802} {\bibfield  {journal}
  {\bibinfo  {journal} {Phys. Rev. Lett.}\ }\textbf {\bibinfo {volume} {124}},\
  \bibinfo {pages} {056802} (\bibinfo {year} {2020})}\BibitemShut {NoStop}%
\bibitem [{\citenamefont {Ezawa}(2019)}]{Ezawa2019Non-Hermitian}%
  \BibitemOpen
  \bibfield  {author} {\bibinfo {author} {\bibfnamefont {M.}~\bibnamefont
  {Ezawa}},\ }\bibfield  {title} {\bibinfo {title} {{Non-Hermitian boundary and
  interface states in nonreciprocal higher-order topological metals and
  electrical circuits}},\ }\href {https://doi.org/10.1103/PhysRevB.99.121411}
  {\bibfield  {journal} {\bibinfo  {journal} {Phys. Rev. B}\ }\textbf {\bibinfo
  {volume} {99}},\ \bibinfo {pages} {121411(R)} (\bibinfo {year}
  {2019})}\BibitemShut {NoStop}%
\bibitem [{\citenamefont {Yang}\ \emph {et~al.}(2019)\citenamefont {Yang},
  \citenamefont {Cao},\ and\ \citenamefont {Zhai}}]{Yang2019Non-Hermitian}%
  \BibitemOpen
  \bibfield  {author} {\bibinfo {author} {\bibfnamefont {X.}~\bibnamefont
  {Yang}}, \bibinfo {author} {\bibfnamefont {Y.}~\bibnamefont {Cao}},\ and\
  \bibinfo {author} {\bibfnamefont {Y.}~\bibnamefont {Zhai}},\ }\bibfield
  {title} {\bibinfo {title} {Non-hermitian weyl semimetals: Non-hermitian skin
  effect and non-bloch bulk-boundary correspondence},\ }\href
  {https://arxiv.org/abs/1904.02492} {\bibfield  {journal} {\bibinfo  {journal}
  {arXiv:1904.02492}\ } (\bibinfo {year} {2019})}\BibitemShut {NoStop}%
\bibitem [{\citenamefont {Goodfellow}\ \emph {et~al.}(2016)\citenamefont
  {Goodfellow}, \citenamefont {Bengio},\ and\ \citenamefont
  {Courville}}]{goodfellow2016deep}%
  \BibitemOpen
  \bibfield  {author} {\bibinfo {author} {\bibfnamefont {I.}~\bibnamefont
  {Goodfellow}}, \bibinfo {author} {\bibfnamefont {Y.}~\bibnamefont {Bengio}},\
  and\ \bibinfo {author} {\bibfnamefont {A.}~\bibnamefont {Courville}},\
  }\href@noop {} {\emph {\bibinfo {title} {{Deep learning}}}}\ (\bibinfo
  {publisher} {MIT press},\ \bibinfo {year} {2016})\BibitemShut {NoStop}%
\bibitem [{\citenamefont {Jordan}\ and\ \citenamefont
  {Mitchell}(2015)}]{Jordan2015Machine}%
  \BibitemOpen
  \bibfield  {author} {\bibinfo {author} {\bibfnamefont {M.}~\bibnamefont
  {Jordan}}\ and\ \bibinfo {author} {\bibfnamefont {T.}~\bibnamefont
  {Mitchell}},\ }\bibfield  {title} {\bibinfo {title} {Machine learning:
  Trends, perspectives, and prospects},\ }\href
  {https://doi.org/10.1126/science.aaa8415} {\bibfield  {journal} {\bibinfo
  {journal} {Science}\ }\textbf {\bibinfo {volume} {349}},\ \bibinfo {pages}
  {255} (\bibinfo {year} {2015})}\BibitemShut {NoStop}%
\bibitem [{\citenamefont {LeCun}\ \emph {et~al.}(2015)\citenamefont {LeCun},
  \citenamefont {Bengio},\ and\ \citenamefont {Hinton}}]{Lecun2015Deep}%
  \BibitemOpen
  \bibfield  {author} {\bibinfo {author} {\bibfnamefont {Y.}~\bibnamefont
  {LeCun}}, \bibinfo {author} {\bibfnamefont {Y.}~\bibnamefont {Bengio}},\ and\
  \bibinfo {author} {\bibfnamefont {G.}~\bibnamefont {Hinton}},\ }\bibfield
  {title} {\bibinfo {title} {Deep learning},\ }\href
  {https://doi.org/10.1038/nature14539} {\bibfield  {journal} {\bibinfo
  {journal} {Nature}\ }\textbf {\bibinfo {volume} {521}},\ \bibinfo {pages}
  {436} (\bibinfo {year} {2015})}\BibitemShut {NoStop}%
\bibitem [{\citenamefont {Dunjko}\ and\ \citenamefont
  {Briegel}(2018)}]{Dunjko2018}%
  \BibitemOpen
  \bibfield  {author} {\bibinfo {author} {\bibfnamefont {V.}~\bibnamefont
  {Dunjko}}\ and\ \bibinfo {author} {\bibfnamefont {H.~J.}\ \bibnamefont
  {Briegel}},\ }\bibfield  {title} {\bibinfo {title} {{Machine learning {\&}
  artificial intelligence in the quantum domain: a review of recent
  progress}},\ }\href {https://doi.org/10.1088/1361-6633/aab406} {\bibfield
  {journal} {\bibinfo  {journal} {Rep. Prog. Phys.}\ }\textbf {\bibinfo
  {volume} {81}},\ \bibinfo {pages} {074001} (\bibinfo {year}
  {2018})}\BibitemShut {NoStop}%
\bibitem [{\citenamefont {Sarma}\ \emph {et~al.}(2019)\citenamefont {Sarma},
  \citenamefont {Deng},\ and\ \citenamefont {Duan}}]{Sarma2019Machine}%
  \BibitemOpen
  \bibfield  {author} {\bibinfo {author} {\bibfnamefont {S.~D.}\ \bibnamefont
  {Sarma}}, \bibinfo {author} {\bibfnamefont {D.-L.}\ \bibnamefont {Deng}},\
  and\ \bibinfo {author} {\bibfnamefont {L.-M.}\ \bibnamefont {Duan}},\
  }\bibfield  {title} {\bibinfo {title} {Machine learning meets quantum
  physics},\ }\href {https://doi.org/10.1063/PT.3.4164} {\bibfield  {journal}
  {\bibinfo  {journal} {Physics Today}\ }\textbf {\bibinfo {volume} {72}},\
  \bibinfo {pages} {48} (\bibinfo {year} {2019})}\BibitemShut {NoStop}%
\bibitem [{\citenamefont {Carleo}\ \emph {et~al.}(2019)\citenamefont {Carleo},
  \citenamefont {Cirac}, \citenamefont {Cranmer}, \citenamefont {Daudet},
  \citenamefont {Schuld}, \citenamefont {Tishby}, \citenamefont
  {Vogt-Maranto},\ and\ \citenamefont {Zdeborov\'a}}]{RevModPhys.91.045002}%
  \BibitemOpen
  \bibfield  {author} {\bibinfo {author} {\bibfnamefont {G.}~\bibnamefont
  {Carleo}}, \bibinfo {author} {\bibfnamefont {I.}~\bibnamefont {Cirac}},
  \bibinfo {author} {\bibfnamefont {K.}~\bibnamefont {Cranmer}}, \bibinfo
  {author} {\bibfnamefont {L.}~\bibnamefont {Daudet}}, \bibinfo {author}
  {\bibfnamefont {M.}~\bibnamefont {Schuld}}, \bibinfo {author} {\bibfnamefont
  {N.}~\bibnamefont {Tishby}}, \bibinfo {author} {\bibfnamefont
  {L.}~\bibnamefont {Vogt-Maranto}},\ and\ \bibinfo {author} {\bibfnamefont
  {L.}~\bibnamefont {Zdeborov\'a}},\ }\bibfield  {title} {\bibinfo {title}
  {{Machine learning and the physical sciences}},\ }\href
  {https://doi.org/10.1103/RevModPhys.91.045002} {\bibfield  {journal}
  {\bibinfo  {journal} {Rev. Mod. Phys.}\ }\textbf {\bibinfo {volume} {91}},\
  \bibinfo {pages} {045002} (\bibinfo {year} {2019})}\BibitemShut {NoStop}%
\bibitem [{\citenamefont {Pasquato}(2016)}]{Pasquato2016Detecting}%
  \BibitemOpen
  \bibfield  {author} {\bibinfo {author} {\bibfnamefont {M.}~\bibnamefont
  {Pasquato}},\ }\bibfield  {title} {\bibinfo {title} {Detecting intermediate
  mass black holes in globular clusters with machine learning},\ }\href
  {https://arxiv.org/abs/1606.08548} {\bibfield  {journal} {\bibinfo  {journal}
  {arXiv:1606.08548}\ } (\bibinfo {year} {2016})}\BibitemShut {NoStop}%
\bibitem [{\citenamefont {Hezaveh}\ \emph {et~al.}(2017)\citenamefont
  {Hezaveh}, \citenamefont {Perreault~Levasseur},\ and\ \citenamefont
  {Marshall}}]{Hezaveh2017Fast}%
  \BibitemOpen
  \bibfield  {author} {\bibinfo {author} {\bibfnamefont {Y.~D.}\ \bibnamefont
  {Hezaveh}}, \bibinfo {author} {\bibfnamefont {L.}~\bibnamefont
  {Perreault~Levasseur}},\ and\ \bibinfo {author} {\bibfnamefont {P.~J.}\
  \bibnamefont {Marshall}},\ }\bibfield  {title} {\bibinfo {title} {Fast
  automated analysis of strong gravitational lenses with convolutional neural
  networks},\ }\href {https://doi.org/10.1038/nature23463} {\bibfield
  {journal} {\bibinfo  {journal} {Nature}\ }\textbf {\bibinfo {volume} {548}},\
  \bibinfo {pages} {555} (\bibinfo {year} {2017})}\BibitemShut {NoStop}%
\bibitem [{\citenamefont {Biswas}\ \emph {et~al.}(2013)\citenamefont {Biswas},
  \citenamefont {Blackburn}, \citenamefont {Cao}, \citenamefont {Essick},
  \citenamefont {Hodge}, \citenamefont {Katsavounidis}, \citenamefont {Kim},
  \citenamefont {Kim}, \citenamefont {Le~Bigot}, \citenamefont {Lee},
  \citenamefont {Oh}, \citenamefont {Oh}, \citenamefont {Son}, \citenamefont
  {Tao}, \citenamefont {Vaulin},\ and\ \citenamefont
  {Wang}}]{Rahul2013Application}%
  \BibitemOpen
  \bibfield  {author} {\bibinfo {author} {\bibfnamefont {R.}~\bibnamefont
  {Biswas}}, \bibinfo {author} {\bibfnamefont {L.}~\bibnamefont {Blackburn}},
  \bibinfo {author} {\bibfnamefont {J.}~\bibnamefont {Cao}}, \bibinfo {author}
  {\bibfnamefont {R.}~\bibnamefont {Essick}}, \bibinfo {author} {\bibfnamefont
  {K.~A.}\ \bibnamefont {Hodge}}, \bibinfo {author} {\bibfnamefont
  {E.}~\bibnamefont {Katsavounidis}}, \bibinfo {author} {\bibfnamefont
  {K.}~\bibnamefont {Kim}}, \bibinfo {author} {\bibfnamefont {Y.-M.}\
  \bibnamefont {Kim}}, \bibinfo {author} {\bibfnamefont {E.-O.}\ \bibnamefont
  {Le~Bigot}}, \bibinfo {author} {\bibfnamefont {C.-H.}\ \bibnamefont {Lee}},
  \bibinfo {author} {\bibfnamefont {J.~J.}\ \bibnamefont {Oh}}, \bibinfo
  {author} {\bibfnamefont {S.~H.}\ \bibnamefont {Oh}}, \bibinfo {author}
  {\bibfnamefont {E.~J.}\ \bibnamefont {Son}}, \bibinfo {author} {\bibfnamefont
  {Y.}~\bibnamefont {Tao}}, \bibinfo {author} {\bibfnamefont {R.}~\bibnamefont
  {Vaulin}},\ and\ \bibinfo {author} {\bibfnamefont {X.}~\bibnamefont {Wang}},\
  }\bibfield  {title} {\bibinfo {title} {Application of machine learning
  algorithms to the study of noise artifacts in gravitational-wave data},\
  }\href {https://doi.org/10.1103/PhysRevD.88.062003} {\bibfield  {journal}
  {\bibinfo  {journal} {Phys. Rev. D}\ }\textbf {\bibinfo {volume} {88}},\
  \bibinfo {pages} {062003} (\bibinfo {year} {2013})}\BibitemShut {NoStop}%
\bibitem [{\citenamefont {Abbott~{\it et al.}}(2016)}]{Abbott2016Observation}%
  \BibitemOpen
  \bibfield  {author} {\bibinfo {author} {\bibfnamefont {B.~P.}\ \bibnamefont
  {Abbott~{\it et al.}}} (\bibinfo {collaboration} {LIGO Scientific
  Collaboration and Virgo Collaboration}),\ }\bibfield  {title} {\bibinfo
  {title} {Observation of gravitational waves from a binary black hole
  merger},\ }\href {https://doi.org/10.1103/PhysRevLett.116.061102} {\bibfield
  {journal} {\bibinfo  {journal} {Phys. Rev. Lett.}\ }\textbf {\bibinfo
  {volume} {116}},\ \bibinfo {pages} {061102} (\bibinfo {year}
  {2016})}\BibitemShut {NoStop}%
\bibitem [{\citenamefont {Deng}(2018)}]{Deng2017MachineBN}%
  \BibitemOpen
  \bibfield  {author} {\bibinfo {author} {\bibfnamefont {D.-L.}\ \bibnamefont
  {Deng}},\ }\bibfield  {title} {\bibinfo {title} {{Machine Learning Detection
  of Bell Nonlocality in Quantum Many-Body Systems}},\ }\href
  {https://doi.org/10.1103/PhysRevLett.120.240402} {\bibfield  {journal}
  {\bibinfo  {journal} {Phys. Rev. Lett.}\ }\textbf {\bibinfo {volume} {120}},\
  \bibinfo {pages} {240402} (\bibinfo {year} {2018})}\BibitemShut {NoStop}%
\bibitem [{\citenamefont {Schoenholz}\ \emph {et~al.}(2016)\citenamefont
  {Schoenholz}, \citenamefont {Cubuk}, \citenamefont {Sussman}, \citenamefont
  {Kaxiras},\ and\ \citenamefont {Liu}}]{Schoenholz2016Structural}%
  \BibitemOpen
  \bibfield  {author} {\bibinfo {author} {\bibfnamefont {S.~S.}\ \bibnamefont
  {Schoenholz}}, \bibinfo {author} {\bibfnamefont {E.~D.}\ \bibnamefont
  {Cubuk}}, \bibinfo {author} {\bibfnamefont {D.~M.}\ \bibnamefont {Sussman}},
  \bibinfo {author} {\bibfnamefont {E.}~\bibnamefont {Kaxiras}},\ and\ \bibinfo
  {author} {\bibfnamefont {A.~J.}\ \bibnamefont {Liu}},\ }\bibfield  {title}
  {\bibinfo {title} {A structural approach to relaxation in glassy liquids},\
  }\href {https://doi.org/10.1038/nphys3644} {\bibfield  {journal} {\bibinfo
  {journal} {Nat. Phys.}\ }\textbf {\bibinfo {volume} {12}},\ \bibinfo {pages}
  {469} (\bibinfo {year} {2016})}\BibitemShut {NoStop}%
\bibitem [{\citenamefont {Kalinin}\ \emph {et~al.}(2015)\citenamefont
  {Kalinin}, \citenamefont {Sumpter},\ and\ \citenamefont
  {Archibald}}]{Kalinin2015Big}%
  \BibitemOpen
  \bibfield  {author} {\bibinfo {author} {\bibfnamefont {S.~V.}\ \bibnamefont
  {Kalinin}}, \bibinfo {author} {\bibfnamefont {B.~G.}\ \bibnamefont
  {Sumpter}},\ and\ \bibinfo {author} {\bibfnamefont {R.~K.}\ \bibnamefont
  {Archibald}},\ }\bibfield  {title} {\bibinfo {title} {Big-deep-smart data in
  imaging for guiding materials design},\ }\href
  {https://doi.org/10.1038/nmat4395} {\bibfield  {journal} {\bibinfo  {journal}
  {Nat. Mater.}\ }\textbf {\bibinfo {volume} {14}},\ \bibinfo {pages} {973}
  (\bibinfo {year} {2015})}\BibitemShut {NoStop}%
\bibitem [{\citenamefont {Zhang}\ and\ \citenamefont
  {Kim}(2017)}]{PhysRevLett.118.216401}%
  \BibitemOpen
  \bibfield  {author} {\bibinfo {author} {\bibfnamefont {Y.}~\bibnamefont
  {Zhang}}\ and\ \bibinfo {author} {\bibfnamefont {E.-A.}\ \bibnamefont
  {Kim}},\ }\bibfield  {title} {\bibinfo {title} {{Quantum Loop Topography for
  Machine Learning}},\ }\href {https://doi.org/10.1103/PhysRevLett.118.216401}
  {\bibfield  {journal} {\bibinfo  {journal} {Phys. Rev. Lett.}\ }\textbf
  {\bibinfo {volume} {118}},\ \bibinfo {pages} {216401} (\bibinfo {year}
  {2017})}\BibitemShut {NoStop}%
\bibitem [{\citenamefont {Zhang}\ \emph {et~al.}(2017)\citenamefont {Zhang},
  \citenamefont {Melko},\ and\ \citenamefont {Kim}}]{PhysRevB.96.245119}%
  \BibitemOpen
  \bibfield  {author} {\bibinfo {author} {\bibfnamefont {Y.}~\bibnamefont
  {Zhang}}, \bibinfo {author} {\bibfnamefont {R.~G.}\ \bibnamefont {Melko}},\
  and\ \bibinfo {author} {\bibfnamefont {E.-A.}\ \bibnamefont {Kim}},\
  }\bibfield  {title} {\bibinfo {title} {{Machine learning ${\mathbb{Z}}_{2}$
  quantum spin liquids with quasiparticle statistics}},\ }\href
  {https://doi.org/10.1103/PhysRevB.96.245119} {\bibfield  {journal} {\bibinfo
  {journal} {Phys. Rev. B}\ }\textbf {\bibinfo {volume} {96}},\ \bibinfo
  {pages} {245119} (\bibinfo {year} {2017})}\BibitemShut {NoStop}%
\bibitem [{\citenamefont {Yoshioka}\ \emph {et~al.}(2018)\citenamefont
  {Yoshioka}, \citenamefont {Akagi},\ and\ \citenamefont
  {Katsura}}]{PhysRevB.97.205110}%
  \BibitemOpen
  \bibfield  {author} {\bibinfo {author} {\bibfnamefont {N.}~\bibnamefont
  {Yoshioka}}, \bibinfo {author} {\bibfnamefont {Y.}~\bibnamefont {Akagi}},\
  and\ \bibinfo {author} {\bibfnamefont {H.}~\bibnamefont {Katsura}},\
  }\bibfield  {title} {\bibinfo {title} {{Learning disordered topological
  phases by statistical recovery of symmetry}},\ }\href
  {https://doi.org/10.1103/PhysRevB.97.205110} {\bibfield  {journal} {\bibinfo
  {journal} {Phys. Rev. B}\ }\textbf {\bibinfo {volume} {97}},\ \bibinfo
  {pages} {205110} (\bibinfo {year} {2018})}\BibitemShut {NoStop}%
\bibitem [{\citenamefont {Zhang}\ \emph {et~al.}(2018)\citenamefont {Zhang},
  \citenamefont {Shen},\ and\ \citenamefont {Zhai}}]{PhysRevLett.120.066401}%
  \BibitemOpen
  \bibfield  {author} {\bibinfo {author} {\bibfnamefont {P.}~\bibnamefont
  {Zhang}}, \bibinfo {author} {\bibfnamefont {H.}~\bibnamefont {Shen}},\ and\
  \bibinfo {author} {\bibfnamefont {H.}~\bibnamefont {Zhai}},\ }\bibfield
  {title} {\bibinfo {title} {{Machine Learning Topological Invariants with
  Neural Networks}},\ }\href {https://doi.org/10.1103/PhysRevLett.120.066401}
  {\bibfield  {journal} {\bibinfo  {journal} {Phys. Rev. Lett.}\ }\textbf
  {\bibinfo {volume} {120}},\ \bibinfo {pages} {066401} (\bibinfo {year}
  {2018})}\BibitemShut {NoStop}%
\bibitem [{\citenamefont {Holanda}\ and\ \citenamefont
  {Griffith}(2020)}]{PhysRevB.102.054107}%
  \BibitemOpen
  \bibfield  {author} {\bibinfo {author} {\bibfnamefont {N.~L.}\ \bibnamefont
  {Holanda}}\ and\ \bibinfo {author} {\bibfnamefont {M.~A.~R.}\ \bibnamefont
  {Griffith}},\ }\bibfield  {title} {\bibinfo {title} {Machine learning
  topological phases in real space},\ }\href
  {https://doi.org/10.1103/PhysRevB.102.054107} {\bibfield  {journal} {\bibinfo
   {journal} {Phys. Rev. B}\ }\textbf {\bibinfo {volume} {102}},\ \bibinfo
  {pages} {054107} (\bibinfo {year} {2020})}\BibitemShut {NoStop}%
\bibitem [{\citenamefont {Narayan}\ and\ \citenamefont
  {Narayan}(2021)}]{narayan2020machine}%
  \BibitemOpen
  \bibfield  {author} {\bibinfo {author} {\bibfnamefont {B.}~\bibnamefont
  {Narayan}}\ and\ \bibinfo {author} {\bibfnamefont {A.}~\bibnamefont
  {Narayan}},\ }\bibfield  {title} {\bibinfo {title} {{Machine learning
  non-Hermitian topological phases}},\ }\href
  {https://doi.org/10.1103/PhysRevB.103.035413} {\bibfield  {journal} {\bibinfo
   {journal} {Phys. Rev. B}\ }\textbf {\bibinfo {volume} {103}},\ \bibinfo
  {pages} {035413} (\bibinfo {year} {2021})}\BibitemShut {NoStop}%
\bibitem [{\citenamefont {Zhang}\ \emph {et~al.}(2021)\citenamefont {Zhang},
  \citenamefont {Tang}, \citenamefont {Huang}, \citenamefont {Zhang},
  \citenamefont {Huang},\ and\ \citenamefont {Zhang}}]{zhang2020machine}%
  \BibitemOpen
  \bibfield  {author} {\bibinfo {author} {\bibfnamefont {L.-F.}\ \bibnamefont
  {Zhang}}, \bibinfo {author} {\bibfnamefont {L.-Z.}\ \bibnamefont {Tang}},
  \bibinfo {author} {\bibfnamefont {Z.-H.}\ \bibnamefont {Huang}}, \bibinfo
  {author} {\bibfnamefont {G.-Q.}\ \bibnamefont {Zhang}}, \bibinfo {author}
  {\bibfnamefont {W.}~\bibnamefont {Huang}},\ and\ \bibinfo {author}
  {\bibfnamefont {D.-W.}\ \bibnamefont {Zhang}},\ }\bibfield  {title} {\bibinfo
  {title} {{Machine learning topological invariants of non-Hermitian
  systems}},\ }\href {https://doi.org/10.1103/PhysRevA.103.012419} {\bibfield
  {journal} {\bibinfo  {journal} {Phys. Rev. A}\ }\textbf {\bibinfo {volume}
  {103}},\ \bibinfo {pages} {012419} (\bibinfo {year} {2021})}\BibitemShut
  {NoStop}%
\bibitem [{\citenamefont {Lian}\ \emph {et~al.}(2019)\citenamefont {Lian},
  \citenamefont {Wang}, \citenamefont {Lu}, \citenamefont {Huang},
  \citenamefont {Wang}, \citenamefont {Yuan}, \citenamefont {Zhang},
  \citenamefont {Ouyang}, \citenamefont {Wang}, \citenamefont {Huang},
  \citenamefont {He}, \citenamefont {Chang}, \citenamefont {Deng},\ and\
  \citenamefont {Duan}}]{PhysRevLett.122.210503}%
  \BibitemOpen
  \bibfield  {author} {\bibinfo {author} {\bibfnamefont {W.}~\bibnamefont
  {Lian}}, \bibinfo {author} {\bibfnamefont {S.-T.}\ \bibnamefont {Wang}},
  \bibinfo {author} {\bibfnamefont {S.}~\bibnamefont {Lu}}, \bibinfo {author}
  {\bibfnamefont {Y.}~\bibnamefont {Huang}}, \bibinfo {author} {\bibfnamefont
  {F.}~\bibnamefont {Wang}}, \bibinfo {author} {\bibfnamefont {X.}~\bibnamefont
  {Yuan}}, \bibinfo {author} {\bibfnamefont {W.}~\bibnamefont {Zhang}},
  \bibinfo {author} {\bibfnamefont {X.}~\bibnamefont {Ouyang}}, \bibinfo
  {author} {\bibfnamefont {X.}~\bibnamefont {Wang}}, \bibinfo {author}
  {\bibfnamefont {X.}~\bibnamefont {Huang}}, \bibinfo {author} {\bibfnamefont
  {L.}~\bibnamefont {He}}, \bibinfo {author} {\bibfnamefont {X.}~\bibnamefont
  {Chang}}, \bibinfo {author} {\bibfnamefont {D.-L.}\ \bibnamefont {Deng}},\
  and\ \bibinfo {author} {\bibfnamefont {L.}~\bibnamefont {Duan}},\ }\bibfield
  {title} {\bibinfo {title} {{Machine Learning Topological Phases with a
  Solid-State Quantum Simulator}},\ }\href
  {https://doi.org/10.1103/PhysRevLett.122.210503} {\bibfield  {journal}
  {\bibinfo  {journal} {Phys. Rev. Lett.}\ }\textbf {\bibinfo {volume} {122}},\
  \bibinfo {pages} {210503} (\bibinfo {year} {2019})}\BibitemShut {NoStop}%
\bibitem [{\citenamefont {Rodriguez-Nieva}\ and\ \citenamefont
  {Scheurer}(2019)}]{rodriguez2019identifying}%
  \BibitemOpen
  \bibfield  {author} {\bibinfo {author} {\bibfnamefont {J.~F.}\ \bibnamefont
  {Rodriguez-Nieva}}\ and\ \bibinfo {author} {\bibfnamefont {M.~S.}\
  \bibnamefont {Scheurer}},\ }\bibfield  {title} {\bibinfo {title}
  {{Identifying topological order through unsupervised machine learning}},\
  }\href {https://www.nature.com/articles/s41567-019-0512-x} {\bibfield
  {journal} {\bibinfo  {journal} {Nat. Phys.}\ }\textbf {\bibinfo {volume}
  {15}},\ \bibinfo {pages} {790} (\bibinfo {year} {2019})}\BibitemShut
  {NoStop}%
\bibitem [{\citenamefont {Scheurer}\ and\ \citenamefont
  {Slager}(2020)}]{PhysRevLett.124.226401}%
  \BibitemOpen
  \bibfield  {author} {\bibinfo {author} {\bibfnamefont {M.~S.}\ \bibnamefont
  {Scheurer}}\ and\ \bibinfo {author} {\bibfnamefont {R.-J.}\ \bibnamefont
  {Slager}},\ }\bibfield  {title} {\bibinfo {title} {{Unsupervised Machine
  Learning and Band Topology}},\ }\href
  {https://doi.org/10.1103/PhysRevLett.124.226401} {\bibfield  {journal}
  {\bibinfo  {journal} {Phys. Rev. Lett.}\ }\textbf {\bibinfo {volume} {124}},\
  \bibinfo {pages} {226401} (\bibinfo {year} {2020})}\BibitemShut {NoStop}%
\bibitem [{\citenamefont {Che}\ \emph {et~al.}(2020)\citenamefont {Che},
  \citenamefont {Gneiting}, \citenamefont {Liu},\ and\ \citenamefont
  {Nori}}]{che2020topological}%
  \BibitemOpen
  \bibfield  {author} {\bibinfo {author} {\bibfnamefont {Y.}~\bibnamefont
  {Che}}, \bibinfo {author} {\bibfnamefont {C.}~\bibnamefont {Gneiting}},
  \bibinfo {author} {\bibfnamefont {T.}~\bibnamefont {Liu}},\ and\ \bibinfo
  {author} {\bibfnamefont {F.}~\bibnamefont {Nori}},\ }\bibfield  {title}
  {\bibinfo {title} {Topological quantum phase transitions retrieved through
  unsupervised machine learning},\ }\href
  {https://doi.org/10.1103/PhysRevB.102.134213} {\bibfield  {journal} {\bibinfo
   {journal} {Phys. Rev. B}\ }\textbf {\bibinfo {volume} {102}},\ \bibinfo
  {pages} {134213} (\bibinfo {year} {2020})}\BibitemShut {NoStop}%
\bibitem [{\citenamefont {Long}\ \emph {et~al.}(2020)\citenamefont {Long},
  \citenamefont {Ren},\ and\ \citenamefont {Chen}}]{PhysRevLett.124.185501}%
  \BibitemOpen
  \bibfield  {author} {\bibinfo {author} {\bibfnamefont {Y.}~\bibnamefont
  {Long}}, \bibinfo {author} {\bibfnamefont {J.}~\bibnamefont {Ren}},\ and\
  \bibinfo {author} {\bibfnamefont {H.}~\bibnamefont {Chen}},\ }\bibfield
  {title} {\bibinfo {title} {{Unsupervised Manifold Clustering of Topological
  Phononics}},\ }\href {https://doi.org/10.1103/PhysRevLett.124.185501}
  {\bibfield  {journal} {\bibinfo  {journal} {Phys. Rev. Lett.}\ }\textbf
  {\bibinfo {volume} {124}},\ \bibinfo {pages} {185501} (\bibinfo {year}
  {2020})}\BibitemShut {NoStop}%
\bibitem [{\citenamefont {Lidiak}\ and\ \citenamefont
  {Gong}(2020)}]{lidiak2020unsupervised}%
  \BibitemOpen
  \bibfield  {author} {\bibinfo {author} {\bibfnamefont {A.}~\bibnamefont
  {Lidiak}}\ and\ \bibinfo {author} {\bibfnamefont {Z.}~\bibnamefont {Gong}},\
  }\bibfield  {title} {\bibinfo {title} {Unsupervised machine learning of
  quantum phase transitions using diffusion maps},\ }\href
  {https://doi.org/10.1103/PhysRevLett.125.225701} {\bibfield  {journal}
  {\bibinfo  {journal} {Phys. Rev. Lett.}\ }\textbf {\bibinfo {volume} {125}},\
  \bibinfo {pages} {225701} (\bibinfo {year} {2020})}\BibitemShut {NoStop}%
\bibitem [{\citenamefont {Fukushima}\ \emph {et~al.}(2019)\citenamefont
  {Fukushima}, \citenamefont {Funai},\ and\ \citenamefont
  {Iida}}]{fukushima2019featuring}%
  \BibitemOpen
  \bibfield  {author} {\bibinfo {author} {\bibfnamefont {K.}~\bibnamefont
  {Fukushima}}, \bibinfo {author} {\bibfnamefont {S.~S.}\ \bibnamefont
  {Funai}},\ and\ \bibinfo {author} {\bibfnamefont {H.}~\bibnamefont {Iida}},\
  }\bibfield  {title} {\bibinfo {title} {Featuring the topology with the
  unsupervised machine learning},\ }\href
  {https://arxiv.org/pdf/1908.00281.pdf} {\bibfield  {journal} {\bibinfo
  {journal} {arXiv preprint arXiv:1908.00281}\ } (\bibinfo {year}
  {2019})}\BibitemShut {NoStop}%
\bibitem [{\citenamefont {Sch\"afer}\ and\ \citenamefont
  {L\"orch}(2019)}]{PhysRevE.99.062107}%
  \BibitemOpen
  \bibfield  {author} {\bibinfo {author} {\bibfnamefont {F.}~\bibnamefont
  {Sch\"afer}}\ and\ \bibinfo {author} {\bibfnamefont {N.}~\bibnamefont
  {L\"orch}},\ }\bibfield  {title} {\bibinfo {title} {Vector field divergence
  of predictive model output as indication of phase transitions},\ }\href
  {https://doi.org/10.1103/PhysRevE.99.062107} {\bibfield  {journal} {\bibinfo
  {journal} {Phys. Rev. E}\ }\textbf {\bibinfo {volume} {99}},\ \bibinfo
  {pages} {062107} (\bibinfo {year} {2019})}\BibitemShut {NoStop}%
\bibitem [{\citenamefont {Balabanov}\ and\ \citenamefont
  {Granath}(2020)}]{PhysRevResearch.2.013354}%
  \BibitemOpen
  \bibfield  {author} {\bibinfo {author} {\bibfnamefont {O.}~\bibnamefont
  {Balabanov}}\ and\ \bibinfo {author} {\bibfnamefont {M.}~\bibnamefont
  {Granath}},\ }\bibfield  {title} {\bibinfo {title} {Unsupervised learning
  using topological data augmentation},\ }\href
  {https://doi.org/10.1103/PhysRevResearch.2.013354} {\bibfield  {journal}
  {\bibinfo  {journal} {Phys. Rev. Research}\ }\textbf {\bibinfo {volume}
  {2}},\ \bibinfo {pages} {013354} (\bibinfo {year} {2020})}\BibitemShut
  {NoStop}%
\bibitem [{\citenamefont {Alexandrou}\ \emph {et~al.}(2020)\citenamefont
  {Alexandrou}, \citenamefont {Athenodorou}, \citenamefont {Chrysostomou},\
  and\ \citenamefont {Paul}}]{alexandrou2020critical}%
  \BibitemOpen
  \bibfield  {author} {\bibinfo {author} {\bibfnamefont {C.}~\bibnamefont
  {Alexandrou}}, \bibinfo {author} {\bibfnamefont {A.}~\bibnamefont
  {Athenodorou}}, \bibinfo {author} {\bibfnamefont {C.}~\bibnamefont
  {Chrysostomou}},\ and\ \bibinfo {author} {\bibfnamefont {S.}~\bibnamefont
  {Paul}},\ }\bibfield  {title} {\bibinfo {title} {The critical temperature of
  the 2d-ising model through deep learning autoencoders},\ }\href
  {https://link.springer.com/article/10.1140/epjb/e2020-100506-5#citeas}
  {\bibfield  {journal} {\bibinfo  {journal} {Eur. Phys. J. B}\ }\textbf
  {\bibinfo {volume} {93}},\ \bibinfo {pages} {1} (\bibinfo {year}
  {2020})}\BibitemShut {NoStop}%
\bibitem [{\citenamefont {Greplova}\ \emph {et~al.}(2020)\citenamefont
  {Greplova}, \citenamefont {Valenti}, \citenamefont {Boschung}, \citenamefont
  {Sch{\"a}fer}, \citenamefont {L{\"o}rch},\ and\ \citenamefont
  {Huber}}]{greplova2020unsupervised}%
  \BibitemOpen
  \bibfield  {author} {\bibinfo {author} {\bibfnamefont {E.}~\bibnamefont
  {Greplova}}, \bibinfo {author} {\bibfnamefont {A.}~\bibnamefont {Valenti}},
  \bibinfo {author} {\bibfnamefont {G.}~\bibnamefont {Boschung}}, \bibinfo
  {author} {\bibfnamefont {F.}~\bibnamefont {Sch{\"a}fer}}, \bibinfo {author}
  {\bibfnamefont {N.}~\bibnamefont {L{\"o}rch}},\ and\ \bibinfo {author}
  {\bibfnamefont {S.~D.}\ \bibnamefont {Huber}},\ }\bibfield  {title} {\bibinfo
  {title} {Unsupervised identification of topological phase transitions using
  predictive models},\ }\href
  {https://iopscience.iop.org/article/10.1088/1367-2630/ab7771/meta} {\bibfield
   {journal} {\bibinfo  {journal} {New Journal of Physics}\ }\textbf {\bibinfo
  {volume} {22}},\ \bibinfo {pages} {045003} (\bibinfo {year}
  {2020})}\BibitemShut {NoStop}%
\bibitem [{\citenamefont {Arnold}\ \emph {et~al.}(2020)\citenamefont {Arnold},
  \citenamefont {Sch{\"a}fer}, \citenamefont {{\v{Z}}onda},\ and\ \citenamefont
  {Lode}}]{arnold2020interpretable}%
  \BibitemOpen
  \bibfield  {author} {\bibinfo {author} {\bibfnamefont {J.}~\bibnamefont
  {Arnold}}, \bibinfo {author} {\bibfnamefont {F.}~\bibnamefont {Sch{\"a}fer}},
  \bibinfo {author} {\bibfnamefont {M.}~\bibnamefont {{\v{Z}}onda}},\ and\
  \bibinfo {author} {\bibfnamefont {A.~U.}\ \bibnamefont {Lode}},\ }\bibfield
  {title} {\bibinfo {title} {Interpretable and unsupervised phase
  classification},\ }\href {https://arxiv.org/abs/2010.04730} {\bibfield
  {journal} {\bibinfo  {journal} {arXiv preprint arXiv:2010.04730}\ } (\bibinfo
  {year} {2020})}\BibitemShut {NoStop}%
\bibitem [{\citenamefont {Kottmann}\ \emph {et~al.}(2020)\citenamefont
  {Kottmann}, \citenamefont {Huembeli}, \citenamefont {Lewenstein},\ and\
  \citenamefont {Ac\'{\i}n}}]{Kottmann2020Unsupervised}%
  \BibitemOpen
  \bibfield  {author} {\bibinfo {author} {\bibfnamefont {K.}~\bibnamefont
  {Kottmann}}, \bibinfo {author} {\bibfnamefont {P.}~\bibnamefont {Huembeli}},
  \bibinfo {author} {\bibfnamefont {M.}~\bibnamefont {Lewenstein}},\ and\
  \bibinfo {author} {\bibfnamefont {A.}~\bibnamefont {Ac\'{\i}n}},\ }\bibfield
  {title} {\bibinfo {title} {Unsupervised phase discovery with deep anomaly
  detection},\ }\href {https://doi.org/10.1103/PhysRevLett.125.170603}
  {\bibfield  {journal} {\bibinfo  {journal} {Phys. Rev. Lett.}\ }\textbf
  {\bibinfo {volume} {125}},\ \bibinfo {pages} {170603} (\bibinfo {year}
  {2020})}\BibitemShut {NoStop}%
\bibitem [{\citenamefont {Beach}\ \emph {et~al.}(2018)\citenamefont {Beach},
  \citenamefont {Golubeva},\ and\ \citenamefont {Melko}}]{Beach2018Machine}%
  \BibitemOpen
  \bibfield  {author} {\bibinfo {author} {\bibfnamefont {M.~J.~S.}\
  \bibnamefont {Beach}}, \bibinfo {author} {\bibfnamefont {A.}~\bibnamefont
  {Golubeva}},\ and\ \bibinfo {author} {\bibfnamefont {R.~G.}\ \bibnamefont
  {Melko}},\ }\bibfield  {title} {\bibinfo {title} {{Machine learning vortices
  at the Kosterlitz-Thouless transition}},\ }\href
  {https://doi.org/10.1103/PhysRevB.97.045207} {\bibfield  {journal} {\bibinfo
  {journal} {Phys. Rev. B}\ }\textbf {\bibinfo {volume} {97}},\ \bibinfo
  {pages} {045207} (\bibinfo {year} {2018})}\BibitemShut {NoStop}%
\bibitem [{\citenamefont {Coifman}\ \emph
  {et~al.}(2005{\natexlab{a}})\citenamefont {Coifman}, \citenamefont {Lafon},
  \citenamefont {Lee}, \citenamefont {Maggioni}, \citenamefont {Nadler},
  \citenamefont {Warner},\ and\ \citenamefont {Zucker}}]{Coifman7426}%
  \BibitemOpen
  \bibfield  {author} {\bibinfo {author} {\bibfnamefont {R.~R.}\ \bibnamefont
  {Coifman}}, \bibinfo {author} {\bibfnamefont {S.}~\bibnamefont {Lafon}},
  \bibinfo {author} {\bibfnamefont {A.~B.}\ \bibnamefont {Lee}}, \bibinfo
  {author} {\bibfnamefont {M.}~\bibnamefont {Maggioni}}, \bibinfo {author}
  {\bibfnamefont {B.}~\bibnamefont {Nadler}}, \bibinfo {author} {\bibfnamefont
  {F.}~\bibnamefont {Warner}},\ and\ \bibinfo {author} {\bibfnamefont {S.~W.}\
  \bibnamefont {Zucker}},\ }\bibfield  {title} {\bibinfo {title} {{Geometric
  diffusions as a tool for harmonic analysis and structure definition of data:
  Diffusion maps}},\ }\href {https://doi.org/10.1073/pnas.0500334102}
  {\bibfield  {journal} {\bibinfo  {journal} {Proc. Natl. Acad. Sci. USA}\
  }\textbf {\bibinfo {volume} {102}},\ \bibinfo {pages} {7426} (\bibinfo {year}
  {2005}{\natexlab{a}})}\BibitemShut {NoStop}%
\bibitem [{\citenamefont {Coifman}\ \emph
  {et~al.}(2005{\natexlab{b}})\citenamefont {Coifman}, \citenamefont {Lafon},
  \citenamefont {Lee}, \citenamefont {Maggioni}, \citenamefont {Nadler},
  \citenamefont {Warner},\ and\ \citenamefont {Zucker}}]{Coifman7432}%
  \BibitemOpen
  \bibfield  {author} {\bibinfo {author} {\bibfnamefont {R.~R.}\ \bibnamefont
  {Coifman}}, \bibinfo {author} {\bibfnamefont {S.}~\bibnamefont {Lafon}},
  \bibinfo {author} {\bibfnamefont {A.~B.}\ \bibnamefont {Lee}}, \bibinfo
  {author} {\bibfnamefont {M.}~\bibnamefont {Maggioni}}, \bibinfo {author}
  {\bibfnamefont {B.}~\bibnamefont {Nadler}}, \bibinfo {author} {\bibfnamefont
  {F.}~\bibnamefont {Warner}},\ and\ \bibinfo {author} {\bibfnamefont {S.~W.}\
  \bibnamefont {Zucker}},\ }\bibfield  {title} {\bibinfo {title} {{Geometric
  diffusions as a tool for harmonic analysis and structure definition of data:
  Multiscale methods}},\ }\href {https://doi.org/10.1073/pnas.0500896102}
  {\bibfield  {journal} {\bibinfo  {journal} {Proc. Natl. Acad. Sci. USA}\
  }\textbf {\bibinfo {volume} {102}},\ \bibinfo {pages} {7432} (\bibinfo {year}
  {2005}{\natexlab{b}})}\BibitemShut {NoStop}%
\bibitem [{\citenamefont {Coifman}\ and\ \citenamefont
  {Lafon}(2006)}]{coifman2006diffusion}%
  \BibitemOpen
  \bibfield  {author} {\bibinfo {author} {\bibfnamefont {R.~R.}\ \bibnamefont
  {Coifman}}\ and\ \bibinfo {author} {\bibfnamefont {S.}~\bibnamefont
  {Lafon}},\ }\bibfield  {title} {\bibinfo {title} {{Diffusion maps}},\ }\href
  {https://www.sciencedirect.com/science/article/pii/S1063520306000546}
  {\bibfield  {journal} {\bibinfo  {journal} {Appl. Comput. Harmon. Anal.}\
  }\textbf {\bibinfo {volume} {21}},\ \bibinfo {pages} {5} (\bibinfo {year}
  {2006})}\BibitemShut {NoStop}%
\bibitem [{Note1()}]{Note1}%
  \BibitemOpen
  \bibinfo {note} {Note that for the diffusion map approach, two samples far
  from each other may have considerable diffusion probability with the
  assistance of symmetry, as discussed in Ref.\cite {PhysRevLett.124.226401}.
  However, in this work we do not assume that the model Hamiltonian has certain
  symmetry and thus the effect of symmetries will not be discussed for
  simplicity.}\BibitemShut {Stop}%
\bibitem [{USM()}]{USMLNonHTopSupp}%
  \BibitemOpen
  \href@noop {} {}\bibinfo {note} {See Supplemental Material at [URL will be
  inserted by publisher] for details on the introduction of diffusion map,
  non-Hermitian topological phases of matter, theoretical analysis and more
  numerical calculations of clustering results.}\BibitemShut {Stop}%
\bibitem [{\citenamefont {Bergholtz}\ \emph {et~al.}(2019)\citenamefont
  {Bergholtz}, \citenamefont {Budich},\ and\ \citenamefont
  {Kunst}}]{Bergholtz2019Exceptional}%
  \BibitemOpen
  \bibfield  {author} {\bibinfo {author} {\bibfnamefont {E.~J.}\ \bibnamefont
  {Bergholtz}}, \bibinfo {author} {\bibfnamefont {J.~C.}\ \bibnamefont
  {Budich}},\ and\ \bibinfo {author} {\bibfnamefont {F.~K.}\ \bibnamefont
  {Kunst}},\ }\bibfield  {title} {\bibinfo {title} {Exceptional topology of
  non-hermitian systems},\ }\href {https://arxiv.org/abs/1912.10048} {\bibfield
   {journal} {\bibinfo  {journal} {arXiv:1912.10048}\ } (\bibinfo {year}
  {2019})}\BibitemShut {NoStop}%
\bibitem [{\citenamefont {Zhang}\ \emph {et~al.}(2020)\citenamefont {Zhang},
  \citenamefont {Yang},\ and\ \citenamefont {Fang}}]{Zhang2020Correspondence}%
  \BibitemOpen
  \bibfield  {author} {\bibinfo {author} {\bibfnamefont {K.}~\bibnamefont
  {Zhang}}, \bibinfo {author} {\bibfnamefont {Z.}~\bibnamefont {Yang}},\ and\
  \bibinfo {author} {\bibfnamefont {C.}~\bibnamefont {Fang}},\ }\bibfield
  {title} {\bibinfo {title} {{Correspondence between Winding Numbers and Skin
  Modes in Non-Hermitian Systems}},\ }\href
  {https://doi.org/10.1103/PhysRevLett.125.126402} {\bibfield  {journal}
  {\bibinfo  {journal} {Phys. Rev. Lett.}\ }\textbf {\bibinfo {volume} {125}},\
  \bibinfo {pages} {126402} (\bibinfo {year} {2020})}\BibitemShut {NoStop}%
\bibitem [{\citenamefont {Kitaev}(2009)}]{Kitaev2009Periodic}%
  \BibitemOpen
  \bibfield  {author} {\bibinfo {author} {\bibfnamefont {A.}~\bibnamefont
  {Kitaev}},\ }\bibfield  {title} {\bibinfo {title} {Periodic table for
  topological insulators and superconductors},\ }\href
  {https://doi.org/10.1063/1.3149495} {\bibfield  {journal} {\bibinfo
  {journal} {AIP Conf. Proc.}\ }\textbf {\bibinfo {volume} {1134}},\ \bibinfo
  {pages} {22} (\bibinfo {year} {2009})}\BibitemShut {NoStop}%
\bibitem [{Note2()}]{Note2}%
  \BibitemOpen
  \bibinfo {note} {Here for simplicity, we did not take into account the effect
  of symmetry in phase clustering. In diffusion map algorithm, two samples far
  from each other may have considerable diffusion probability with the
  assistance of symmetry. For details, see {\protect \em
  PRL.124.226401}.}\BibitemShut {Stop}%
\bibitem [{Note3()}]{Note3}%
  \BibitemOpen
  \bibinfo {note} {For non-Hermitian topological systems with PBC, there also
  exist gapless phases. For example, the 2D non-Hermitian Qi-Wu-Zhang(QWZ)
  model has gapless phases.}\BibitemShut {Stop}%
\end{thebibliography}%

\clearpage
\onecolumngrid
\makeatletter
\setcounter{figure}{0}
\setcounter{equation}{0}
\renewcommand{\thefigure}{S\@arabic\c@figure}
\renewcommand \theequation{S\@arabic\c@equation}
\renewcommand \thetable{S\@arabic\c@table}

\begin{center} 
	{\large \bf Supplementary Material for:  Unsupervised Learning of Non-Hermitian Topological Phases}
\end{center}

Diffusion map is a sort of typical manifold learning algorithm \cite{Coifman7426,Coifman7432,coifman2006diffusion}, which provides non-linear dimensionality reduction and unsupervised clustering of raw data  without any priori knowledge. 
It combines the heat diffusion with the random walk Markov chain.  Concretely, given a set of input data ${\bf x } = \{{\bf x}^{(1)},\,{\bf x}^{(2)},\,\cdots {\bf x}^{(L)} \}$, where ${\bf x}^{(i)}$ represents the $i$-th data point in complex space $\mathbb{C}^d$.   The connectivity between two points ${\bf x}^{(l)}$ and ${\bf x}^{(l')}$ is described by the local similarity, which is required to be positive definite and symmetric.  For example,  the Gaussian kernel
\begin{equation}
\mathcal{K}_{l,l'} = \exp\left(-\frac{\|{\bf x}^{(l)}-{\bf x}^{(l')}\|_{\mathbb{L}_p}^2}{2\epsilon}\right),
\end{equation}
where $\|{\bf x}^{(l)}-{\bf x}^{(l')}\|_{\mathbb{L}_p}$ represents the $\mathbb{L}_p$-norm distance between two points ${\bf x}^{(l)}$ and ${\bf x}^{(l')}$, variance $\epsilon$ is a small quantity to be adjusted. Recently, applications of  $p=1,\, 2, \infty$ cases in unsupervised clustering topological phases have been reported \cite{rodriguez2019identifying,PhysRevLett.124.226401,che2020topological,PhysRevLett.124.185501}. When $p=2$, the distance is the familiar Euclidean distance. With such kernel, the one-step transition matrix $\mathcal{P}$ of Markovian random walk  between two points ${\bf x}^{(l)}$ and ${\bf x}^{(l')}$ can be defined as follows
\begin{equation}
\mathcal{P}_{l,l'}=\frac{\mathcal{K}_{l,l'}}{\sum_{l'}\mathcal{K}_{l,l'}},
\end{equation}
where $\mathcal{P}_{l,l'}$ obeys the probability conservation condition $\sum_{l}\mathcal{P}_{l,l'}=1$.  Then after $2t$ steps of random walk, the connectivity between ${\bf x}^{(l)}$ and ${\bf x}^{(l')}$ is given by the diffusion distance
\begin{equation}
\begin{aligned}
D_{t}(l, l')=D_{t}({\bf x}^{(l)}, {\bf x}^{(l')})= \sum_{k=1}^{L}\frac{\left(\mathcal{P}^t_{l,k}-\mathcal{P}^t_{l',k}\right)^2}{\sum_{j}\mathcal{K}_{k,j}}= \sum_{k=1}^{L-1}\lambda_{k}^{2t}[(\psi_k)_l - (\psi_k)_{l'}]^2\geq 0,
\end{aligned}
\end{equation}
where  $\{\psi_k\}$ are the right eigenvectors of $\mathcal{P}$, $\mathcal{P}\psi_k =\lambda_k \psi_k$, $k=0,1,... L-1$, the corresponding eigenvalues rank in descending order, i.e. $\lambda_0=1\geq \lambda_1\geq\cdots \geq \lambda_{L-1}$.  $k=0$ term does not contribute because the corresponding right eigenvector is constant with all vector elements equivalent. 

Under the mapping 
\begin{equation}\label{diff_map_pbc}
{\bf x}^{(l)}\rightarrow \Psi_t^{(l)} : = [\lambda_1^t(\psi_1)_l, \lambda_2^t(\psi_2)_l, \cdots , \lambda_{L-1}^t(\psi_{L-1})_l],
\end{equation}
the distance between samples ${\bf x}^{(l)}$ and ${\bf x}^{(l')}$ can be recast as the Euclidean distance in $\Psi$ space
\begin{equation}
D_{t}({\bf x}^{(l)}, {\bf x}^{(l')}) =\|\Psi_l - \Psi_{l'}\|^2_{\mathbb{L}_2}.
\end{equation}
After $t\rightarrow \infty$ steps, only the first few components  with largest $|\lambda_k|\approx 1$ are dominant due to the term $\lambda_k^t$ in $\Psi_t$. Hence almost all the distance information is encoded in such few components. Then the original samples ${\bf x}^{(l)}$ with higher dimension are reduced to the lower ones, and the clustering method (e.g. $k$-means) can be applied in $\Psi$ space to cluster the corresponding samples with no prior knowledge.  Specifically, in clustering the topological phases of quantum models, the number of $|\lambda_k|\approx1$ equals to the number of topological clusters without prior labels. Hence it is possible for such algorithm to detect unknown topological phases.
\section{Phases of non-Hermitian topological models}
\subsection{1D non-Hermitian Su-Schrieffer-Heeger model}

\subsubsection{Periodic boundary condition}

With periodic boundary condition(PBC), the system obeys translational symmetry,  and the bulk Hamiltonian  takes the form
\begin{equation}
\begin{aligned}
\hat{H}_p^{1D}(k)&=\vec{{\bf d}}\cdot {\bf \sigma} = d_x\sigma_x + d_y\sigma_y 
%d_x &= t_1+(t_2+t_3)\cos k,\\
%d_y &= (t_2-t_3)\sin k+ i\gamma
\end{aligned}
\end{equation} 
in the momentum bases $\{C_{k,A}, C_{k,B}\}$,  where $C_{k,A}$ and $C_{k,B}$ represent the fermionic sublattice sites, $\vec{{\bf d}} = (d_x,\, d_y)$, $d_x = t_1+t_2\cos k$, $d_y = t_2\sin k+ i\gamma$. %Without loss of generality, let us suppose  $t_3=0$. %This model holds the sublattice symmetry(subtle different from chiral symmetry\cite{PhysRevX.8.031079}) $\mathcal{S}=\sigma_z$, which obeys $\mathcal{S}H_p\mathcal{S}^{-1}=-H_p$.  
The 1D Brillouin zone is given by $[-\pi,\, \pi]$. %the unit vector $\hat{\bf d}(k) = \frac{\vec{\bf d}}{\sqrt{d_x^2+d_y^2}}$ maps the compact Brillouin zone $\mathbb{S}^1$ to the unit circle $\mathbb{S}^1$. 
 Winding number $W=\frac{1}{\pi}\int_{-\pi}^{\pi} \tfrac{d_x\partial_k d_y-d_y\partial_k d_x}{d_x^2+d_y^2} dk$ counts the times of the mapping wrapping around the original point. Different winding numbers indicate different topological phases of matter:
\begin{equation}
W = \left\{ \begin{aligned} 
&0,\quad |t_1\pm \gamma| >|t_2|;\\ &\frac{1}{2},\quad |t_1\pm\gamma|<|t_2| \& |\gamma|>|t_2|; \\ &1,\quad |t_1\pm\gamma|<|t_2| \& |\gamma|<|t_2|.\end{aligned}\right.
\end{equation}
The above half-winding number has a geometrical explanation, see Ref.~\cite{PhysRevA.97.052115} for details. For the non-Hermitian system with PBC, the sublattice symmetry \cite{PhysRevX.8.031079} $\mathcal{S}=\sigma_z$, $\sigma_z \hat{H}_p^{(1D)}(k) \sigma_z = -\hat{H}_p^{(1D)}(k)$ ensures that the  bulk bands are in pairs $E_{\pm}(k)=\pm\sqrt{(t_1+t_2\cos k)^2+(t_2 \sin k +i\gamma)^2}$.  The topological phase transition occurs at the exceptional points $t_1= -t_2\pm \gamma$ ($k=0$) and $t_1= t_2\pm \gamma$ ($k=\pi$), which coincide with the change of winding numbers. 
\subsubsection{Open boundary condition}
The 1D non-Hermitian SSH model in the real space takes the following form:
\begin{equation}\label{app_nhssh_obc}
\begin{aligned}
\hat{H}_{o}^{1D}=\sum_{i=1}^N(t_1+\gamma)C^\dag_{i,A}C_{i,B}+(t_1-\gamma)C^\dag_{i,B}C_{i,A}+t_2C^\dag_{i,B}C_{i+1,A}+t_2C^\dag_{i+1,A}C_{i,B}.
\end{aligned}
\end{equation}
With OBC, the conventional bulk-boundary correspondence(BBC) breaks down, and the non-Hermitian skin effect arises. Theoretically, one straightforward method of detecting the topological phase transition is to calculate its ground state degeneracy. Direct numerical calculation of Eq.~(\ref{app_nhssh_obc}) with a large enough chain length shows that the 2-fold ground state degeneracy holds in the interval $|t_1|< \sqrt{t_2^2+\gamma^2}$. Hence the phase transition occurs at $t_1 = \pm \sqrt{t_2^2+\gamma^2}$. This phase boundary does no-longer correspond to the exceptional point in Brillouin zone, but one can reconstruct the generalized BBC based on a similarity transformation $\Gamma$ on the Hamiltonian matrix represented in bases $\{C_{1,A},C_{1,B}, C_{2,A},\cdots, C_{N,B}\}^T$, 
\begin{equation}
\bar{H}_o^{1D} = \Gamma^{-1}\hat{H}_o^{1D} \Gamma,
\end{equation}
where $\Gamma$  is a $2N\times 2N$ diagonal matrix with the diagonal elements $\{1, r , r,r^2, r^2, ..., r^{N-1}, r^{N-1}, r^{N}\}$, and $r =\sqrt{|(t_1-\gamma)/(t_1+\gamma)|} $. Then the non-Hermitian matrix $\hat{H}^{1D}_o$ is transformed into a Hermitian matrix $\bar{H}_o^{1D}$ by $S$, with the trade-off that the bases are transformed to be non-orthonormal as $\{C_{1,A},r^{-1}C_{1,B}, r^{-1}C_{2,A},\cdots, r^{-N}C_{N,B}\}^T$. When $|t_1|>|\gamma|$, i.e. $r$ is real, the  Hermitian $\bar{H}_o^{1D}$ is exactly the SSH matrix.  
Fourier transformation maps the real space  SSH matrix  $\bar{H}_o^{1D}$ to the momentum space,
\begin{equation}\label{HK_OBC}
\bar{H}_o^{1D}(k) = (\bar{t}_1+t_2 \cos k)\sigma_x + t_2 \sin k \sigma_y,\quad \bar{t}_1=\sqrt{(t_1+\gamma)(t_1-\gamma)}.
\end{equation} 
Clearly,  the topological phase transition of $\bar{H}_o^{1D}(k)$ occurs at $|\bar{t}_1| = |t_2|$, i.e. $t_1=\pm\sqrt{t_2^2+\gamma^2}$, with the change of topological winding numbers.  Hence to discuss the BBC for such non-Hermitian model, one approach is to transform it into the Hermitian matrix formalism under the similarity transformation.

Another equivalent approach to define the non-Hermitian BBC is to introduce the concept of generalized Brillouin zone $H(e^{ik}\rightarrow \beta)$, which transforms the Bloch phase factor $e^{ik}$ into  $|\beta|\neq1$ for the non-Hermitian case,
\begin{equation}
e^{ik}:= \beta = \sqrt{\frac{t_1-\gamma}{t_1+\gamma}}e^{iq}, \quad q\in \mathbb{R}.
\end{equation} 
Under such transformation,  the matrix form of  $H'(\beta)$ in generalized Brillouin zone is exactly the same as Eq.~(\ref{HK_OBC}). Then the generalized BBC holds.

\subsection{2D non-Hermitian Qi-Wu-Zhang model}
\subsubsection{Periodic boundary condition}
The  non-Hermitian Qi-Wu-Zhang (QWZ) model takes the following form\cite{PhysRevLett.121.136802},
\begin{equation}\label{QWZreal}
\begin{aligned}
\hat{H}^{2D}=& [\sum_{\bf y}\sum_{j=x,y}c_{\bf y}^\dag(-\frac{i}{2} v_j\sigma_j-\frac{1}{2}t_j\sigma_z)c_{\bf y+e_j}+h.c.] +\sum_{\bf y}c_{\bf y}^\dag(M\sigma_z+i\sum_{j=x,y,z}\gamma_j\sigma_j)c_{\bf y}, 
\end{aligned}
\end{equation}
where $c_{\bf y} = (c_{{\bf y},A}, c_{{\bf y},B})^T$ represents the unit cell fermionic annihilation operator, ${\bf y}$  labels the 2D lattice number of the unit cell, and ${\bf e}_j$ is the unit vector in $j=x, y$ directions. Here we also consider two types of boundary conditions: PBC and OBC.

With PBC, the non-Hermitian QWZ model in Eq.~(\ref{QWZreal}) is Fourier transformed into the bulk Hamiltonian in Brillouin zone, 
\begin{equation}\label{2DNHQWZ}
\begin{aligned}
H_p({\bf k})=&\vec{{\bf d}}\cdot \vec{\sigma} = d_x\sigma_x+d_y\sigma_y+d_z\sigma_z\\
=&(v_x\sin k_x +i\gamma_x)\sigma_x+(v_y\sin k_y +i\gamma_y)\sigma_y\\
&+(M-t_x\cos k_x-t_y\cos k_y +i\gamma_z)\sigma_z. 
\end{aligned}
\end{equation}
For simplicity, let $\gamma_z=0$, $t_x = t_y =0.5$, then the theoretical phase transition boundaries are 
\begin{equation}\begin{aligned}
&M^{(1)}_\pm = 1\pm \sqrt{\gamma_x^2+\gamma_y^2} &&\, (k_x, k_y)=(0,0),\\
&M^{(2)}_\pm =\pm \sqrt{\gamma_x^2+\gamma_y^2}&& \,(k_x, k_y)=(0,\pi),\, (\pi,0),\\
&M^{(3)}_\pm = -1\pm \sqrt{\gamma_x^2+\gamma_y^2}&& \, (k_x, k_y)=(\pi,\pi),
\end{aligned}
\end{equation}
One unique feature of such model is that in  regions $M\in (M_{-}^{(\nu)},\,M_{+}^{(\nu)})$ for $\nu=1,2,3$ the Hamiltonian is gapless and the topological indices are not well-defined  in  these regions. 
\subsubsection{Open boundary condition}
With OBC, the conventional BBC breaks down. We have to consider the influence of non-Hermitian skin effect. In OBC (independent of the geometrical configuration, such as square or disk, {\em etc.} \cite{PhysRevLett.121.136802}),  it has been shown from  both the analytical and numerical aspects that one topological phase boundary is approximately expressed as
 \begin{equation}
 M = t_x + t_y + \frac{t_x\gamma_x^2}{2v_x^2} + \frac{t_y\gamma_y^2}{2v_y^2}
 \end{equation}
for small $\frac{\gamma_{x(y)}}{v_{x(y)}}$.  When $M < t_x + t_y + \frac{t_x\gamma_x^2}{2v_x^2} + \frac{t_y\gamma_y^2}{2v_y^2}$, the corresponding Chern number in generalized Brillouin zone for the valence band ($Re(E)<0$) is $C=1$.  While for $M > t_x + t_y + \frac{t_x\gamma_x^2}{2v_x^2} + \frac{t_y\gamma_y^2}{2v_y^2}$, the topological index is $C=0$. Hence the  boundary separates the system into two distinct topological regions. For more details, see Ref. \cite{PhysRevLett.121.136802}.

\section{The applicability of diffusion map in non-Hermitian topological phase transition}
\label{sec:iii}
Here we  theoretically analyze the applicability of diffusion map in classifying the non-Hermitian topological phases.
\subsection{Periodic boundary condition}
With PBC, the 2-level topological band model reads
\begin{equation}
\hat{H} = \vec{\bf d}\cdot{\bf \sigma} = d_x\sigma_x+d_y\sigma_y + d_z\sigma_z
\end{equation}
with the free fermionic bases $\{C_{{\bf k},A},C_{{\bf k},B}\}$, where $\{d_x,d_y,d_z\}$ can be complex for the non-Hermitian model, the corresponding energy levels $E_{\pm}=\pm\sqrt{d_x^2+d_y^2+d_z^2}$. Then the Hamiltonian are equivalently described by the vector $\vec{{\bf d}}$ in Pauli space.
In the $d$-dimensional lattice model, the momentum vectors ${\bf k}$ are discrete in Brillouin zone $[-\pi,\pi]^d$.  To apply the diffusion map algorithm to such models, the  vectors $\hat{\bf d} = \frac{ \vec{\bf d}}{\sqrt{d_x^2+d_y^2+d_z^2}}$ are chosen as the raw data,  e.g. for the 1D model with length $N$,  the data sample ${\bf x}^{(l)} = \{\hat{\bf d}(k_i),| k_i =\frac{2i-N-2}{N}\pi, \, i\in[1,N] \}$. By varying the parameters $\vec{t} = (t_1,t_2,...)$ in $\hat{\bf d}$, one obtains the data set $\{{\bf x}^{(l)}\}$.

In our diffusion map algorithm, the diffusion probability $\mathcal{P}_{l,l'}=\frac{\mathcal{K}_{l,l'}}{\sum_{l'}\mathcal{K}_{l,l'}}$  between samples ${\bf x}^{(l)}$ and ${\bf x}^{(l')}$ is defined by choosing the Gaussian kernel function with the  $\mathbb{L}_1$-norm
\begin{equation}
\mathcal{K}_{l,l'} = \exp\left(-\frac{\|{\bf x}^{(l)}-{\bf x}^{(l')}\|_{\mathbb{L}_1}^2}{2\epsilon N^2}\right),
\end{equation}
where the variance is controlled by  $0 < \epsilon\ll 1$.

Importantly, without the assistance of symmetric operators \footnote{Here for simplicity, we did not take into account the effect of symmetry in phase clustering. In diffusion map algorithm, two samples far from each other  may have considerable diffusion probability with the assistance of  symmetry.  For details, see {\em PRL.124.226401}.},  it is easy to find that the prominent contributions of the one-step diffusion probability $P_{l, l'}$ are from those nearest samples $l'=l+\delta l$, i.e. the corresponding parameters $\vec{t}\,'=\vec{t}+\delta \vec{t}$,  then the  $\mathbb{L}_1$-norm distance  between $l$ and $l+\delta l$ can be approximately recast to
\begin{equation}\label{L1_approx}
\begin{aligned}
\|{\bf x}^{(l)}-{\bf x}^{(l+\delta l)}\|_{\mathbb{L}_1} &= \sum_{i=1}^N \sum_{\alpha=x,y,z} \left(\|\hat{d}_\alpha^{(l)}(k_i)-\hat{d}_\alpha^{(l+\delta l)}(k_i)\|_{\mathbb{L}_1}\right)\approx \sum_{i=1}^N\sum_{\alpha=x,y,z} \left(\|\nabla_{\vec{t}}\,(\hat{d}_\alpha^{(l)}(k_i))\|_{\mathbb{L}_1}\right)\cdot \delta{\vec{t}}.
\end{aligned}
\end{equation}
Then the Gaussian kernel $\mathcal{K}_{l,l+\delta l}$ reads
\begin{equation}
\mathcal{K}_{l,l+\delta l}\approx \exp\left(-\frac{\left(\sum_{i=1}^N\sum_{\alpha=x,y,z} \left(\|\nabla_{\vec{t}}\,(\hat{d}_\alpha^{(l)}(k_i))\|_{\mathbb{L}_1}\right)\cdot \delta{\vec{t}}\,\right)^2}{2\epsilon N^2}\right).
\end{equation}
In case of confusion, we note  that $\|\nabla_{\vec{t}} \, \hat{d}\|_{\mathbb{L}_1} = \left(\|\partial_{t_1} \hat{d}\|_{\mathbb{L}_1}, \|\partial_{t_2} \hat{d}\|_{\mathbb{L}_1},\dots \|\partial_{t_n} \hat{d}\|_{\mathbb{L}_1}\right)$. 

As long as  ${\sum_{i=1}^N \left(|\nabla_{\vec{t}}\,(\hat{d}_x^{(l)}(k_i))|+|\nabla_{\vec{t}}\,(\hat{d}_y^{(l)}(k_i))|+|\nabla_{\vec{t}}\,(\hat{d}_z^{(l)}(k_i))|\right)}$ is finite, the constant $\delta {\vec{t}\,}^{2}/\epsilon$ can always be adjusted so as to keep $K_{l,l+\delta l}\approx 1$.  Hence the connectivity between $l$ and $l'$ depends on the derivability of the  vector $\hat{\bf d}^{(l)} = \frac{ \vec{\bf d}}{\sqrt{d_x^2+d_y^2+d_z^2}}=\frac{ {\bf d}^{(l)}}{E_+^{(l)}}$ on $\vec{t}$ for all $k_i \in [-\pi,\pi]$.  The gap closure points $E_{\pm}^{(l)}=0$ usually break the constraint. Hence two nearest data samples divided by the gap closure point should have  $\mathcal{K}_{l,l'}\approx 0$, i.e. no one-step diffusion probability between such samples. Combined with the approximation that only the nearest samples prominently contribute to the diffusion,   as a consequence, the  gap closure points divide the diffusion matrix into blocks. In most cases, different blocks usually correspond to different topological phases both for Hermitian and non-Hermitian cases \footnote{For non-Hermitian topological systems with PBC, there also exist gapless phases. For example, the 2D non-Hermitian Qi-Wu-Zhang(QWZ) model has gapless phases.}.

{\noindent \bf Example: 1D non-Hermitian SSH model with PBC.}-- For illustration, we focus on the 1D non-Hermitian SSH model with PBC
\begin{equation}
\begin{aligned}
H_p(k)=\vec{\bf d}\cdot \vec{\bf \sigma} = d_x\sigma_x + d_y\sigma_y, \quad d_x = t_1+t_2\cos k,\, d_y = t_2\sin k+ i\gamma.
\end{aligned}
\end{equation} 
The data set $\{{\bf x}^{(l)}| {\bf x}^{(l)} = \{\hat{\bf d}(k_i),| k_i =\frac{2i-N-2}{N}\pi, \, i\in[1,N] \}\}$ for clustering is obtained by varying only one parameter $t_1$, while fixing $t_2$ and the non-Hermitian term $\gamma$. Correspondingly, the $\mathbb{L}_1$-norm term in Eq.~(\ref{L1_approx}) reads
\begin{equation}\label{L1_approx_SSH_PBC}
\begin{aligned}
\frac{\|{\bf x}^{(l)}-{\bf x}^{(l+\delta l)}\|_{\mathbb{L}_1}}{\delta t_1} &\approx \sum_{i=1}^N \left(|\partial_{t_1}\,(\hat{d}_x^{(l)}(k_i))|+|\partial_{t_1}\,(\hat{d}_y^{(l)}(k_i))|\right)= \sum_{i=1}^N \left(\left|\frac{{d_y^{(l)}}^2}{{E_+^{(l)}}^3}\right|_{k_i}+\left|\frac{d_x^{(l)}d_y^{(l)}}{{E_+^{(l)}}^3}\right|_{k_i}\right).
\end{aligned}
\end{equation}
It is easy to verify that Eq.~(\ref{L1_approx_SSH_PBC}) tends to be infinite at the gap closure points $E_{\pm}=0$, i.e. the critical cases $t_1 = t_2\pm\gamma \,(k_i=-\pi)$ and $t_1 = -t_2\pm\gamma \,(k_i=0)$. Hence the kernel value $\mathcal{K}_{l,l+\delta l}\approx 0$ around such points and the kernel matrix becomes block diagonal. 
As a consequence, the diffusion map algorithm  successfully classifies the topological phases of the  1Dnon-Hermitian SSH model with PBC.

\subsection{Open boundary condition}

Here we show how to apply the diffusion map method to classify phases of non-Hermitian topological models with OBC.

For the case of OBC, the raw data is no longer the Hamiltonian vector in momentum space. Instead, we choose the real space projective matrix elements as the raw data.   Generally, a projective matrix of such topological model is defined as
\begin{equation}
P = \sum_{{\rm Re}[E_{m}]<0} |m_R\rangle\langle m_L|,
\end{equation} 
where $ |m_R\rangle$ and $\langle m_L|$ are the right and left eigenstates of the non-Hermitian model, $m$ covers the continuum bulk spectrum leaving out the discrete zero modes. 

We reiterate that the diffusion between the nearest samples contributes prominently. To illustrate the applicability of the diffusion map method in clustering real space data samples,  we consider the projective matrix $P^{(l+\delta l)} = P(\vec{t}^{\,(l+\delta l)})$ in first-order perturbation
\begin{equation}
P ^ {(l+\delta l)} = \sum_{{\rm Re}[E_{m}]<0}|m_R'\rangle\langle m_L'| \approx \sum_{{\rm Re}[E_{m}]<0}\left(|m_R\rangle+\sum_{n\neq m}\frac{\langle n_L|\delta \hat{H}|m_R\rangle }{E_m-E_n} |n_R\rangle\right)\left(\langle m_L|+\sum_{n\neq m}\frac{\langle m_L|\delta \hat{H}|n_R\rangle }{E_m-E_n} \langle n_L|\right),
\end{equation}
where $\delta \hat{H}=\hat{H}^{(l+\delta l)}-\hat{H}^{(l)}$ and then
\begin{equation}
\delta P = P ^ {(l+\delta l)} -P^{(l)} = \sum_{{\rm Re}[E_{m}]<0}(|m_R'\rangle\langle m_L'|-|m_R\rangle\langle m_L|) \approx \sum_{\substack{{\rm Re}[E_{m}]<0 \\ {n\neq m}}}\left(\frac{\langle n_L|\delta \hat{H}|m_R\rangle }{E_m-E_n} |n_R\rangle\langle m_L|+\frac{\langle m_L|\delta \hat{H}|n_R\rangle }{E_m-E_n}|m_R\rangle \langle n_L|\right).
\end{equation}
Hence the Gaussian kernel can be reexpressed as
\begin{equation}
\mathcal{K}_{l,l+\delta l}= \exp\left( -\frac{\|P^{(l)}-P ^ {(l+\delta l)}\|^2_{\mathbb{L}_1}}{2\epsilon N^2}\right)= \exp\left( -\frac{\|\delta P\|^2_{\mathbb{L}_1}}{2\epsilon N^2}\right)= \exp\left( -\frac{(\|\nabla_{\vec{t}} P\|_{\mathbb{L}_1}\cdot \delta \vec{t}\,)^2}{2\epsilon N^2}\right),
\end{equation}
where $\|\nabla_{\vec{t}} \,P\|_{\mathbb{L}_1} = \left(\|\partial_{t_1} P\|_{\mathbb{L}_1}, \|\partial_{t_2} P\|_{\mathbb{L}_1},\dots \|\partial_{t_n} P\|_{\mathbb{L}_1}\right)$. The singularity of $\|\nabla_{\vec{t}} P\|_{\mathbb{L}_1}$ is crucial to the kernel values.

\medskip

{\noindent \bf Example: 1D non-Hermitian SSH model with OBC.}-- Now we take the  1D non-Hermitian SSH model in Eq.~(\ref{app_nhssh_obc})) as an example. The data set $\{{\bf x}^{(l)}\}$ is obtained by varying $t_1$. %Due to the non-Hermiticity, Eq.~\ref{nabla_p} is no longer valid and the term $\|\nabla_{t_1} P\|_{\mathbb{L}_1}$ can be  rewritten as 
It is well known that the non-Hermitian SSH matrix $\hat{H}_o^{1D}$  in the orthonormal bases can be transformed into a Hermitian SSH matrix $\bar{H}_o^{1D}$ in the non-orthonormal bases for $|t_1|>|\gamma|$,
\begin{equation}
\bar{H}_o^{1D} = \Gamma^{-1}\hat{H}_{o}^{1D}\Gamma, \quad \Gamma^{-1}|n\rangle = |n_R\rangle, \quad \langle n|\Gamma = \langle n_L|,\quad \bar{H}_o^{1D}|n\rangle = E_n|n\rangle.
\end{equation}
where $\Gamma = {\rm Diag} (1, r, r, r^2,r^2,\cdots, r^{N-1}, r^N)$, $r=\sqrt{|(t_1-\gamma)/(t_1+\gamma)|}$. Then the term $\|\nabla_{\vec{t}} P\|_{\mathbb{L}_1}$ can be expressed as

\begin{equation}\label{nabla_p_nh}
\begin{aligned}
\|\nabla_{\vec{t}} P\|_{\mathbb{L}_1} =\|\partial_{t_1} P\|_{\mathbb{L}_1}=& \left\|\sum_{\substack{{\rm Re}[E_{m}]<0 \\ {n\neq m}}}\left(\frac{\langle n_L|\partial_{t_1} \hat{H}_{o}^{1D}|m_R\rangle }{E_m-E_n} |n_R\rangle\langle m_L|+\frac{\langle m_L|\partial_{t_1} \hat{H}_{o}^{1D}|n_R\rangle }{E_m-E_n}|m_R\rangle \langle n_L|\right)\right\|_{\mathbb{L}_1}\\
=&\left\|\sum_{\substack{{\rm Re}[E_{m}]<0 \\ {n\neq m}}}\left(\frac{\langle n|\Gamma\partial_{t_1} \hat{H}_{o}^{1D}\Gamma^{-1}|m\rangle }{E_m-E_n} \Gamma^{-1}|n\rangle\langle m|\Gamma+\frac{\langle m|\Gamma\partial_{t_1} \hat{H}_{o}^{1D}\Gamma^{-1}|n\rangle }{E_m-E_n}\Gamma^{-1}|m\rangle \langle n|\Gamma\right)\right\|_{\mathbb{L}_1}.
\end{aligned}
\end{equation}
Here for convenience, we calculate the $\mathbb{L}_1$-norm in the fermionic bases $\{C_{1,A},C_{1,B},C_{2,A},..., C_{N,B}\}$ instead of the eigenstate bases $\{|n\rangle\}$ of $\bar{H}$.  Similar to the Hermitian SSH model, the singularity of $\|\nabla_{\vec{t}} P\|_{\mathbb{L}_1}$ still occurs in condition $|E_{-1}-E_{1}|\rightarrow0$, which corresponds to the phase transition points $|t_1|=\sqrt{t_2^2+\gamma^2}$. %In Sec.~\ref{app_sec_detailed}
 In the next section,  we show in detail that the terms $\frac{\langle 1|\Gamma\partial_{t_1} \hat{H}\Gamma^{-1}|-1\rangle }{E_{-1}-E_1}$ and $\frac{\langle -1|\Gamma\partial_{t_1} \hat{H}\Gamma^{-1}|1\rangle }{E_{-1}-E_{1}}$ are both infinite, while all other associated parameters are finite.  Here to remove the possible exponential infinity of NHSE, we choose part of the $P$-matrix elements  as the raw input data: $\{P_{iA,iB}|i\in[1,N]\}$, i.e. the parameters of $C_{i,B}^\dag C_{i,A}$-terms in $\Gamma^{-1}|-1\rangle \langle 1|\Gamma$ and $\Gamma^{-1}|1\rangle \langle -1|\Gamma$ are finite numbers $\pm r$. Hence for each matrix element $P_{iA,iB}$, the value of the $\mathbb{L}_1$-norm tends to be infinite near the phase transition points $|t_1|=\sqrt{t_2^2+\gamma^2}$. As a consequence, the corresponding Gaussian kernel value tends to be zero, and there is not diffusion probability between the two samples in different phases.  It indicates the applicability of the diffusion map algorithm in non-Hermitian model with OBC.

\section{Detailed calculations of $\langle n | \Gamma \partial_{t_1}\hat{H}_o^{1D}\Gamma^{-1}|m\rangle/N$ in 1D non-Hermitian SSH model}\label{app_sec_detailed}

To manifest that the only singularity originates from the gap closure, we need to show that  the term $\langle n | \Gamma \partial_{t_1}\hat{H}_o^{1D}\Gamma^{-1}|m\rangle/N$ is finite for all bulk  eigenstates $\{|n\rangle\}$  of the 1D non-Hermitian SSH model($|t_1|> |\gamma|$) with OBC in Eq.~(\ref{app_nhssh_obc}).

In single fermion system, the eigenstate $|n\rangle$ can be expressed as  $|n\rangle=\sum_{i} \left(u_{ni}^AC^\dag_{i,A}+u_{ni}^BC^\dag_{i,B}\right)|{\rm Vac}\rangle$, $\sum_{i} \left(|u_{ni}^A|^2+|u_{ni}^B|^2\right)=1$. Then the term $\langle n | \Gamma \partial_{t_1}\hat{H}_o^{1D}\Gamma^{-1}|m\rangle/N$ takes the form 
\begin{equation}
\begin{aligned}
\frac{1}{N}\langle n | \Gamma \partial_{t_1}\hat{H}_o^{1D}\Gamma^{-1}|m\rangle &= \frac{1}{N}\sum_{i=1}^{N}\langle n |(r C^\dag_{i,A}C_{i,B}+r^{-1} C^\dag_{i,B}C_{i,A})|m\rangle,\\
&=\frac{1}{N}\sum_{i=1}^{N} r(u_{ni}^A)^*u_{mi}^B+r^{-1}(u_{ni}^B)^*u_{mi}^A
\end{aligned}
\end{equation}
which  is of order $r^{\pm1}$. Hence the term  $\langle n | \Gamma \partial_{t_1}\hat{H}_o^{1D}\Gamma^{-1}|m\rangle/N$ is finite for finite $r$.

Then we focus on the singularity of the term $\tfrac{\langle -1|\Gamma\partial_{t_1} \hat{H}^{1D}_o\Gamma^{-1}|1\rangle \Gamma^{-1}|-1\rangle\langle 1|\Gamma+ \langle 1|\Gamma\partial_{t_1} \hat{H}^{1D}_o\Gamma^{-1}|-1\rangle \Gamma^{-1}|1\rangle\langle -1|\Gamma }{N(E_{-1}-E_{1})}$.  Let the system be initially in topological trivial phase. Without loss of generality, we suppose the chain length $N$ is large enough, and the lowest excitation (annihilation) energy $E_{1}(E_{-1})$ of $\bar{H}_{o}^{1D}$ with OBC  is almost equivalent to that of  $\bar{H}_{o}^{1D}(k)$ with PBC, and the corresponding eigenstates can be approximately expressed by $$|\pm1\rangle\approx\bar{C}_{k_0,A(B)}^\dag |{\rm Vac}\rangle$$ for large $N$, where $k_0$ corresponds to the lowest excitation momentum of  $\bar{H}_{o}^{1D}(k)$ with PBC in Eq.~(\ref{HK_OBC}). By diagonalizing $\bar{H}_{o}^{1D}$ with Fourier transformation in the momentum bases $\{C^{\dag}_{k,A},\, C^{\dag}_{k,B}\}$, one obtains
\begin{equation}\label{SSH_bulk_state}
\bar{H}_k= \sum_{k=-\pi}^{\pi}E_k\bar{C}^\dag_{k,A}\bar{C}_{k,A}-E_k\bar{C}^\dag_{k,B}\bar{C}_{k,B}, \quad C_{k,A(B)}^\dag=\frac{1}{\sqrt{N}}\sum_{j=1}^N e^{-ikj}C_{j,A(B)}^{\dag},
\end{equation}
where $\bar{C}^\dag_{k,A}=\frac{1}{\sqrt{2}}(\xi_k C_{k,A}^\dag+C_{k,B}^\dag)$ and $\bar{C}^\dag_{k,B}=\frac{1}{\sqrt{2}}(-\xi_k^* C_{k,A}^\dag+C_{k,B}^\dag)$ are the fermionic quasiparticle operators, $\xi_k = \sqrt{(\bar{t}_1+t_2 e^{-ik})/(\bar{t}_1+t_2 e^{ik})}$ is just a phase for real $\bar{t}_1$ in Eq.~(\ref{HK_OBC}).

We first estimate the value of the term $\langle -1|\Gamma\partial_{t_1} \hat{H}^{1D}_o\Gamma^{-1}|1\rangle/N$,
\begin{equation}
\begin{aligned}
&\frac{1}{N}\langle -1|\Gamma\partial_{t_1} \hat{H}^{1D}_o\Gamma^{-1}|1\rangle \\
=&\frac{1}{2N}\sum_{k=-\pi}^{\pi}\langle{\rm Vac}|\bar{C}_{k_0,B}\left((\frac{r}{\xi_k}+\frac{\xi_k}{r}) (\bar{C}^\dag_{k,A}\bar{C}_{k,A}-\bar{C}^\dag_{k,B}\bar{C}_{k,B})+(\frac{r}{\xi_k}-\frac{\xi_k}{r}) (\bar{C}^\dag_{k,A}\bar{C}_{k,B} - \bar{C}^\dag_{k,B}\bar{C}_{k,A}) \right)\bar{C}_{k_0,A}^\dag |{\rm Vac}\rangle\\
=& \frac{1}{2N} (\frac{\xi_{k_0}}{r}-\frac{r}{\xi_{k_0}}).
\end{aligned}
\end{equation}
Correspondingly,  the term  $\langle 1|\Gamma\partial_{t_1} \hat{H}^{1D}_o\Gamma^{-1}|-1\rangle/N= \frac{1}{2N} (\frac{r}{\xi_{k_0}}-\frac{\xi_{k_0}}{r})$.

Then we analyze the singularity of the term $\Gamma^{-1}|-1\rangle\langle 1|\Gamma$. Recalling the definition of the $\mathbb{L}_1$-norm for the matrix $\|P\|_{\mathbb{L}_1}=\sum_{i,j}|P_{ij}|$, here we choose the  fermionic bases in real space $\{C_{1,A},C_{1,B},C_{2,A},..., C_{N,B}\}$. The term can be expressed as 
\begin{equation}
\Gamma^{-1}|-1\rangle\langle 1|\Gamma=\Gamma^{-1}\bar{C}_{k_0,B}^\dag |{\rm Vac}\rangle\langle {\rm Vac}|\bar{C}_{k_0,A}\Gamma=\frac{1}{2}\Gamma^{-1}(-\xi_{k_0}^* C_{k_0,A}^\dag+C_{k_0,B}^\dag) |{\rm Vac}\rangle\langle {\rm Vac}|(\xi_{k_0}^* C_{k_0,A}+C_{k_0,B})\Gamma.
\end{equation}
 It is easy to find that in the  fermionic bases,  $\Gamma$-matrix would contribute the exponential infinite term $\sim r^{N}$, hence we  choose the element set $\{C_{i,A}^\dag  |{\rm Vac}\rangle\langle {\rm Vac}|C_{i,B}|i\in[1,N]\}$ as the raw data instead of the full projective matrix $P$ to overcome the obstacle. In $\Gamma^{-1}|-1\rangle\langle 1|\Gamma$,   the term $C_{i,A}^\dag  |{\rm Vac}\rangle\langle {\rm Vac}|C_{i,B}$  has parameter $-\frac{\xi_{k_0}^*r}{2N}$, while in $\Gamma^{-1}|1\rangle\langle -1|\Gamma$, the $C_{i,A}^\dag  |{\rm Vac}\rangle\langle {\rm Vac}|C_{i,B}$ has parameter $\frac{\xi_{k_0} r}{2N}$. Thus for each matrix element $C_{i,A}^\dag  |{\rm Vac}\rangle\langle {\rm Vac}|C_{i,B}$, the total parameter contributed from the two energy levels $E_{\pm1}$  is 
 \begin{equation}\label{gs_correlation}
 \begin{aligned}
 &\left.\frac{\langle -1|\Gamma\partial_{t_1} \hat{H}^{1D}_o\Gamma^{-1}|1\rangle \Gamma^{-1}|-1\rangle\langle 1|\Gamma+ \langle 1|\Gamma\partial_{t_1} \hat{H}^{1D}_o\Gamma^{-1}|-1\rangle \Gamma^{-1}|1\rangle\langle -1|\Gamma }{N(E_{-1}-E_{1})}\right|_{C_{i,A}^\dag  |{\rm Vac}\rangle\langle {\rm Vac}|C_{i,B}}\\
 =& \frac{-\frac{\xi_{k_0}^*r}{2N}\langle -1|\Gamma\partial_{t_1} \hat{H}^{1D}_o\Gamma^{-1}|1\rangle + \frac{\xi_{k_0} r}{2N}\langle 1|\Gamma\partial_{t_1} \hat{H}^{1D}_o\Gamma^{-1}|-1\rangle  }{N(E_{-1}-E_{1})}\\
  =& \frac{-\frac{\xi_{k_0}^*r}{2N}\frac{1}{2N} (\frac{\xi_{k_0}}{r}-\frac{r}{\xi_{k_0}}) -\frac{\xi_{k_0} r}{2N}\frac{1}{2N} (\frac{\xi_{k_0}}{r}-\frac{r}{\xi_{k_0}})  }{N(E_{-1}-E_{1})}\\
  =& \frac{r(\xi_{k_0}+\xi_{k_0}^*)}{4N^3(E_{-1}-E_1)}(\frac{r}{\xi_{k_0}}-\frac{\xi_{k_0}}{r}).
 \end{aligned}
 \end{equation}
 The above term is non-zero when $\xi_{k_0}+\xi_{k_0}^*\neq0$ and $\frac{r}{\xi_{k_0}}-\frac{\xi_{k_0}}{r}\neq0$, which is usually satisfied by the non-Hermitian case($r\neq1$).

Eq.~(\ref{gs_correlation}) tends to be infinite when the system approximates to the phase boundary ($(E_{-1}-E_1)\rightarrow 0$). 

In conclusion, by analyzing the singularity of the values on $\{C_{i,A}^\dag  |{\rm Vac}\rangle\langle {\rm Vac}|C_{i,B}|i\in[1,N]\}$, we show that  the gap closure point (phase transition point) where $E_{\pm1}=0$ corresponds to the infinity of the $\mathbb{L}_1$-norm term in Gaussian kernel, which leads to the zero kernel value as well as the non-diffusion probability. In such sense, the choice of the raw data we have made here is reasonable and can be used for classifying the topological phases via the diffusion map method.
\section{More detailed numerical results of Fig. 2 and Fig. 3 in the main manuscript}
\begin{figure}
\centering
\includegraphics[width=\linewidth]{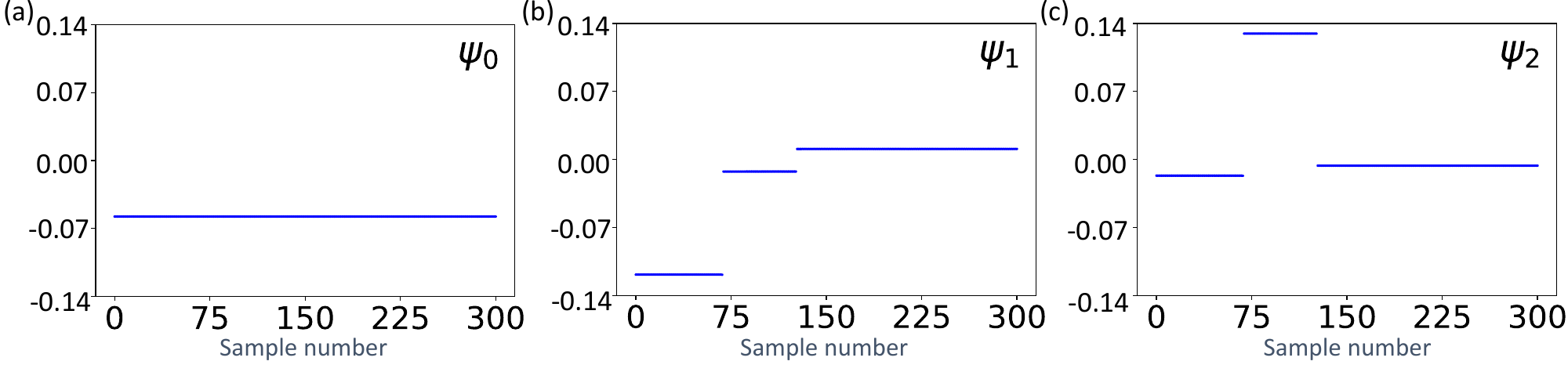}
\caption{{ 1D Non-Hermitian SSH model with PBC.} Parameters: number of unit cells $N=160$, $t_2=1$, non-Hermitian term $\gamma = 0.3$.  $t_1\in [0, 3.0695]$, samples interval $\delta t_1= 1/\pi^4$,  $\epsilon=0.001$. 
(a-c) Three right eigenvectors $\psi_{0,1,2}$ with the  corresponding eigenvalues $\lambda_{0,1,2} \approx 1$. The horizontal axis denotes the sample number,  and the vertical axis denotes the coefficients of each sample site in eigenvectors. It is easy to verify that the two jumping points in the figures exactly match with the theoretically predicted phase transition points $t_1 =0.7$ and $t_1 =1.3$.  The three types of topological phases  are clustered around  the points $(-0.0577, -0.1148, 0.0338)$ with 69 samples,  $(-0.0577, -0.0348, -0.1262)$ with 58 samples, $(-0.0577, 0.0120, -0.0049)$ with 173 samples %$(0.0577, -0.1148, 0.0328)$,  $(0.0577, -0.0340, -0.1261)$, $(0.0577, 0.0118, 0.0328)$ 
 in $\mathbb{R}^3$ respectively, which can be easily projected onto a 2D plane for visualization. The result matches the theoretically predicted phase transition points $t_1=0.7$ and $t_1=1.3$.  }
\label{NH_SSH_PBC_Evals}
\end{figure}

Here for details, we plot the eigenvectors of the one-step diffusion probability matrix $\mathcal{P}$ with the corresponding eigenvalues $\lambda_i\approx1$. We have the following four cases in total: the 1D non-Hermitian SSH and the 2D non-Hermitian QWZ models with PBC and OBC respectively, see Figs.~(\ref{NH_SSH_PBC_Evals}, \ref{NHSSH: OBC}, \ref{NH_QWZ_OBC_Evals}, \ref{NH_QWZ_PBC_M_Average}).

%\bibliographystyle{apsrev4-1.bst}
%%\bibliographystyle{unsrt.bst}
%\bibliography{DMNH_Man_R2}
%%\subsection{1D non-Hermitian SSH model in PBC and OBC}
%\newpage

%\newpage

\begin{figure}
\centering
\includegraphics[width=\linewidth]{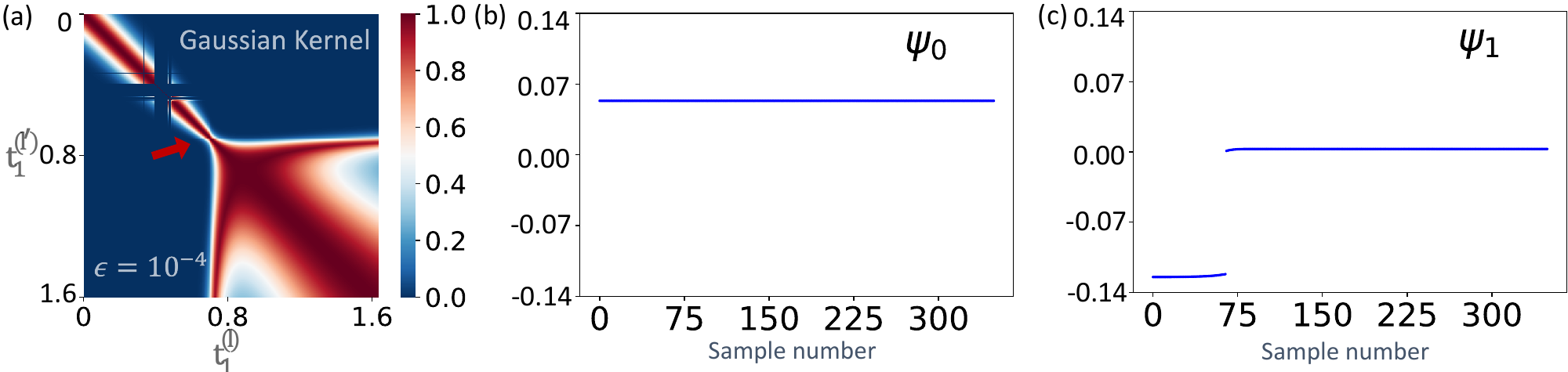}
\caption{
{ 1D Non-Hermitian SSH model with OBC. }Parameters: number of unit cells $N=40$,  $t_2=0.6$, non-Hermitian term $\gamma = 0.4$, samples interval $\delta t_1= 1/\pi^5$. (a) Heatmap for  Gaussian kernel matrix of samples ${\bf x}^{(l)}$ with varying $t_1\in [0,1.6306)$, hyper parameter $\epsilon=1\times 10^{-4}$. The non-diffusion zone around $t_1\approx 0.3921$  originates from the singularity  of parameter $r^{-1}=\sqrt{|(t_1+\gamma)/(t_1-\gamma)|}\rightarrow \infty$.  (b, c) Hyper parameter $\epsilon=1\times 10^{-6}$, for samples $\{{\bf x}^{(l)}\}$ with $t_1^{(l)}$ varying from 0.4902 to 1.6306, two right eigenvectors $\psi_{0,1}$ of $\mathcal{P}$  with the eigenvalues $\lambda_{0,1}\approx1$, which indicate the two different topological phases in non-Hermitian SSH model in OBC. In (b, c), the horizontal axis denotes the sample number,  and the vertical axis denotes the coefficients of each sample site in eigenvectors.  Each sample ${\bf x}^{(l)}$ with $N=80$ features can be mapped to the reduced two dimensional feature space $((\psi_0)_l,(\psi_1)_l)$. The 350 samples $\{{\bf x}^{(l)}\}$ with varying $t_1^{(l)}\in  [0.4902,1.6306]$ are clustered into two parts around the points (0.0535, -0.1223) with 65 samples and (0.0535, 0.0025) with 285 samples. The phase transition point can be directly observed from the eigenvectors. }
 %(a) Heatmap for  Gaussian kernel value distribution samples ${\bf x}^{(l)}$ with varying $t_1\in [0,1.6306)$. The non-diffusion zone around $t_1\approx 0.3921$  originates from the singularity  of parameter $r^{-1}=\sqrt{|(t_1+\gamma)/(t_1-\gamma)|}\rightarrow \infty$.   (b) Heatmap for  Gaussian kernel value distribution between samples ${\bf x}^{l}$ with varying $t_1\in [0.4902,1.6306)$. Red arrows point the phase boundaries. (c) Eigenvalues of one-step random walk matrix $\mathcal{P}$ for $t_1\in [0.4902,1.6306)$. Eigenvalues $\lambda _{0,1}\approx1$ indicate the two different topological phases in non-Hermitian SSH model in OBC.  (d) Scatter diagram of eigenvectors $\{\psi_0,\, \psi_1\}$ with the corresponding eigenvalues $\lambda_{0,1}\approx1$, where the samples are clustered into two topological phases.%(d), (e) Two right eigenvectors $\psi_{0,1}$ of $\mathcal{P}$ with the eigenvalues $\lambda_{0,1}\approx1$. Each sample ${\bf x}^{(l)}$ with $N=80$ features can be mapped to the reduced two dimensional feature space $((\psi_0)_l,(\psi_1)_l)$.    1D Non-Hermitian SSH model in PBC.  (a) Heatmap for Gaussian kernel value distribution between samples with varying $t_1$. (b) Eigenvalues of the one-step random walk matrix $\mathcal{P}$. (c) Scatter diagram of eigenvectors $\{\psi_1,\, \psi_2\}$ with the corresponding eigenvalues $\lambda_{1,2}\approx1$, where the samples are clustered into three topological phases. 
\label{NHSSH: OBC}
\end{figure}
%\subsection{2D non-Hermitian QWZ model}
%\begin{figure}
%\centering
%\includegraphics[width=\linewidth]{nh_qwz_pbc_evals.pdf}
%\caption{{ 2D non-Hermitian QWZ model with PBC.} Parameters: number of unit cells {\color{red}$N=30\times30=900$}, $t_x=t_y=0.5$, $v_x=v_y =0.5$, non-Hermitian terms $\gamma_x = 0.2$, $\gamma_y = 0.15$, $\gamma_z = 0$.  $M\in[-1.9607, 1.9574]$ with the interval $\delta t_1= 1/\pi^5$,  number of samples: 1200,  $\epsilon=0.00005$. (a) Heatmap of the Gaussian kernel  values. It is hard to identify the clustering blocks by eyes, but in (b-h), seven right eigenvectors $\psi_{0\sim6}$ of $\mathcal{P}$ with the eigenvalues $\lambda_{0\sim6}\approx1$ indicate the seven different topological phases in non-Hermitian QWZ model with PBC.  Each sample ${\bf x}^{(l)}$ with {\color{red}$3N =2700$} features can be mapped to the reduced seven dimensional feature space $((\psi_0)_l,(\psi_1)_l,\dots ,(\psi_6)_l)$. With such seven eigenvectors, the topological phases are clustered into seven parts. Numerical results show that  the samples are clustered into seven parts labeled by different $M$ domains:  $[-1.9607, -1.2516]$,  $[-1.2483,  -0.7516]$,  $[-0.7483,  -0.2516]$,  $[-0.2484, 0.2484]$,  $[0.2516, 0.7483]$,  $[0.7516, 1.2483]$, $[1.2516,  1.9574]$.}%$k$-means method can be applied on the reduced space, samples with the same topological phases are clustered together.}
%\label{NH_QWZ_PBC_Evals}
%\end{figure}
\begin{figure}
\centering
\includegraphics[width=\linewidth]{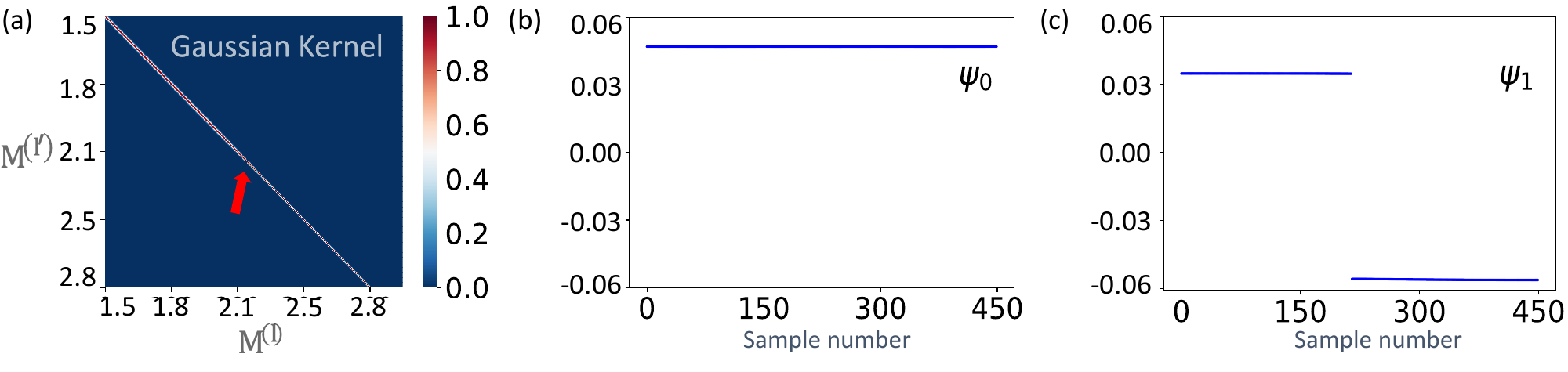}
\caption{{2D non-Hermitian QWZ model with OBC.} Parameters: number of unit cells $N=24\times24=576$, $t_x=t_y=1$, $v_x=v_y=1$, non-Hermitian term $\gamma_x =\gamma_y=\sqrt{5}/5$,  samples interval $\delta M= 1/\pi^5$, $M\in[ 1.4705, 2.9377)$, $\epsilon=5\times 10^{-9}$.   (a) Heatmap of the Gaussian kernel  values. The red arrow indicates the phase transition point.   (b, c) Two right eigenvectors $\psi_{0,1}$ of $\mathcal{P}$ with the eigenvalues $\lambda_{0,1}\approx1$, which indicate the two different topological phases in non-Hermitian QWZ model in OBC.  In (b, c), the horizontal axis denotes the sample number,  and the vertical axis denotes the coefficients of each sample site in eigenvectors. Each sample ${\bf x}^{(l)}$ with $N = 576$ features can be mapped to the reduced two dimensional feature space $((\psi_0)_l,(\psi_1)_l)$.   With such two eigenvectors, the 450 samples $\{{\bf x}^{(l)}\}$ with varying $M$ are clustered into two parts around two points: (0.0471, 0.0348) with 215 samples and (0.0471, -0.0560) with 235 samples. Each part corresponds to one topological phase. % $k$-means method is applied on the reduced space, samples with the same topological phases are clustered together.  
 }
\label{NH_QWZ_OBC_Evals}
\end{figure}

\section{Diffusion map in classifying phases with gapless band spectra}
\label{sec:vi}

In this work, we have studied the classification of the 2D NH QWZ model with PBC based on the diffusion map algorithm. Theoretically, for PBC the diffusion map cannot be directly utilized for clustering topological phases with the gapless bulk spectra, since the input data $\hat{d}(k)=\frac{\vec{\bf d}}{\sqrt{\vec{\bf d}^2}}$ could be singular at the gapless momentum points. The reason has been shown  in Sec.~\ref{sec:iii}, which demonstrates that the samples with the gapless point $(\vec{\bf d})^2=0$ lead to the zero diffusion probability, and the kernel matrix is separated into diagonal blocks by such samples. Adding
samples at exactly or near the gapless point could result in artificial zero diffusion probability, and consequently lead to more predicted phases than the true number of phases hosted by the system. % To correctly classify the phases with gapless bulk spectra, we choose judiciously proper parameter values to avoid the gapless momentum $\{k_x, k_y\}$ in the main manuscript. 
%One shortcoming of the above approach is that one need {\it a priori} knowledge of the system, and the advantage of unsupervised learning would be weakened. 

%To circumvent the obstacle posed by the samples with the gapless momenta, the diffusion map algorithm in classifying topological phases with the gapless bulk spectra,  here we provide a method that can circumvent the obstacle, which is posed by the samples within the gapless regions,  on diffusion map then propose an approach that can possibly resolve such issue. 
 
\begin{figure}
\centering
\includegraphics[width=\linewidth]{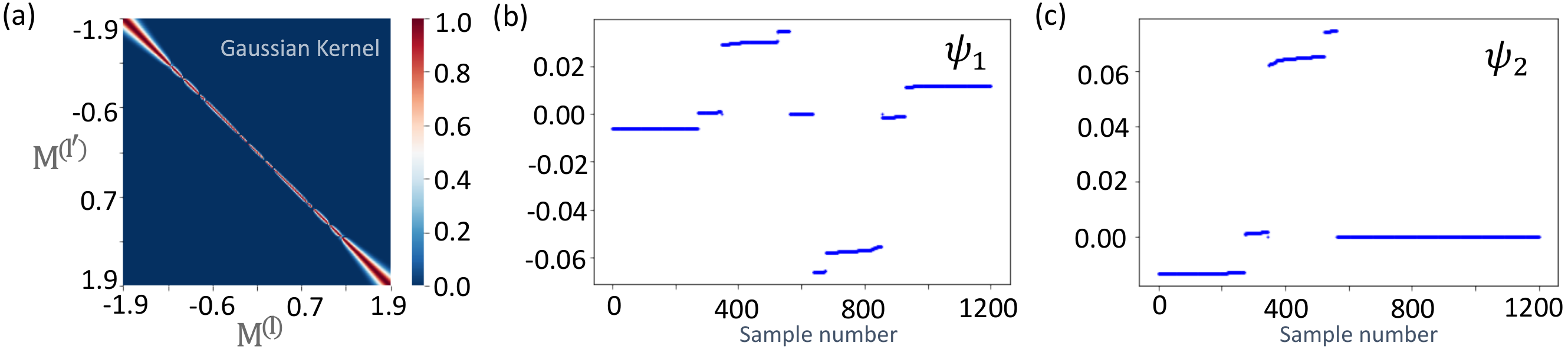}
\caption{2D non-Hermitian QWZ model with PBC.  Failed classification of samples with gapless momenta in the gapless regions of Hamiltonian. Parameters: number of unit cells $N=20\times20=400$, $t_x=t_y=0.5$, $v_x=v_y =0.5$, non-Hermitian terms $\gamma_x = 1/\sqrt{32}$, $\gamma_y =  1/\sqrt{32}$, $\gamma_z = 0$.  $M\in[-1.9607, 1.9574]$ with the interval $\delta t_1= 1/\pi^5$,  number of samples: 1200,  hyper parameter $\epsilon=0.001$.  The six theoretical phase boundaries should be $\{\pm 1.25,\,\pm 0.75, \pm 0.25 \}$.  (a) Heatmap of the Gaussian kernel values.  It is obvious that there are more than seven blocks, which is beyond the theoretical predicted clusters of phases. This is due to the existence of gapless momenta in discrete momentum configuration within the gapless phase regions.  (b-c) Two eigenvectors $\psi_{1,2}$ of diffusion matrix $\mathcal{P}$ with eigenvalues $\lambda_{1,2}\approx 1$. From the diagrams of eigenvectors, one observes that the samples are clustered into at least nine parts labeled by different $M$ domains: $[-1.9607, -1.0751]$, $[-1.0718, -0.8300]$, $[-0.8267, -0.2516]$, $[-0.2484, -0.1209]$, $[-0.1176, 0.1209]$, $[0.1242, 0.2484]$, $[0.2516, 0.8300]$, $[0.8333, 1.0718]$, $[1.0751, 1.9574]$. 
%More accurately, the samples can be classified into more clusters.
Owing  to the accuracy of the figures,  one can not directly observe the real phase boundaries $\pm1.25$ and $\pm0.75$ from the figures. Such phase boundaries can be located by checking the numerical data of the two eigenvectors $\psi_{1,2}$.
The artificial phase boundaries originate from the samples with the gapless momenta in the gapless phase regions.} 
\label{failure_detection_qwz}
\end{figure}

 We first introduce how to identify the artificial phase boundaries owing to the samples with the gapless spectra. Given the parameter region of a Bloch Hamiltonian, one can obtain different sample sets by choosing different lattice sizes. Although different lattice size leads to the different feature dimension of an input sample, the matrix dimensions of kernels for different lattice sizes are the same due to that the number of samples in each set are the same. The mechanism for detection is simple: If the gapless bulk spectrums only occur at the phase boundaries (usually with the gapless spectra at the momentum 0 or $\pi$), then the block boundaries of the kernel matrices should be same for different lattice sizes,  hence the diffusion map approach applies directly. While there exist phases with gapless spectra, like the 2D NH QWZ model, the appearance of momenta for the gapless bulk spectra would depend on the choice of lattice sizes (note that for different lattice sizes, the discrete momentum points are different after the Fourier transformation), i.e., whether the momenta for gapless bulk spectra is in the discrete momentum set. Hence the blocks of the corresponding kernel matrices may change with the varying lattice sizes, then the failure of the machine learning method  can be detected by directly comparing the classification results in different lattice sizes. At the phase transition point, the gapless momentum $k_0$ is usually 0 or $\pi$, whose appearance in discrete momentum configuration is independent of the lattice sizes. Those invariant block boundaries with varying lattice sizes are the real phase boundaries.  We take the 2D NH QWZ model as an example, detailed numerical calculations in Fig.~\ref{failure_detection_qwz} show that the artificial phases boundaries appear owing to the samples with the gapless momenta within the gapless phases region.

%One possible approach to resolve the issue mentioned above is to generate a set of kernel matrices by 
 
\begin{figure}
\centering
\includegraphics[width=\linewidth]{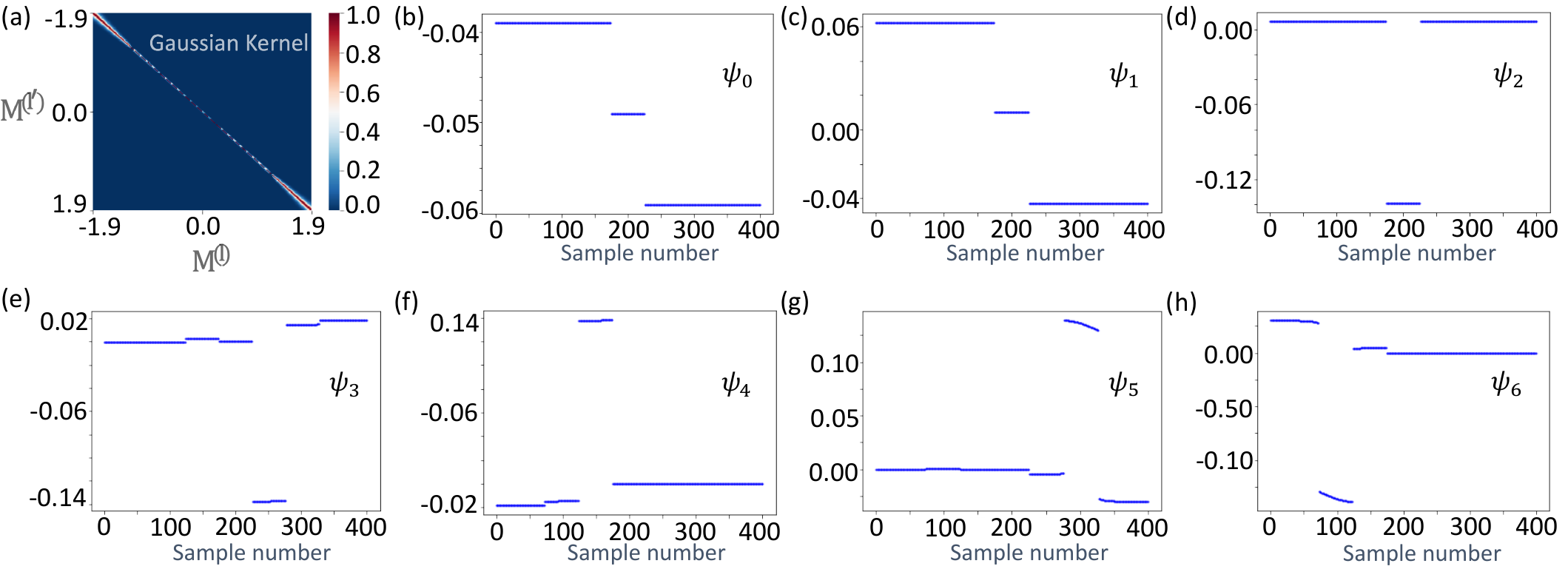}
\caption{2D non-Hermitian QWZ model with PBC. Successful classification of samples with  gapless momenta in the gapless regions of Hamiltonian. Parameters: number of unit cells $N=20\times20=400$, $t_x=t_y=0.5$, $v_x=v_y =0.5$, non-Hermitian terms $\gamma_x = 1/\sqrt{32}$, $\gamma_y =  1/\sqrt{32}$, $\gamma_z = 0$.  $M\in[-1.9607, 1.9509]$ with the interval $\delta t_1= 3/\pi^5$,  number of samples: 400,  hyper parameter $\epsilon=0.00005$.   (a)  Heatmap of the ``effective'' Gaussian kernel matrix, as an average of a set of seven kernel matrices with varying number of unit cells $N=\{14\times14, 16\times16, 18\times18, 20\times20, 22\times22, 24\times24, 26\times26\}$. (b-h) Seven eigenvectors $\psi_{0\sim6}$ of diffusion matrix $\mathcal{P}$ with the largest seven eigenvalues $\lambda_{0\sim6}\approx 1$. From the diagrams of eigenvectors, one obtains that the samples are clustered into seven parts labeled by different $M$ domains: $[-1.9607, -1.2548]$, $[-1.2450, -0.7549]$, $[-0.7451,-0.2549]$, $[-0.2451,  0.2451]$, $[0.2549,  0.7451]$, $[0.7549 ,  1.2450]$,  $[1.2548,  1.9509]$.
%More accurately, the samples can be classified into more clusters.
In comparison with Fig.~\ref{failure_detection_qwz}, the artificial phase boundaries here are eliminated by averaging the kernel matrices with different lattice sizes.} 
\label{NH_QWZ_PBC_M_Average}
\end{figure}

 To circumvent the above obstacle,  one can construct an ``effective'' Gaussian kernel matrix by simply averaging the kernel matrices of different input sample sets with varying lattice sizes (different discrete momentum configurations), so that the artificial zero diffusion probabilities become nonzero after the average.  The example of the 2D NH QWZ model is shown in Fig.\ref{NH_QWZ_PBC_M_Average}.
 
 %All of the sample sets share the same parameter regions, i.e., have the same number of samples. Each sample set corresponds to one kernel matrix. By varying the lattice size, we obtain a set of kernel matrices with the same matrix dimension. To circumvent the obstacle of those momenta for gapless bulk spectra on phase classification, one can take an average on the set of kernel matrices with different lattice size. This is supported by the fact that only the real phase transition points (usually with gapless spectra at momentum 0 or $\pi$) are independent of the lattice sizes, while the appearance of other momenta with gapless spectra in discrete momentum configurations would depend on the lattice size. 
 
 \section{Discrepancy of the diffusion map algorithm in predicting phase boundaries}
\begin{figure}
\centering
\includegraphics[width=0.5\linewidth]{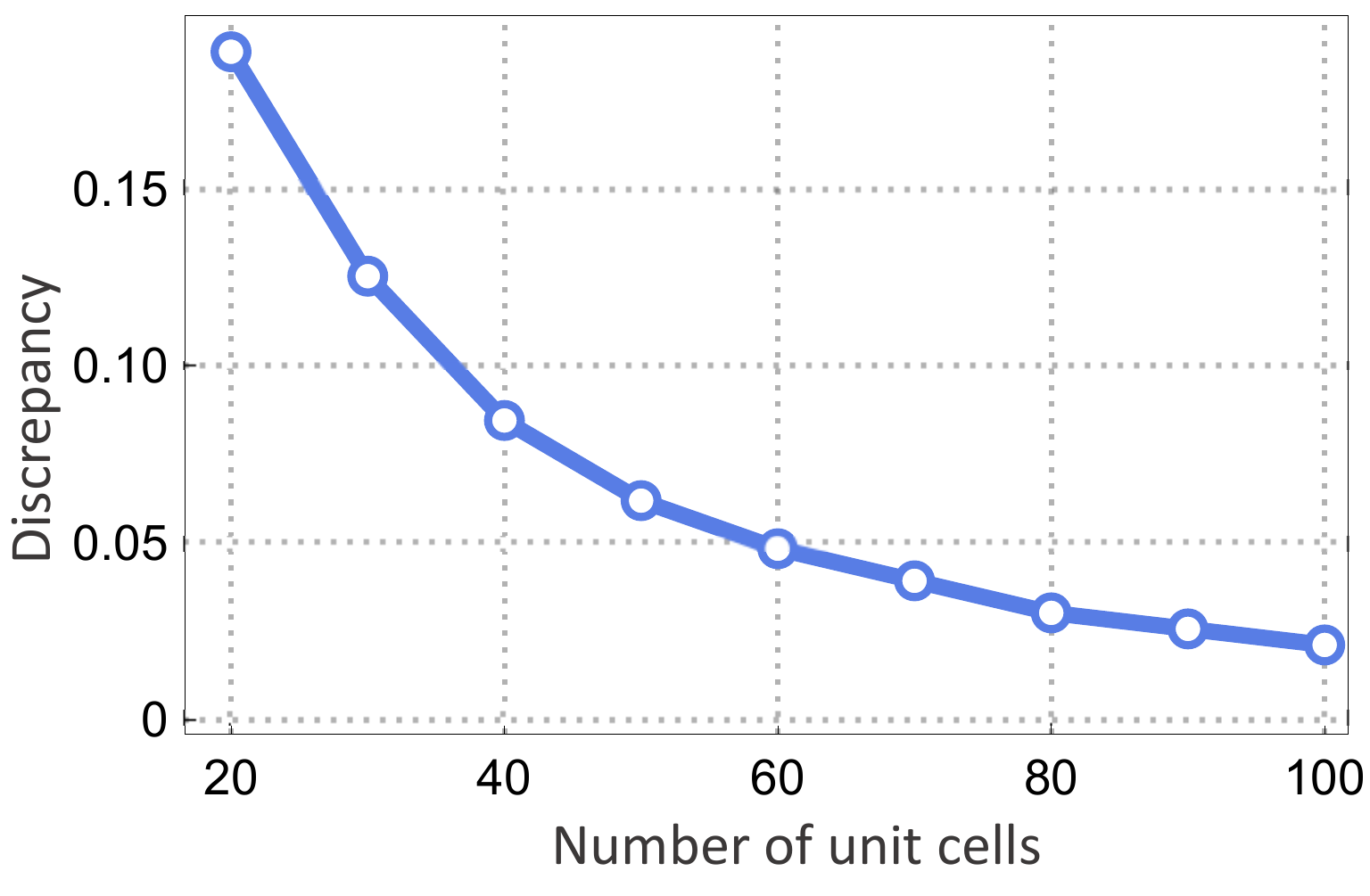}
\caption{Discrepancy between the learned phase transition point and the theoretical prediction of  the 1D NH SSH model with OBC. The discrepancy tends to be smaller with the number of unit cells increasing.  } 
\label{Discrepancy_SSH_OBC}
\end{figure}

 In locating the phase boundaries of models with PBC, the diffusion map algorithm performs high accuracy, i.e., the discrepancy between the learned phase boundaries and the theoretical ones is very small. In the case of OBC, the phase boundaries predicted by the unsupervised method behave  $\sim 1\%$ discrepancy. We conclude that the following reasons lead to the discrepancy. 
 \begin{itemize}
 \item[1)] One prominent reason for the discrepancy should be the finite size of the model we study here. The numerical calculated phase boundary approximates to the theoretical one when the size tends to be infinite. We carry out numerical calculations to support this point. For the 1D NH SSH model with OBC, we numerically obtain the input data with different model sizes for the diffusion map algorithm. We choose the same parameters as in the main manuscript with the theoretical phase transition point 0.7211. We obtain that when the number of unit cells N=60, the learned phase transition point is 0.6862; When N=80, the learned phase transition point is 0.6993; When N=100, the learned phase transition point is 0.7058. Fig.~\ref{Discrepancy_SSH_OBC} demonstrates that with the model size increasing, the discrepancy between the learned phase transition point and the theoretical prediction tends to be smaller.

\item[2)] 	Another reason should be that we choose only part of the projective matrix as the input raw data. For example, in the 1D NH SSH model with N unit cells, to circumvent the obstacle of NHSE, we just choose a small part, i.e., $N$ out of $4N^2$ projective matrix elements for each sample as the input data. This can be regarded as a trade-off between the compression of features and the training precision.  The similar trade-off also exists in another work \cite{PhysRevLett.122.210503} of the CNN-based supervised learning topological phases based on the experimental data. In that work, the authors showed that the trained CNN could successfully identify (with a probability $> 90\%$) different topological phases with less than 
 $10\%$ of the experimental data. 
 
\item[3)] In addition, for the 2D NH QWZ model with OBC, the theoretical predicted phase boundary itself is an approximation, which is based on the perturbation theory. In the original paper of the 2D NH QWZ model \cite{PhysRevLett.121.136802}, there already exists discrepancy ($\sim0.6\%$ for $M\approx2.2$) between the theoretical phase transition point and the numerical results. This also contributes to the discrepancy which appears in our learned results. 
\end{itemize}
 
 \section{Unsupervised learning of 2D NH QWZ model with PBC by varying other  parameters}
In the main manuscript, we successfully classify the non-Hermitian topological phases in an unsupervised fashion. For the 2D NH QWZ model with PBC,  the samples are obtained by varying the parameter $M$. Here for completeness, we utilize the diffusion map to classify samples generated by varying other parameters. To circumvent the obstacle posed by the samples with the gapless momenta within the gapless regions, here we choose the ``effective'' kernel matrix mentioned in Sec.~\ref{sec:vi} for the diffusion map algorithm.

\begin{figure}
\centering
\includegraphics[width=\linewidth]{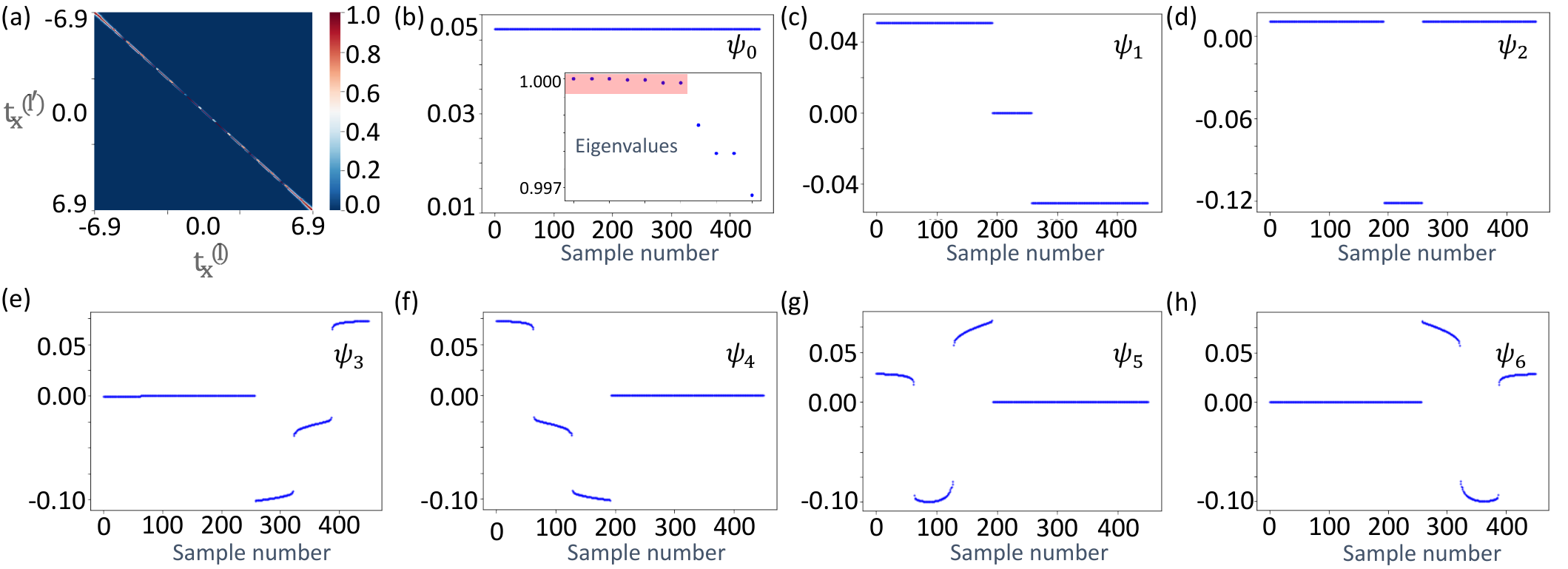}
\caption{2D non-Hermitian QWZ model with PBC, samples are generated by varying parameter $t_x$. Parameters:  $v_x=v_y=1$, $\gamma_x=0.6$, $\gamma_y=0.8$, $\gamma_z=0$, $M=2$ and $t_y=2$.  $t_x\in[-6.9295, 6.9295]$ with the interval $\delta t_1= 3/\pi^4$,  number of samples: 450,  hyper parameter $\epsilon=0.00015$.  (a)  Heatmap of the ``effective'' Gaussian kernel matrix, as an average of a set of five kernel matrices with varying number of unit cells $N=\{18\times18, 20\times20, 22\times22, 24\times24, 26\times26\}$. (b-h) Seven eigenvectors $\psi_{0\sim6}$ of the $P$-matrix with the largest eigenvalues $\lambda_{0\sim6}\approx 1$. (b) also contains the figure of  seven largest eigenvalues (red area) of the ``effective'' diffusion matrix $\mathcal{P} $. From the diagrams of eigenvalues and eigenvectors, one obtains that the samples are clustered into seven parts labeled by different $t_x$ domains: $[-6.9295,-5.0201]$, $[-4.9893,-3.0182]$, $[-2.9874,-1.0163]$, $[-0.9855,0.9855]$, $[1.0163,2.9874]$, $[3.0182,4.9893]$, $[5.0201,6.9295]$. The numerical calculated phase boundaries  coincide with the theoretical ones $t_x=\{-5,-3,-1,1,3,5\}$ with a low discrepancy.} 
\label{NH_QWZ_PBC_Tx}
\end{figure}

Firstly, we discuss the case of varying the parameter $t_x$ while fixing other parameters. Without loss of generality, one can set $v_x=v_y=1$, $\gamma_x=0.6$, $\gamma_y=0.8$, $\gamma_z=0$, $M=2$ and $t_y=2$.  Then from Eq.~\ref{2DNHQWZ} one can theoretically obtain seven phases, with the theoretical phase boundaries $t_x=\{-5,-3,-1,1,3,5\}$. Detailed numerical calculations based on the diffusion map method show that the seven different phases are successfully classified, see Fig.~\ref{NH_QWZ_PBC_Tx} for details.

\begin{figure}
\centering
\includegraphics[width=\linewidth]{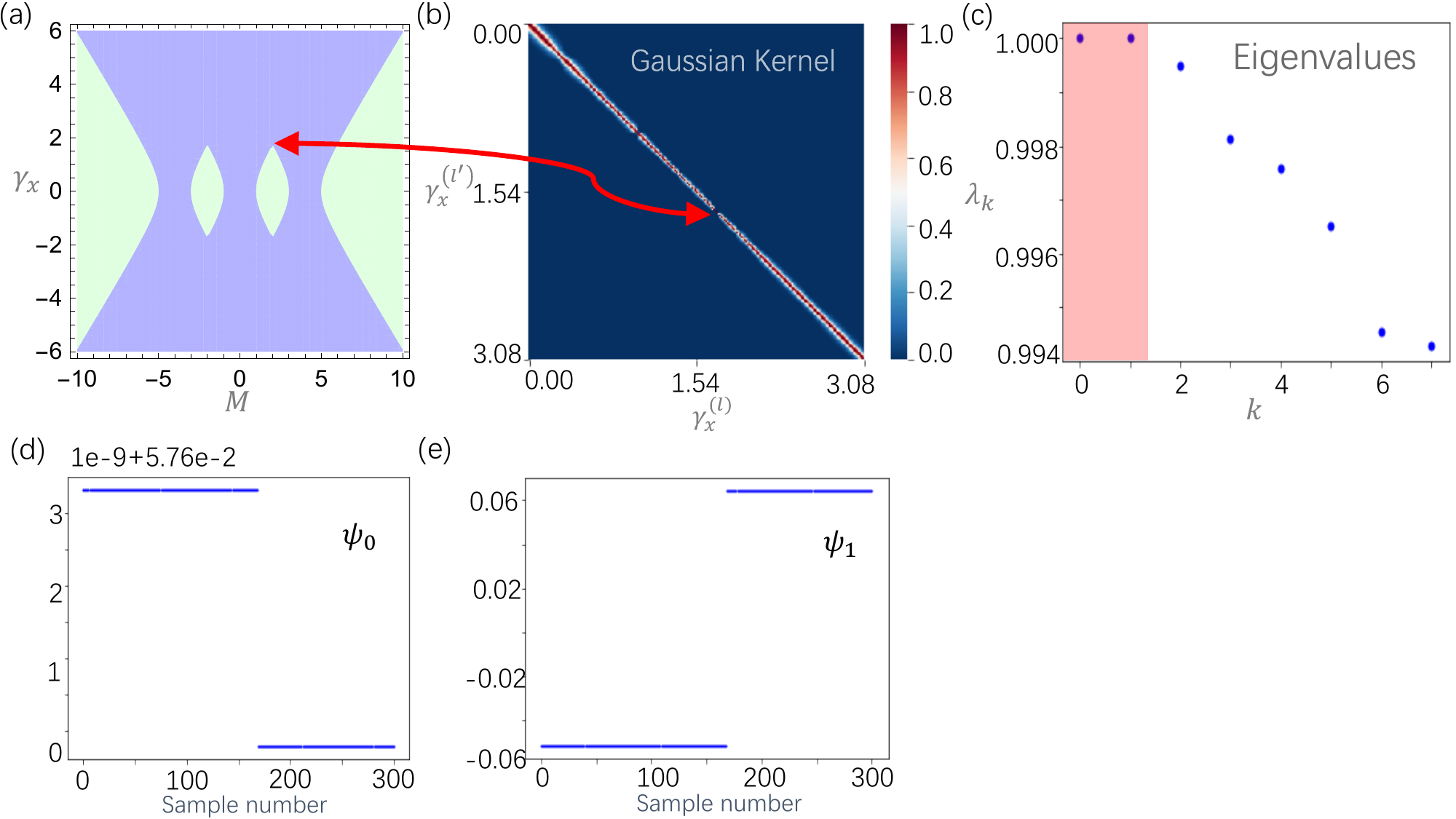}
\caption{2D non-Hermitian QWZ model with PBC, samples are generated by varying parameter $\gamma_x$. Parameters:  $t_x=t_y=2$, $v_x=v_y =0.5$, non-Hermitian terms $\gamma_y =  1$, $\gamma_z = 0$, $M=2$.  $\gamma_x\in[0, 3.0798]$ with the interval $\delta t_1= 1/\pi^4$,  number of samples: 300,  hyper parameter $\epsilon=0.001$.  (a) Theoretical phase diagram of 2D NH QWZ model with PBC, parameters $t_x=t_y=2$, $v_x=v_y =0.5$, $\gamma_y =  1$, $\gamma_z = 0$. The horizontal axis represents the value of parameter $M$, and the vertical axis represents the value of parameter $\gamma_x$. The green area represents the gapped phases with well defined topological indices, whereas the purple area  represents the gapless phase without well defined topological indices. In the parameter region $\{M=2, \, \gamma_x\in[0, 3.0798]\}$, there is only one phase boundary at $\gamma_x=\sqrt{3}\approx1.7321$. (b) Heatmap of the ``effective'' Gaussian kernel matrix, as an average of a set of six kernel matrices with varying number of unit cells $N=\{20\times20, 22\times22, 24\times24, 26\times26, 28\times28, 30\times30\}$. The red arrow line connects the corresponding phase boundary between the theoretical (a) and numerical results (b). (c) The largest eigenvalues of ``effective'' diffusion matrix $\mathcal{P}$. (d-e) Two eigenvectors $\psi_{0,1}$ of  $P$-matrix with the largest eigenvalues $\lambda_{0,1}\approx 1$. From the diagrams of eigenvectors, the samples are clustered into two parts labeled by different $\gamma_x$ domains: phase with the gapped spectra $[0, 1.7350]$ and phase with the gapless spectra $[1.7452, 3.0798]$. The numerical calculated phase boundary $\gamma_x\approx 1.7350$ coincides with the theoretical one $\gamma_x\approx 1.7321$ with a low discrepancy.} 
\label{NH_QWZ_PBC_Gamma_x}
\end{figure}

Then  we discuss the case of varying the parameter $\gamma_x$ and keeping other values fixed.   Numerical results show that the phases are successfully classified, see Fig.~\ref{NH_QWZ_PBC_Gamma_x} for details.

 \section{Discussion about the obstacle posed by NHSE in machine learning methods}

In this section, we would like to make a discussion about the universality of obstacle owing to NHSE for different machine learning methods. Recently, a number of unsupervised learning methods have been proposed to classify topological phases, such as clustering, variational autoencoders, divergence-based predictive method, learning by confusion, topological data augmentation, and so on \cite{rodriguez2019identifying,PhysRevLett.124.226401,che2020topological,PhysRevLett.124.185501,lidiak2020unsupervised,fukushima2019featuring,PhysRevE.99.062107,PhysRevResearch.2.013354,alexandrou2020critical,greplova2020unsupervised,arnold2020interpretable,Kottmann2020Unsupervised}. However, most of these methods are based on convolutional neural network (CNN) and have only been applied to Hermitian systems. For non-Hermitian systems with NHSE, since all the eigenstates are exponentially localized at the boundaries, the relevant features could be dramatically suppressed and consequently are hard to extract. This leads to a notable obstacle owing to the NHSE, independent of which learning algorithm is utilized. % A recent work on CNN-based learning of non-Hermitian topological phases supports this conclusion.
 In a recent work \cite{zhang2020machine}, the authors successfully predicted the topological phases of non-Hermitian SSH model in the momentum space with PBC based on the supervised CNN machine learning method. However, they noted that the CNN algorithm for learning the non-Hermitian models in the momentum space (with PBC) could not carry over to the non-Hermitian topological phases with the skin effect in the real space (with OBC), and further studies on NHSE and the classification of non-Hermitian topological phases with OBCs based on machine learning algorithms would be conducted. The unsupervised methods based on CNN would suffer the same obstacle owning to NHSE as for the diffusion map method.

\end{document}